 \documentclass[twocolumn,preprint2,trackchanges]{aastex62}
\hypersetup{linkcolor=red,citecolor=blue,filecolor=cyan,urlcolor=magenta}
\usepackage{aas_macros}
\usepackage{grffile}
\usepackage{amsmath}
\usepackage{amsfonts}
\usepackage{bm}
\graphicspath{{./}{figures/}}

\received{---}
\revised{---}
\accepted{---}
\submitjournal{ApJ}

\shorttitle{Wide-band Rotation Measure Synthesis}
\shortauthors{Pratley \& Johnston-Hollitt}
\begin{document}
\title{Wide-band Rotation Measure Synthesis}

\correspondingauthor{Luke Pratley}
\email{Luke.Pratley@gmail.com}

\author{Luke Pratley}
\affiliation{Dunlap Institute for Astronomy and Astrophysics, \\ University of Toronto, ON M5S 3H4, Canada}
\affiliation{Mullard Space Science Laboratory (MSSL),\\ 
University College London (UCL),\\ 
Holmbury St Mary, Surrey RH5 6NT, UK}

\author{Melanie Johnston-Hollitt}
\affiliation{International Centre for Radio Astronomy Research (ICRAR), \\
Curtin University,\\
1 Turner Ave., Technology Park, Bentley, 6102,	WA, Australia}

\begin{abstract}
	 Rotation measure synthesis allows the estimation of Faraday dispersion via a Fourier transform and is the primary tool to probe cosmic magnetic fields. We show this can be considered mathematically equivalent to the one dimensional interferometric intensity measurement equation, albeit in a different Fourier space. As a result, familiar concepts in two dimensional intensity interferometry designed to correctly account for a range of instrumental conditions can be translated to the analysis of Faraday dispersion. In particular, we show how to model the effect of channel averaging during Faraday reconstruction, which has to date limited the progress of polarimetic science using wide-band measurements. Further, we simulate 1d sparse reconstruction with channel averaging for realistic frequency coverages, and show that it is possible to recover signals with large rotation measure values that were previously excluded from possible detection. This is especially important for low-frequency and wide-band polarimetry. We extended these ideas to introduce mosaicking in Faraday depth into the channel averaging process. This work, thus provides the first framework for correctly undertaking wide-band rotation measure synthesis, including the provision to add data from multiple telescopes, a prospect that should vastly improve the quality and quantity of polarimetric science. This is of particular importance for extreme environments which generate high magnetic fields such as those associated with pulsars and Fast Radio Bursts (FRBs), and will allow such sources to be accurately used as probes of cosmological fields.	
\end{abstract}

% look up list of allowable keywords for this section

\section{INTRODUCTION}
\label{sec:intro}
Since the advent of radio interferometry in the 1940s \citep{Pawsey46,Ryle48} radio astronomers have built an impressive suite of interferomteric imaging techniques to allow signals from collections of antennas to be used collectively to image astronomical sources. As successive generations of interferometric arrays were built and operated, techniques were developed to obtain an estimate of the true sky brightness distribution and to correct for different instrumental effects inherent in the process. 
Among these methods are algorithms for deconvolution of the interferometric array response, so-called `CLEANing' techniques \citep{hog74,sch78,ste84}, algorithms to calibrate atmospheric and other phase rotations, and methods to account for wide-field and other direction dependent effects (DDEs) such as $w$-projection  \citep{Cornwell08,PJM19a} and $a$-projection \citep{Bhat08}. The focus of many of these developments has been total intensity imaging and the associated understanding of the Fourier transforms which are fundamental to aperture synthesis. 

While total intensity techniques have been well explored for decades, the situation for polarimetric imaging and signal interpretation has received considerably less attention. However, with cosmic magnetism as a unique science driver for next generation radio telescopes such as the Square Kilometre Array (SKA) \citep{mjh15}, as well as a focus for current instruments such as the Low Frequency Array (LOFAR; \citealp{vh13, Beck13}), the Murchison Widefield Array (MWA; \citealp{Tingay13,Lenc17}), and Jansky Very Large Array (JVLA; \citealp{VLA80,Mao14}), there has been renewed interest in polarimetry imaging algorithms \citep{Us16}. Additionally, the advent of wide-band correlators on instruments such as the Australia Telescope Compact Array (ATCA), JVLA and the Australian SKA Pathfinder (ASKAP) have presented both unique opportunities for polarimetric analysis \citep{Kac18,Pas18,And19} whilst at the same time presenting considerable challenges in interpretation of complex vectors which vary across these wide bands. 

It is well known that radio interferometry and the Faraday dispersion relation \citep{bur66} have the same measurement equation, with only minor differences of context. By interchanging three-dimensional $uvw$-coverage with one-dimensional channel coverage in $\lambda^2$ space, and sky coordinates with Faraday depth we can recast the standard interferometry measurement equation as the familiar expression for rotation measure synthesis. However, it is not well understood that many of the standard instrumental correction techniques which have been developed for interferometric imaging have analogous concepts in the context of rotation measure synthesis. As a result instrumental corrections and techniques like gridding and degridding, that are common place in radio intensity imaging, are not currently applied in the context of rotation measure synthesis \citep{bre05}. By using degridding and gridding with the projection algorithm from interferometric imaging, we show that it is possible to model and accurately correct for depolarization due to channel averaging in wide-band (also known as broadband) polarimetry. This greatly improves the accuracy of the recovered linear polarization intensity and rotation measure structure for sources with large rotation measure values. 

In this work, we review aspects of the radio interferometric measurement equation, and describe their analogous application in the context of rotation measure synthesis. 
We find that these concepts are key to understanding a wide variety of effects that are not yet corrected for in wide-band polarimetry. Hence we present the first framework for accurate wide-band rotation measure synthesis. Including the extension of the  framework to co-add data from multiple telescopes, a prospect that should vastly improve the quality and quantity of polarimetric science. 

In Section \ref{sec:measurement_equations} we link the interferometry and Faraday dispersion measurement equations. With Section \ref{sec:gridding} we introduce the degridding and gridding algorithm from interferometric imaging where it is described for calculating the Faraday dispersion measurement equation used in rotation measure synthesis. Section \ref{sec:chan_averaging} describes the $\delta \lambda^2$-projection algorithm for simulating and correcting channel averaging during Faraday dispersion reconstruction, and also verifies the accuracy of the method. In Section \ref{sec:simulations} we simulate wide-band and low frequency Faraday dispersion reconstruction with channel averaging, and show the impact of applying the corrections. We then suggest the possibility of mosaicking in Faraday depth in Section \ref{sec:mosaic} and conclude the work in Section \ref{sec:conclusion}.

\section{Interferometric measurement equation}
\label{sec:measurement_equations}
The interferometric measurement equation for a radio telescope can be represented by the following integral  
\begin{equation}
	y(\bm{u}) = \int x(\bm{l}) a(\bm{l}, \bm{u}){\rm e}^{-2\pi i \bm{l}\cdot \bm{u}}\, {\rm d}l\, ,
\end{equation} 
where $x$ is the sky brightness. $\bm{l}$ and $\bm{u}$ are the sky coordinates and baseline coordinates, respectively. $a$ includes direction dependent effects such as the primary beam and field of view (FoV) \citep{tho08}. The measurement equation is a mathematical model of the measurement operation that allows one to calculate model measurements $y$ when provided with a sky model $x$. Having such a measurement equation allows one to find a best fit model of the sky brightness, for a given set of (incomplete) measurements. Many techniques are available for inverting a measurement equation in an attempt to find a best fit model. These include traditional methods such as CLEAN and Maximum Entropy \citep{Ables74,Cornwell85}, and state of the art deconvolution methods such as Sparse Regularization algorithms \citep{LBCH11, Li11, Dabbech15, Garsden15, Sun15, LP18, Dabbech18,PJM19a}.

This measurement equation is typically approximated by a non-uniform fast Fourier transform (NUFFT type 2; \citealp{FS03}), since it reduces the computational complexity from $\mathcal{O}(MN)$ to $\mathcal{O}(MJ + N\log N)$. This process is traditionally known as degridding (gridding for the reverse process). The version of the measurement equation relevant in this work is represented by the following linear operations 
\begin{equation}
	\bm{y} = \bm{\mathsf{W}}\bm{\mathsf{G}}\bm{\mathsf{A}}\bm{\mathsf{F}}\bm{\mathsf{Z}}\bm{\mathsf{S}} \bm{x} + \bm{n}\, , 
\end{equation}
where $\bm{n}$ is Gaussian distributed noise and $\bm{y}$ are the weighted measurements.
Here $\bm{\mathsf{S}}$ represents a gridding correction (correction of apodisation due to averaging (convolving) visibilities off of a grid), $\bm{\mathsf{Z}}$ represents zero padding of the image, $\bm{\mathsf{F}}$ is a fast Fourier transform (FFT), $\bm{\mathsf{A}}$ represents baseline dependent effects such as variations in the primary beam and $w$-component, $\bm{\mathsf{G}}$ represents a sparse circular convolution matrix that interpolates measurements off the grid, and $\bm{\mathsf{W}}$ are weights applied to the measurements. This linear operator represents the application of the measurement equation, so is typically called a measurement operator $\bm{\mathsf{\Phi}} \in \mathbb{C}^{M \times N}$. 

In this case, $\bm{x}_i = x(\bm{l}_i)$ and $\bm{y}_i = y(\bm{u}_i)$ are discrete vectors in $\mathbb{C}^{N \times 1}$ and $\mathbb{C}^{M \times 1}$ of the sky brightness and visibilities, respectively.

Since the measurement operator is linear it has an adjoint operator $\bm{\mathsf{\Phi}}^\dagger$, which practically speaking, consists of applying these operators in reverse. Additionally, it is possible to represent these operators in matrix form, however, this is not always efficient or practical.

The back-projection, or `dirty map', which in the case of a radio interferometer is the convolution of the sky and instrument response (FoV and point spread function \footnote{Known in radio astronomy as the primary beam and synthesised beam, respectively.}), can be calculated from $\bm{\mathsf{\Phi}}^\dagger\bm{y}$, and the residuals from $\bm{\mathsf{\Phi}}^\dagger(\bm{\mathsf{\Phi}}\bm{x} - \bm{y})$.

\subsection{Faraday dispersion measurement equation}
\label{sec:RM}
Passing a linearly polarized electromagnetic wave through a magnetised plasma will result in rotation of the electric vector by an amount propositional to the strength of the magnetic field through which it passes. This effect, known as Faraday rotation, was first proposed for astronomical sources by \cite{CP62}, who used it to explain the observed wavelength dependence of the polarization position angel seen in Centaurus A. The phenomena can be described by: 
\begin{equation}
	\label{eq:rm_equation}
	\chi = \chi_{\rm int} + (\rm{RM})\lambda^2\, ,
\end{equation}
where the intrinsic position angle of the radiation $\chi_{\rm int}$ is in radians, $\lambda$ is the wavelength in metres and RM is the rotation measure in radians per metre squared. Additionally, the degree of rotation is given by the relation between the RM, line-of-sight magnetic field and the electron density such that:
\begin{equation}
	\rm{RM} = 8.1 \times 10^5 \int_0^L {\rm B}_{\|}(l) n_e(l) {\rm dl},
\end{equation}
where B$_{\|}$ is the line-of-sight component of the magnetic field in Gauss, n$_e$ is the electron density in cm$^{-3}$ and L is the path length in pc with the differential path length {\rm dl}. The integral can be separated for different Faraday screens, implying that the position angle measured will be the linear sum of all rotations along the line-of-sight. Thus, a more general form of Equation \ref{eq:rm_equation} is: 
\begin{equation}
	\chi = \chi_{\rm int} + \sum_i({\rm RM})_i\lambda^2.
\end{equation}

While it was initially common place to undertake rotation measure analysis in the $\lambda^2 -\chi$ plane, it has since been recognized that rotation measure synthesis \citep{bre05} is a better approach. This starts from the Faraday dispersion measurement equation
\begin{equation}
	y(-\lambda^2/\pi, -\delta \lambda^2/\pi) = \int x(\phi) a( \phi, -\delta \lambda^2/\pi){\rm e}^{2i \lambda^2 \phi} {\rm d}\phi
\end{equation}
where $y$ is the linear polarization as a function of wavelength squared, $x$ is the signal (linear polarized intensity) as a function of Faraday depth, and $a(\phi, -\delta \lambda^2/\pi)$ is the Faraday depth dependent instrumental effects. The main possibilities for Faraday depth dependent effects include the main focus of this paper, channel averaging and the frequency averaging response of the correlator known as spectral leakage. Each channel will have different amounts of vector averaging in $\lambda^2$ for a linearly polarized Faraday rotation measure signal. This results in different Faraday depth dependent sensitivity windows for each channel.

If we perform a change of coordinates, so that $u = -\lambda^2/\pi$ and $\delta u = -\delta \lambda^2 /\pi$, then we find
\begin{equation}
	y(u, \delta u) = \int x(\phi) a(\phi, \delta u){\rm e}^{-2\pi i u \phi} {\rm d}\phi \,
\end{equation}
which is identical to a 1d interferometric imaging measurement equation. We expect that $a(\phi, -\delta u) = a(\phi, |\delta u|)$.

Similarly, the above equation has the same series of operations when written as a linear operator
\begin{equation}
	\bm{y} = \bm{\mathsf{W}}\bm{\mathsf{G}}\bm{\mathsf{A}}\bm{\mathsf{F}}\bm{\mathsf{Z}}\bm{\mathsf{S}} \bm{x}\, , 
\end{equation}
however, now each operation is performed on a 1 dimensional Faraday depth signal $\bm{x}$ to map back to complex valued spectral channels $\bm{y}$. $\bm{\mathsf{A}}$ represents effects that occur in $\lambda^2$, such as varying channel widths. As such it should be clear that many concepts that have been well developed for imaging with interferometric arrays are also translatable to rotation measure synthesis and the analysis of Faraday dispersion. In the following sections, each of these operations are described in detail, and then placed in the context of rotation measure synthesis.

\section{Gridding and Degridding}
\label{sec:gridding}
In standard interferometry, measurements do not lie on a regular grid. As as result visibilities must be interpolated off and onto a grid when calculating the measurement equation during signal reconstruction. Degridding, also known as the NUFFT, is the process of applying the linear operators $\bm{\mathsf{G}}\bm{\mathsf{F}}\bm{\mathsf{Z}}\bm{\mathsf{S}}$. There are many works in the literature describing this process (see Section 4 of \cite{LP18} for a brief review). The zero padding, $\bm{\mathsf{Z}}$, (normally by a factor of 2) is to increase accuracy of degridding/gridding of visibilities, by up sampling in the Fourier domain. The choice of interpolation weights in $\bm{\mathsf{G}}$, known as the gridding kernel, affects the aliasing error, where ghost periodic structures can appear in the dirty map from outside the imaged region. An ideal gridding kernel would be a sinc interpolation kernel, which would cut any ghosting from the imaged region with a box function, but this has a large support \footnote{The support of a function is the subset of the domain containing those elements which are not mapped to zero.}. Well known kernels, such as prolate spheroidal wave functions (PSWF) and Kaiser-Bessel functions, are known to suppress the ghosting through apodisation while having minimal support on the Fourier grid \citep{LP18}. This apoidisation is then corrected for with the gridding correction $\bm{\mathsf{S}}$.

Importantly, the size of the cell in a grid is inversely proportional to the field of view, and the number of cells in a grid determine the resolution of the image. A formula to describe this can be found in \cite{PJM19a}. 

\subsection{Gridding and Degridding in Rotation Measure}
\begin{figure*}
\center
	\includegraphics[width=16cm]{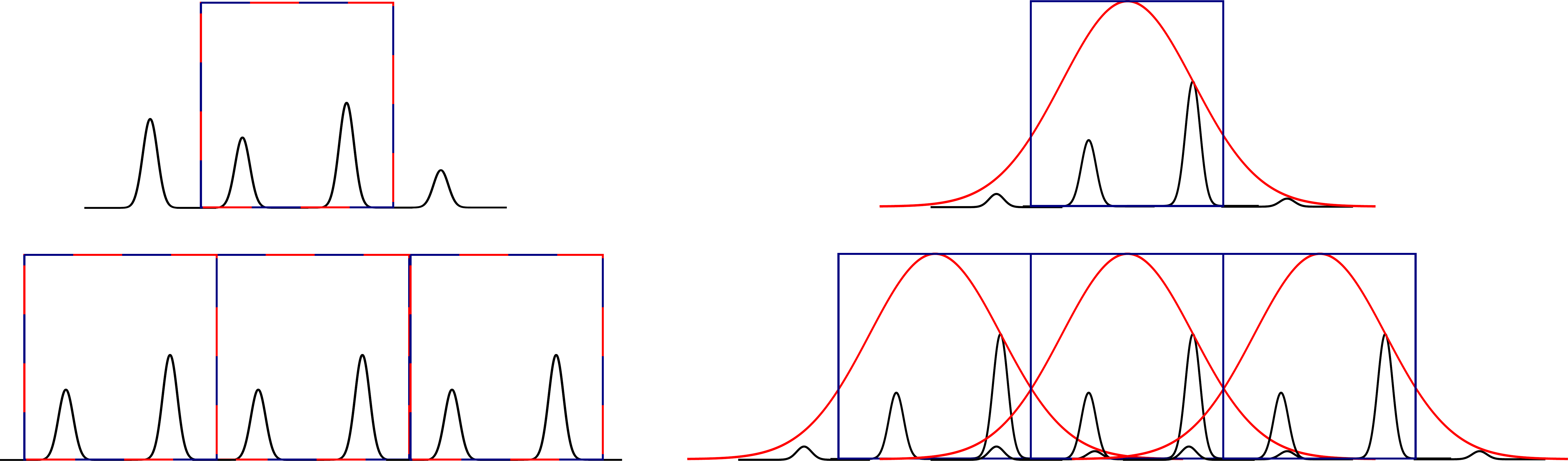}
	\caption{Schematic of the effect of aliasing in the Faraday depth due measurements in $\lambda^2$ not lying on a regular grid and restricting the region in Faraday depth. In both panels blue is region in Faraday depth of scientific interest, red is the window determined by the ideal anti-aliasing filter (gridding kernel). 
	(Left) shows a spectrum in Faraday depth over which only the window drawn is considered, where the window matches the reconstructed region. This window has a sharp cutoff. Performing the FFT enforces periodic boundary conditions that match the width of the window. However, enforcing this window via the Fourier domain when interpolating measurements off the grid is computationally intensive, since it would mean convolving the signal with a sinc function of large support. (Right) shows that if don't apply a sinc function as an anti-aliasing filter, the window does not match the region of reconstruction, allowing for sources and noise outside the region to add to the signal when interpolating measurements off the grid as denoted by the additional components in Faraday depth overlapping. The process of gridding allows a compromise to replace the ideal computationally expensive sinc function with an less expensive anti-aliasing filter to minimize the ghosting effects. More indepth discussion of aliasing error can be found in \citet{tho08}.}
	\label{fig:aliasing}
\end{figure*}

Degridding in the context of rotation measure is exactly the same as with interferometric imaging. In particular, if the region of Faraday depth imaged does not extend to the full sensitivity determined by the channel widths in $\lambda^2$, it is possible to see ghosting of structures from signal outside the Faraday depth region reconstructed \citep{tho08}. Figure \ref{fig:aliasing} provides a schematic to illustrate the effect. Ghosting results in both artificial signals and sidelobes in Faraday depth, which in practice increase the noise level of the complex polarization intensity as a function of radians per metre squared, decreasing sensitivity to physically meaningful RM structures. To minimise this, one can image the entire Faraday depth region determined by the channel sensitivity, or zero pad in Faraday depth and apply more apodization from the gridding kernel to reduce aliasing error. This is equivalent to applying an anti-aliasing filter to suppress the aliasing error. 

The coordinates of the FFT grid and wavelength for each dimension are determined by $\lambda^2 = \lambda^2_{\rm pix}\Delta \lambda^2$ with 
\begin{equation}
    \Delta \lambda^2 = -\frac{\pi}{\alpha  N  \Delta \phi}
\end{equation} 
where $\alpha$ is the oversampling ratio, $N$ is the image width in pixels, and $\Delta \phi$ is the size of a pixel in the appropriate units for $u$. $\Delta \phi$ has the units of rad/m$^2$ and $\lambda^2$ has units of $m^2$. 
With this relation, we find that $u = u_{\rm pix} \Delta u$, where $u_{\rm pix}$ has coordinates of the Fourier grid and 
\begin{equation}
    \Delta u = \frac{1}{\alpha  N  \Delta \phi}\, .
\end{equation} 
We have chosen the definitions such that $ |u_{\rm pix}| = |\lambda^2_{\rm pix}|$.

If we want the so-called `dirty map' to be Nyquist sampled, we required that $\alpha N \Delta \lambda^2 \geq 2\lambda^2_{\rm max}$. We can use this formula above to estimate a value of $\Delta \phi$ by combining it with the equations above. The largest $\Delta \phi$ that is expected to work based on the wavelength coverage is 
\begin{equation}
    \Delta \phi = -\frac{\pi}{2\lambda^2_{\rm max}}\, ,
\end{equation}
or
\begin{equation}
    \Delta \phi = \frac{1}{2u_{\rm max}}\, .
\end{equation}

In this work, we use the Kaiser-Bessel gridding kernel with normalization constants that have the relation $\mathcal{F}^{-1}\{ G(u)\}(x) s(x) = 1$. For the Kaiser-Bessel kernel, this is ensured with the equations
\begin{equation}
    G(u_{\rm pix}) = \frac{I_0\left [\beta \sqrt{1 - 4 u_{\rm pix}^2 / J^2} \right ]}{I_0(\beta)}
    \label{eq:normed_kb_kernel}
\end{equation}
with $G(u_{\rm pix}) = 0$ when $|u_{\rm pix}| > J /2$ and
\begin{equation}
    s(x) = \left[\frac{J}{I_0(\beta)}\frac{\sin \sqrt{(\pi x J)^2 - \beta^2 }}{\sqrt{(\pi x J)^2 - \beta^2 }}\right]^{-1}
\end{equation} 
where $J$ is the support size of the kernel in pixels, $I_0$ is the modified Bessel function, and $u_{\rm pix}$ is in units of pixels, and we typically determine the width of the kernel by choosing $\beta = 2.34 J$, as it has shown to perform well \citep{FS03}.

Because we are working with 1d Fourier space, the degridding matrix $\bm{\mathsf{G}}$ is constructed from a kernel $G(u)$ by \citet{LP18}:
\begin{equation}
	\bm{\mathsf{G}}_{i, \{k_i + j\}_K} = G({u_{\rm pix}}_i - (k_i + j)) {\rm e}^{- \pi i (k_i + j)} \, ,
\end{equation}
where $i$ is the index of the measurement $\bm{y}_i$, $k_i$ is the closest integer to visibility coordinate $u_i - J/2$ (in units of pixels and can be found using $k_i = {\rm floor}(u_i - J/2)$), and $j = 1 \dots J$ are the possible non-zero entries of the kernel. The modulo-$K$ function is denoted by $\{ \cdot \}_K$, where $K = \alpha N$ is the dimension of the Fourier grid in one dimension. The complex exponential is a phase shift that may or may not be needed to shift the centre of the signal so that $\phi = 0$ occurs at $N/2$ (this is discussed in the context of imaging in \cite{PJM19a}).

Finally, the gridding correction operator $\bm{\mathsf{S}}$ is calculated using
\begin{equation}
	\bm{\mathsf{S}}_{i,i} = s\left(\frac{i}{K} - \frac{1}{2}\right)\, ,
\end{equation}
where $s(x)$ is the reciprocal of the inverse Fourier transform of $G(u)$. 
	
\section{Channel Averaging and Measurement Dependent Effects}
\label{sec:chan_averaging}
In the imaging domain it is increasingly common to consider measurement dependent effects such as different primary beams, calibration errors, and the $w$-component. All of these can be modeled as a multiplication in the image domain, which is correspondingly a convolution in the Fourier domain. Methods such as the $w$-projection  \citep{Cornwell08,PJM19a} and $a$-projection \citep{Bhat08} take the baseline and direction dependent effects from the image domain, and linearly convolve them with the gridding kernel in the Fourier domain. This allows linear mathematical modeling of baseline dependent DDEs inside the measurement operator.

One particular example that is relevant to this work is different primary beams across an array. The sensitivity of an antenna in the far field is determined by the averaging process of the incoming electric field across the antenna. Simply put, the primary beam pattern is the Fourier transform of the complex gains across the aperture of the antenna, therefore a smaller aperture will have a wider field of view. If the response across the aperture is constant (i.e. a box or a top hat function), the primary beam will be a sinc function. However, if the array uses different antennae patterns (such as the case for arrays comprised of different sizes antennas like the Atacama Large Millimeter Array (ALMA)), or if the antennae rotates relative to the source as a function of time, the $a$-projection can be used to model these effects in the measurement equation.

The analogous situation with rotation measure is to consider channel dependent effects in Faraday depth. The most critical example is the different channel widths when averaging in $\lambda^2$ for wide-band observations. For example Faraday depth dependent effects from the instrument or ionosphere, or from calibration errors, may be present in the data. At present these are left untreated in the data, however, it is entirely possible to model such effects in the analogous way as done in interferometric imaging.

\subsection{The projection algorithm for rotation measure synthesis}
In this section we describe the implementation of an analogous concept to the $a$-projection algorithm in interferometric imaging to rotation measure synthesis. Just as the $a$-projection algorithm models the averaging of visibility data due to the physical size of each antenna, the new $\delta \lambda^2$-projection algorithm models the averaging in $\lambda^2$ due to the spectral averaging of the correlator. It follows that channel averaging in $\lambda^2$ creates a Faraday dispersion sensitivity pattern analogous to the primary beam pattern of an antenna in interferometric imaging.

The core idea of the projection family of algorithms is that an image domain multiplication can be performed using convolution in the Fourier domain. The interpolation in degridding is a convolution for each measurement, this allows a different image domain multiplication to be performed for each measurement, and is efficient when the projection kernel has a small support size in the Fourier domain. 

Since the convolution with gridding kernels is already measurement dependent we find
\begin{equation}
		y(u, \delta u) = \int \left[\frac{x(\phi)}{g(\phi)}\right] 
		 g(\phi)a(\phi; \delta u){\rm e}^{-2\pi i \phi u}\,  {\rm d}\phi \, ,
\end{equation}
 where $g$ is the window function for the anti-aliasing interpolation kernel and $a(\phi; \delta u)$ is the Faraday depth sensitivity window determined by the channel width $\delta u$. We denote the Fourier transform of $g(\phi)$ and $a(\phi;\delta u)$ as $G(u)$ and $A(u, \delta u)$, respectively. This suggests that we can define a new convolutional kernel that includes the channel averaging effects in Faraday depth when interpolating measurements in the Fourier domain. This convolutional kernel reads
\begin{equation}
	\left[GA\right](u, \delta u)= G(u) \star A(u, \delta u)\,
\end{equation}
and we can rewrite the measurement equation as
\begin{equation}
	y(u, \delta u) = \tilde{y}(u, 0) \star \left[GA\right](u, \delta u)\, ,
	\label{eq:projection_gridding}
\end{equation}
where $\tilde{y}(u, 0)$ is now the Fourier transform of the gridding corrected Faraday depth  
\begin{equation}
	\tilde{y}(u, 0) = \int \frac{x(l, m)}{g(l, m)}
	 {\rm e}^{-2\pi i u \phi}\,  {\rm d}\phi \, .
\end{equation} 

\subsection{Channel averaging}
Due to frequency averaging of the correlator for each channel $\delta\nu$, we also have averaging in $\lambda^2$, $\delta \lambda^2$.
A finite channel width provides the following relations 
\begin{equation}
	\lambda_i^2 =  \frac{1}{2}\left(\left[ \frac{c}{\nu_i - \delta\nu_i/2}\right]^2 +\left[\frac{c}{\nu_i + \delta\nu_i/2} \right]^2 \right)
\end{equation} 
 and
 \begin{equation}
	\delta\lambda_i^2 =  \left[ \frac{c}{\nu_i - \delta\nu_i/2}\right]^2  - \left[\frac{c}{\nu_i + \delta\nu_i/2} \right]^2  \, .
\end{equation} 
For a correlator with no spectral leakage\footnote{The true frequency response is determined by the type of correlator, see \citep{pri16} for examples.}, the ideal window function in $\lambda^2$ is a box function of width $\delta \lambda^2$
\begin{equation}
	A(u, \delta u) = 
	\begin{cases} 
		1/\delta u           & |u| < \delta u /2 \\
		0\,
	\end{cases}
\end{equation}
We can then calculate $GA = G \star A$ either directly in the Fourier domain or in the image domain by using the convolution theorem and a Fourier transform. In this case the Fourier domain is likely to be very efficient when using adaptive quadrature, since $G$ is smooth and $A$ is a change of integration bounds. The total support of $GA$ is $J + |\delta \lambda^2/\Delta \lambda^2|$, since $G$ has a total support of $J$ in pixels. 

However, in this work, we use the convolution theorem to perform the calculation of $GA$ using the image domain representations. It follows that
\begin{equation}
	\begin{split}
		[GA](u_{\rm pix}, \delta u, \Delta u) =
		\int_{-\alpha/(2\Delta u)}^{\alpha/(2\Delta u)}
		g(\Delta u \phi)a(\phi, \delta u)\\
		\times{\rm e}^{-2\pi i \Delta u u_{\rm pix}l} {\rm d}\phi\, .
	\end{split}
	\label{eq:kernel_calculation}
\end{equation}
where $a$ is a sinc function in the image domain with a width determined by $\delta \lambda^2$
\begin{equation}
	a(\phi, \delta u) = \frac{\sin(\pi \delta u \phi)}{\pi \delta u \phi}\, .
	\label{eq:chan_sensivity}
\end{equation} 

From this equation, we find $a$ the sensitivity in $\phi$ as a function of channel width, which can be used to calculate how large a region in Faraday depth we need to be image to reach the full sensitivity, i.e. the value of $N \Delta \phi$. This can be done by determining the largest value of $\delta u$, and then imaging to an acceptable value of $a(N \Delta \phi, (\delta u)_{\rm max})$. This could be the full width at half maximum (FWHM) of the sinc function or the first null or first side-lobe. The solution of $\sin(x)/x = 0.5$ is $x \approx 1.895\dots$ in radians, and it follows the total region of sensitivity is determined by $\phi = \pm \frac{1.895}{\pi \delta u}$ which is $\phi = \pm \frac{1.895}{\delta \lambda^2}$. For a small channel width the FWHM is larger, making it clear that there is more sensitivity to large values of $\phi$. However, the first null for the sensitivity of a channel in Faraday depth can be calculated from $\phi = \pm \frac{\pi}{\delta \lambda^2}$.

The model of $a$ as a box function in $\lambda^2$ is simplified and used for some approximate calculations in \citet{bre05}. Note that the box function for averaging is still only an approximation and improvements could be made by using a more accurate mode of the correlator response. This is akin to improvements in the primary beam model in interferometric imaging.

\subsection{Kernel calculation and verification}
In this section we verify the accuracy of using the projection kernels calculated by integrating with adaptive quadrature. We use the same calculation method that has been previously used for wide-field interferometric imaging to generate $w$-projection kernels in \citet{PJM19a}. We use the Cubature software package\footnote{This can be found at \url{https://github.com/stevengj/cubature}.} to perform $p$-adaptive quadrature to an absolute and relative accuracy of $10^{-6}$. For more details on calculating projection kernels, see \citet{PJM19a}.

\begin{figure}
	\center
	\includegraphics[trim=30 0 0 0,clip, width=9cm]{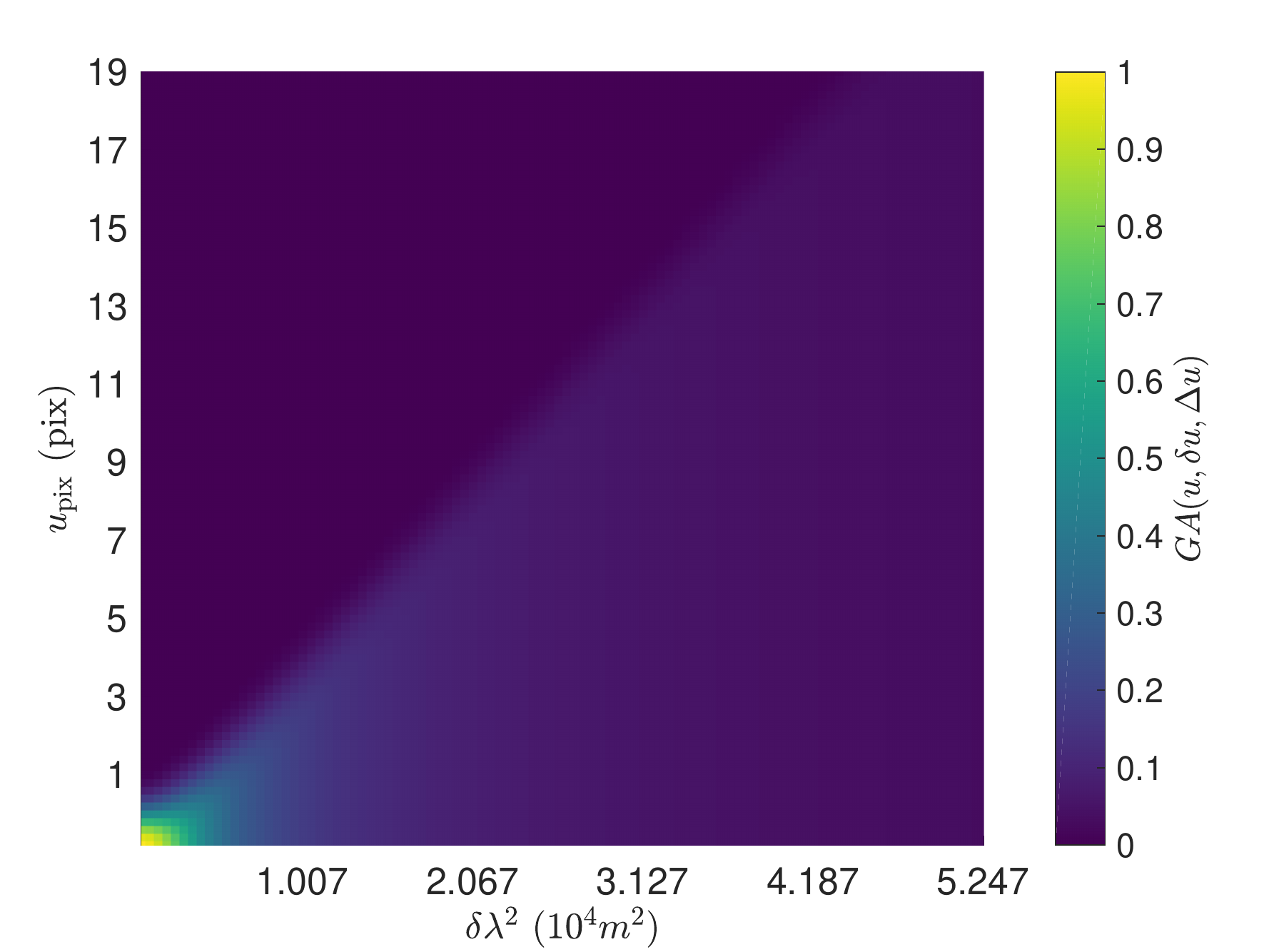}
	\includegraphics[trim=30 0 0 0,clip, width=9cm]{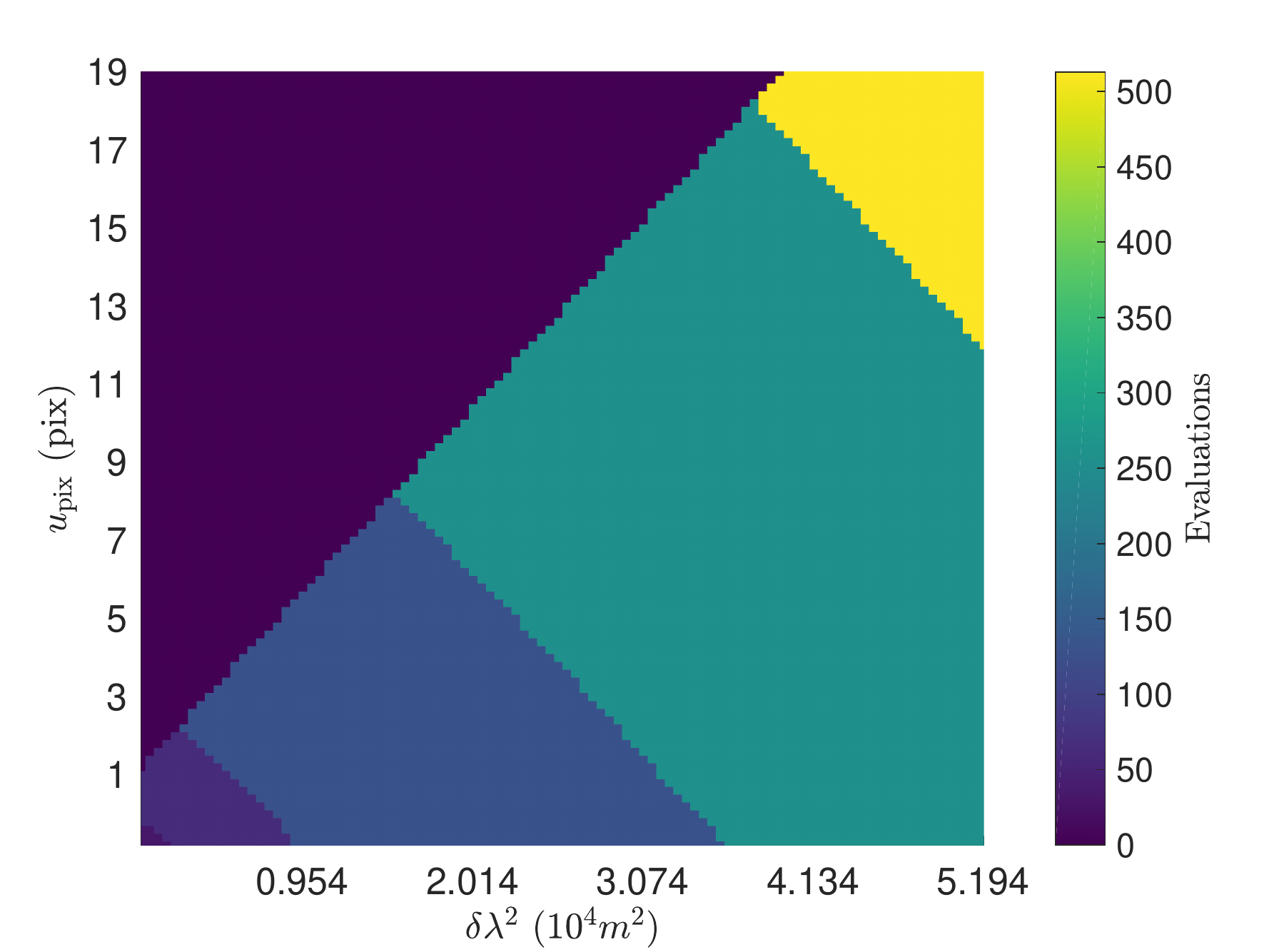}
	\caption{Projection kernel calculations were performed using the Kaiser-Bessel kernel with a support size of $J = 4$ and a $\Delta u$ calculated from the equivalent of using $N = 16,384$ and $\Delta \phi = 8.56395$ rad/$m^2$. The kernel size is determined by $J + \delta u/\Delta u$. (Top) shows the kernel calculations for different channel widths (with $|\delta \lambda^2| = \pi |\delta u|$) as a function of pixel coordinates for a kernel centred at zero. (Bottom) shows the number of function evaluations used in numerical quadrature for each value, we find that the computation required increases with support size and channel width size.}
	\label{fig:kernel_plots}
\end{figure}

In Figure \ref{fig:kernel_plots} we calculate $[GA](u_{\rm pix}, \delta u, \Delta u)$ up to $J + |\delta \lambda^2/\Delta \lambda^2|$ as a function of $\delta u$ and $u_{\rm pix}$ for fixed $\Delta u$. We can see how the support size of the kernel increases as a function of both $\delta u$ (i.e. large channel width) and small $\Delta u$ (i.e. imaging a large region in Faraday depth, $\phi$). We also see how many function evaluations are required for each value calculated using adaptive quadrature, and find it increases with $\Delta u$, $\delta u$, and $u_{\rm pix}$. The amplitude of the kernel decreases as a function of $\delta u$, this is due to $1/\delta u$ in the box function. The kernel is expected to be real valued, and we numerically find that the imaginary component is zero.

We find that the calculation time is negligible, especially when compared to projection kernels calculated for interferometric imaging. This is because of: i) number of channels ranges in the 100's to 10,000's, ii) the 1d support, and iii) the limited support size. Furthermore, memory taken up by the kernels is negligible compared to interferometric imaging for the same reasons. However, kernel support size and required calculations increases with channel width.

\begin{figure}
\center
	\includegraphics[width=9cm]{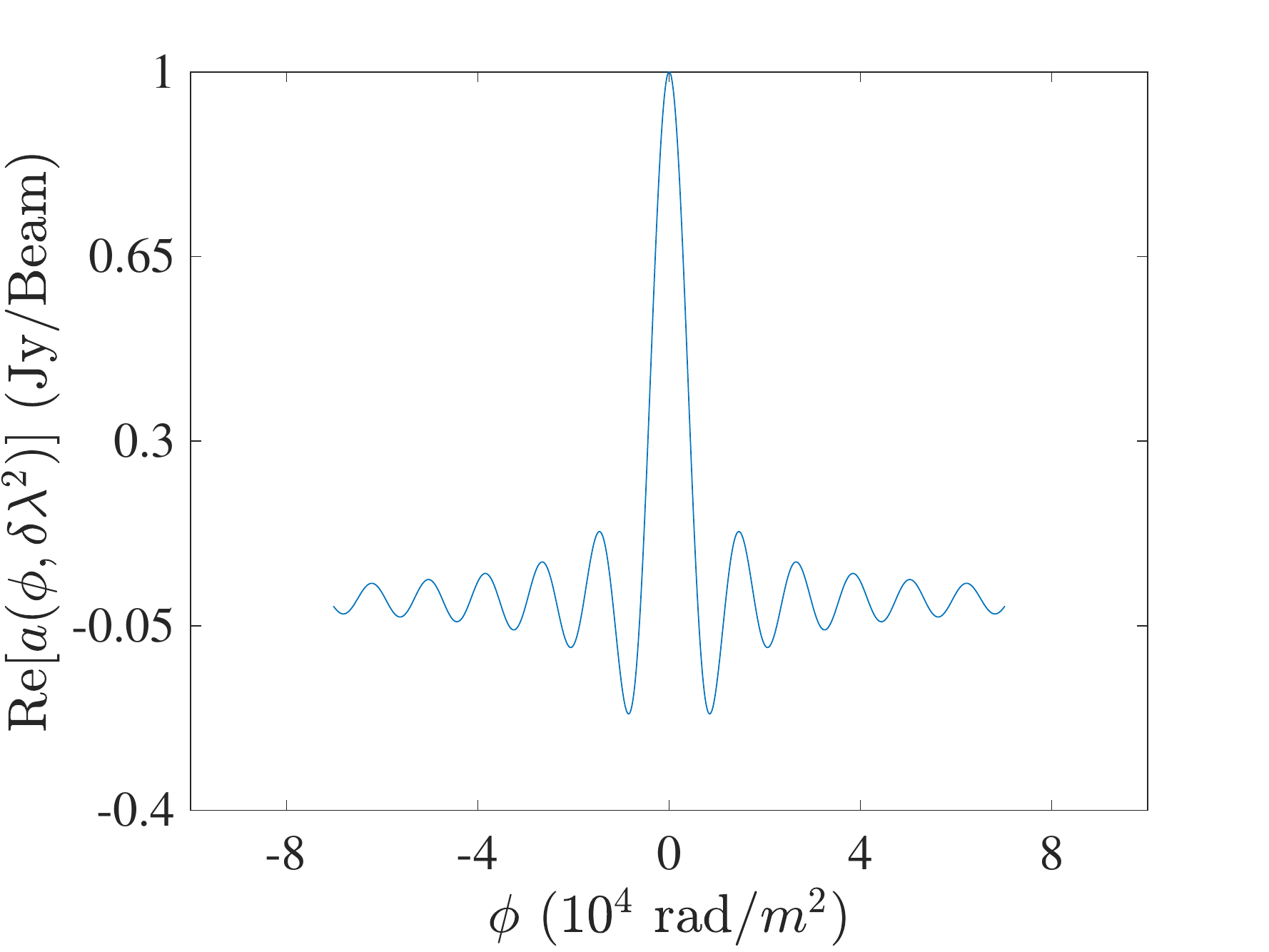}
	\includegraphics[width=9cm]{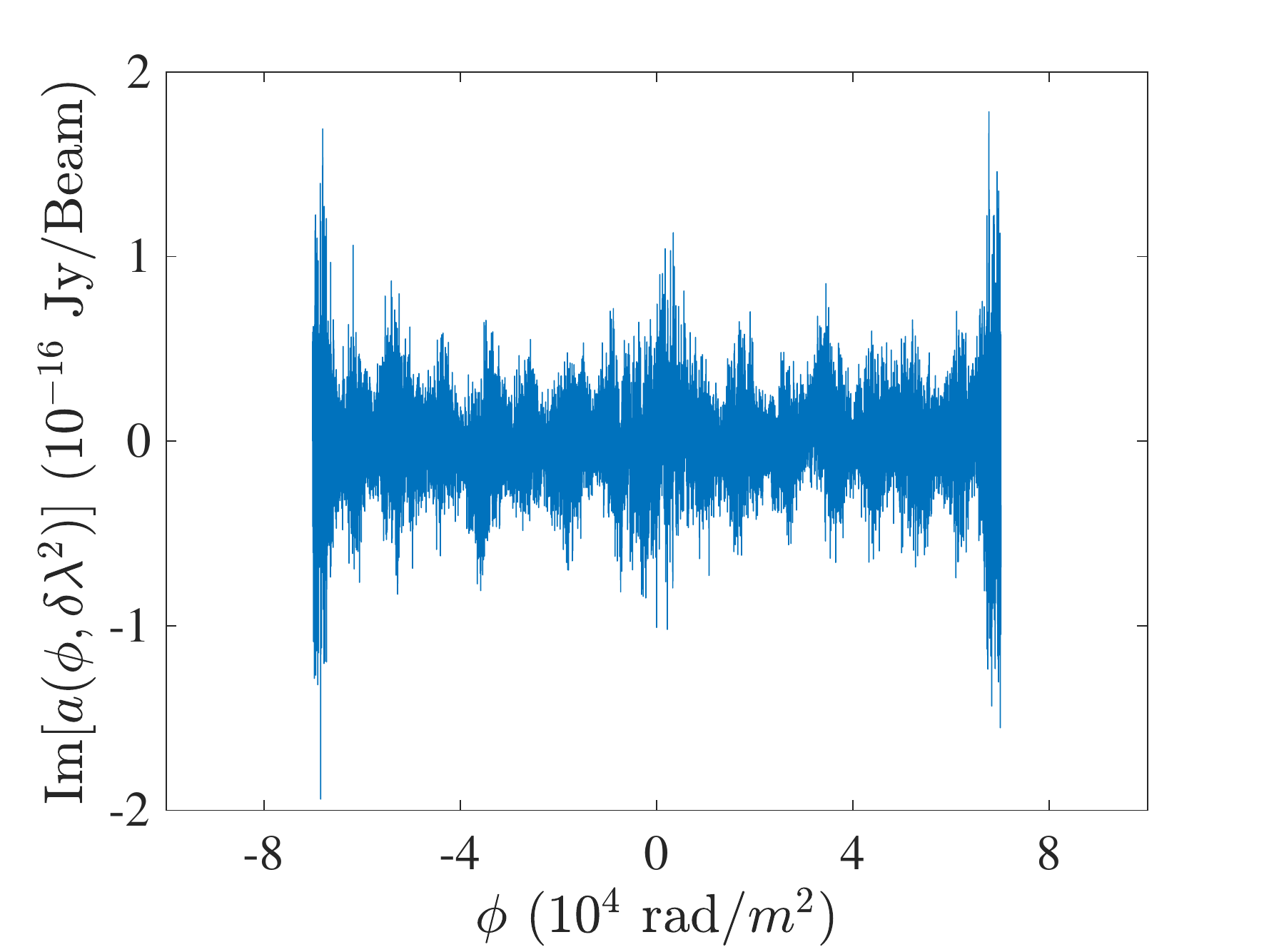}
	\includegraphics[width=9cm]{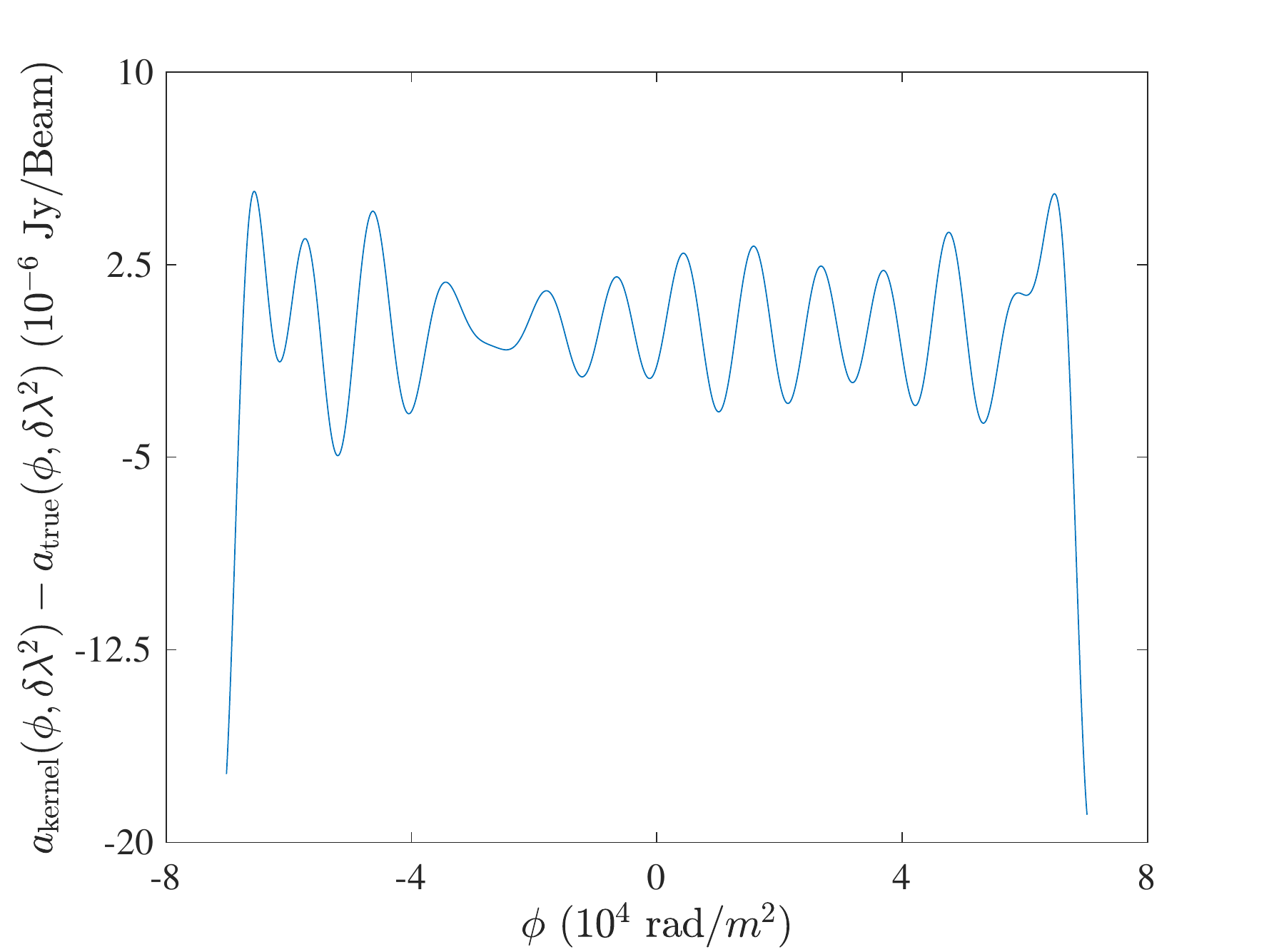}
	\caption{Above we see the projection kernel imaged in Faraday depth for a channel width $\delta \lambda^2$. This image was generated using the same method as \citet{PJM19a} where ${\bm{\mathsf{\Phi}}}_{\lambda^2 = 0,\, \delta \lambda^2}^\dagger$ is used to image a directional dependent effect in Faraday depth. (Top) we show that the real part is a sinc function for a channel width of $\delta \lambda^2 = 0.00053$ $m^2$, this has a FWHM of $\pm 3,575.5$ rad/$m^2$. (Middle) we find that the imaginary component is numerically close to zero as expected.  (Bottom) we show that the difference between the sinc functions is on the order of $10^{-5}$. This shows that the projection kernel is correctly modelling the Faraday depth effect of channel averaging in $\lambda^2$.}
	\label{fig:window_plots}
\end{figure}

Additionally, in Figure \ref{fig:window_plots} we image the kernel using the adjoint of the measurement operator $\bm{\mathsf{\Phi}}^\dagger$ on a single measurement at $\lambda^2 = 0$ with a fixed channel width $\delta \lambda^2$. We denote the result of this calculation as $a_{\rm kernel}(\phi, \delta \lambda^2)$. The analogous mathematics is explained in detail in \cite{PJM19a} for the interferometric imaging case. We expect to only obtain the sinc function ${\rm sinc} \left (\delta \lambda^2 \phi\right)$ in Faraday depth when applying the kernel to the zero $\lambda^2$ Fourier mode, we denote the evaluation of the sinc function as $a_{\rm true}(\phi, \delta \lambda^2)$. We verify this relation and find that the absolute difference is on the order of $10^{-5}$ between the real parts of  $a_{\rm kernel}(\phi, \delta \lambda^2)$ and  $a_{\rm true}(\phi, \delta \lambda^2)$. We also find that the imaginary component is close to zero (within precision), as expected.

\section{Wide-band rotation measure synthesis with the $\delta \lambda^2$-projection}
\label{sec:simulations}
Using the projection kernels developed in the previous section, we can now simulate both observations and perform 1d image (Faraday dispersion) reconstruction, while including the effects of channel averaging in the linear inverse problem. While there are many methods for solving inverse problems to reconstruct the signal in Faraday depth, we demonstrate the potential of using sparse regularization and convex optimization algorithms to invert the process of channel averaging. We also expect that a CLEAN style algorithm can also be applied using the same gridding and degridding process, but it is not clear if the CLEAN algorithm is designed for this type of inverse problem because CLEAN is not typically used to correct for the primary beam response but only to reconstruct point sources.

\subsection{Construction of the measurement operator}
In this section we detail the construction of the measurement operator $\bm{\mathsf{\Phi}}$. For $M$ channels, we have the vectors of channel frequencies $\bm{\nu}\in \mathbb{R}^{M}$ and channel widths $\bm{\delta \nu} \in \mathbb{R}^{M}$ that correspond to where we have sampled. We then calculate $\bm{\lambda^2}\in \mathbb{R}^{M}$ and $\bm{\delta\lambda^2}\in \mathbb{R}^{M}$ for each measurement, which is needed for calculating the gridding kernels. We can then choose a pixel (cell) size and image size in Faraday depth to determine $\Delta \lambda^2$ and the total region of Faraday depth imaged (the previous sections suggest how to choose the pixel size and image size based on the sampled $\bm{\lambda^2}$ and $\bm{\delta\lambda^2}$). The oversampling factor $\alpha$ is required to reduce aliasing error while degridding, typically $\alpha = 2$. We choose a support size of $J = 4$ for the Kaiser-Bessel kernel described in the previous sections. The construction of $\bm{\mathsf{FZS}}$ is straightforwardly done as a function that performs multiplication, followed by zero padding and an FFT. The construction of $\bm{\mathsf{GA}}$ is performed by constructing values of a sparse circular convolution matrix, where a kernel $[GA](u_{\rm pix}, \delta u, \Delta u)$ of support size $J + \delta \lambda^2/\Delta \lambda^2$ is constructed for each measurement using adaptive quadrature to a relative and absolute accuracy of $10^{-6}$. We follow the same process described in the previous sections used to construct $\bm{\mathsf{G}}$. We then use the power method to remove arbitrary scale during image reconstruction (Pratley et al in prep). 

\subsection{Simulation}
There are two important regimes in which these corrections will make critical differences: i) the case for wide-band polarimetry, and ii) low frequency observations. We provide two illustrative examples of these scenarios below. 

For both scenarios we start with a ground truth model in Faraday depth, and simulate the observed measurements using the measurement operator
\begin{equation}
    \bm{y}_0 = \bm{\mathsf{\Phi}} \bm{x}_{\rm Ground Truth}\, .
\end{equation}
Note that in the definition of the measurement operator we are using the $\delta\lambda^2$-projection kernels in the degridding process to simulate channel averaging in $\lambda^2$.

We can then add noise at an input signal to noise ratio (ISNR), using the formula 
\begin{equation}
    \sigma = \frac{\left\| \bm{y}_0 \right\|_{\ell_2}}{\sqrt{M}} \times 10^{-\frac{\rm ISNR}{20}}\, ,
    \label{eq:insr}
\end{equation}
where $\sigma$ is the RMS Gaussian noise of each of the $M$ measurements. We then generate the noise $\bm{n}$ and simulate the observation
\begin{equation}
    \bm{y} = \bm{y}_0 + \bm{n}\, .
\end{equation}
We can then image the noisy dirty map using
\begin{equation}
    \bm{x}_{\rm dirty} = \bm{\mathsf{\Phi}}^\dagger \bm{y}\, ,
\end{equation}
and the residual Faraday map can be generated through
\begin{equation}
    \bm{x}_{\rm residual} = \bm{\mathsf{\Phi}}^\dagger \left(\bm{\mathsf{\Phi}} \bm{x} - \bm{y}\right)\, .
\end{equation}

\subsubsection{Case I: Wide-band Polarization}
\begin{figure}
\center
	\includegraphics[width=9cm]{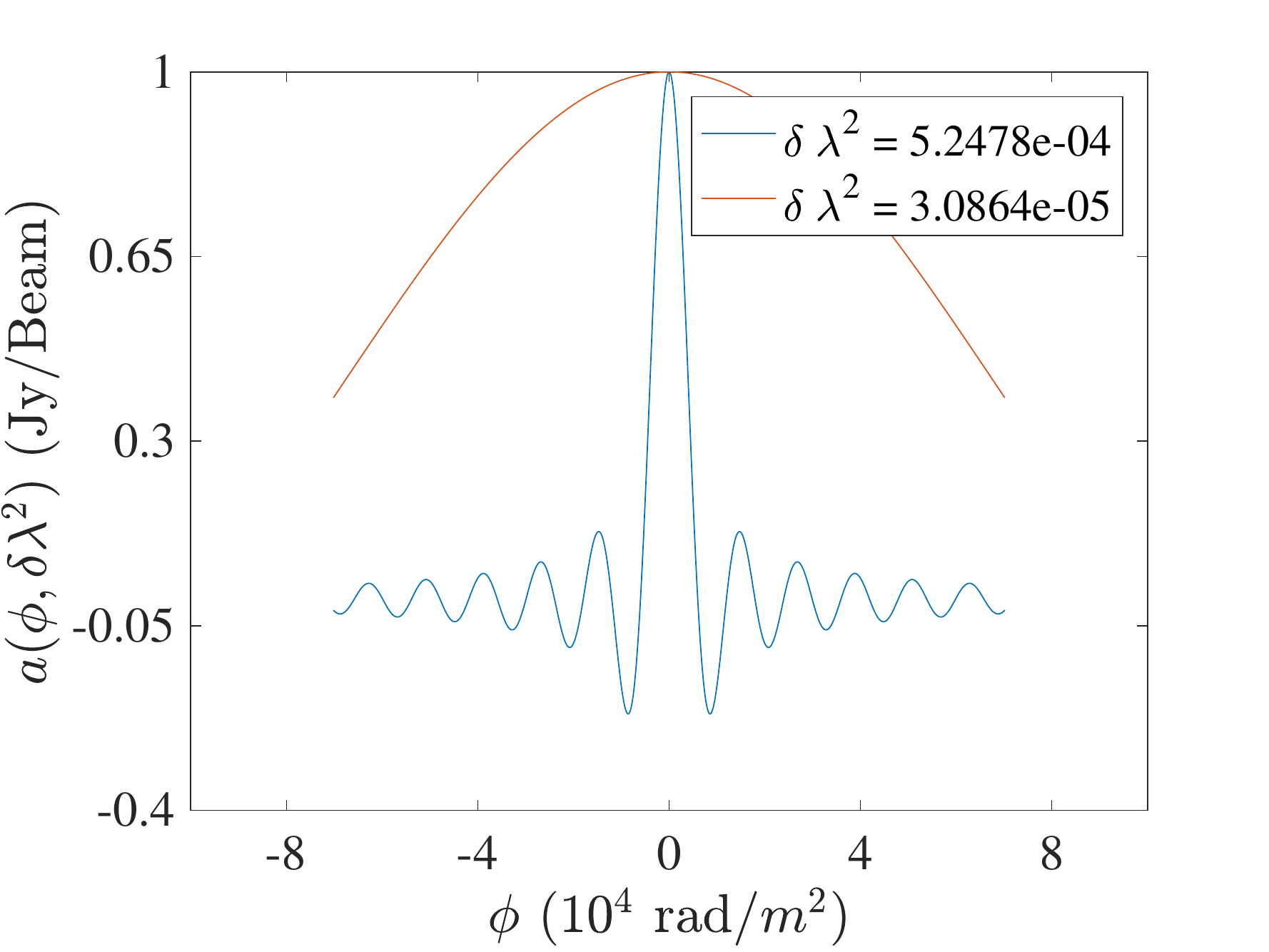}
	\caption{The sensitivity in Faraday depth between two extremes of the POSSUM band due to channel averaging, at 700 MHz (red) and 1800 MHz (blue) with 1 MHz channel width. This sensitivity was calculated using Equation \ref{eq:chan_sensivity}. There is a factor of 20 difference in range between the widest and thinnest channel widths.}
	\label{fig:windows_over_range}
\end{figure}

We simulate the frequency range of 700 MHz to 1800 MHz, with a channel width of 1 MHz which is based on the observational bandwidth of the Polarization Sky Survey of the Universe's Magnetism (POSSUM; \citealp{gae10}) which will be undertaken on ASKAP \citep{hotan14}. We use this frequency coverage to construct $\bm{\lambda^2}$ and $\bm{\delta \lambda^2}$ used in the simulations. The FWHM of the sinc functions (see Equation \ref{eq:chan_sensivity}) in Faraday depth for each channel ranges from $\pm 61,483$ rad/$m^2$ to $\pm 3,616$ rad/$m^2$, the two extremes can be seen in Figure \ref{fig:windows_over_range}.

We simulate observation of 15 Faraday-thin sources evenly spread in Faraday depth, over a range of $\pm 70,1156$ rad/$m^2$ with an image size of 16,384 pixels and pixel size of 8.56395 rad/$m^2$, which we define as the ground truth $\bm{x}_{\rm Ground Truth}$. Each of the sources has a complex phase added, which determines the intrinsic polarization angle of the source.

\begin{figure*}
	\center
	\includegraphics[width=5.75cm]{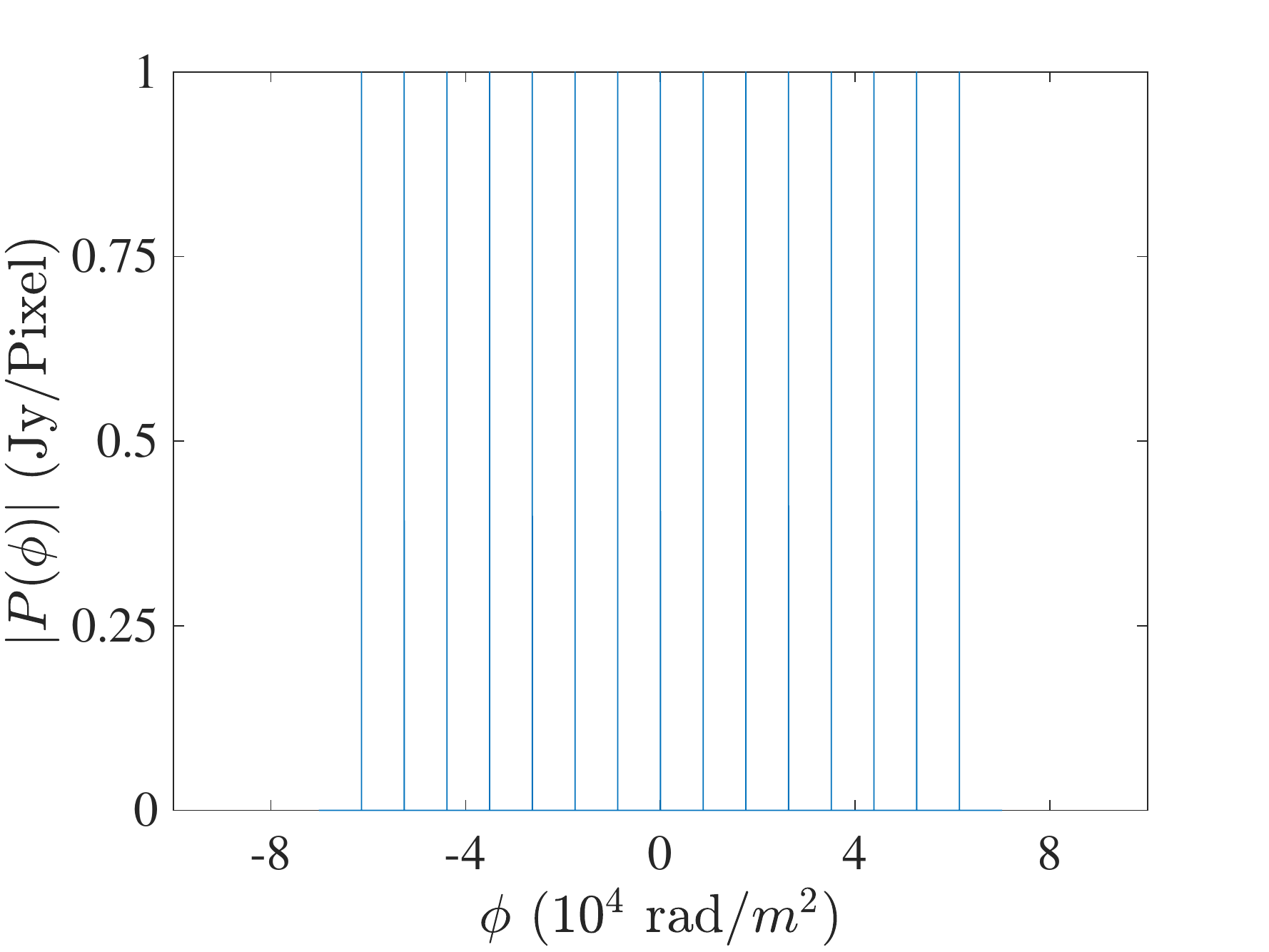}
	\includegraphics[width=5.75cm]{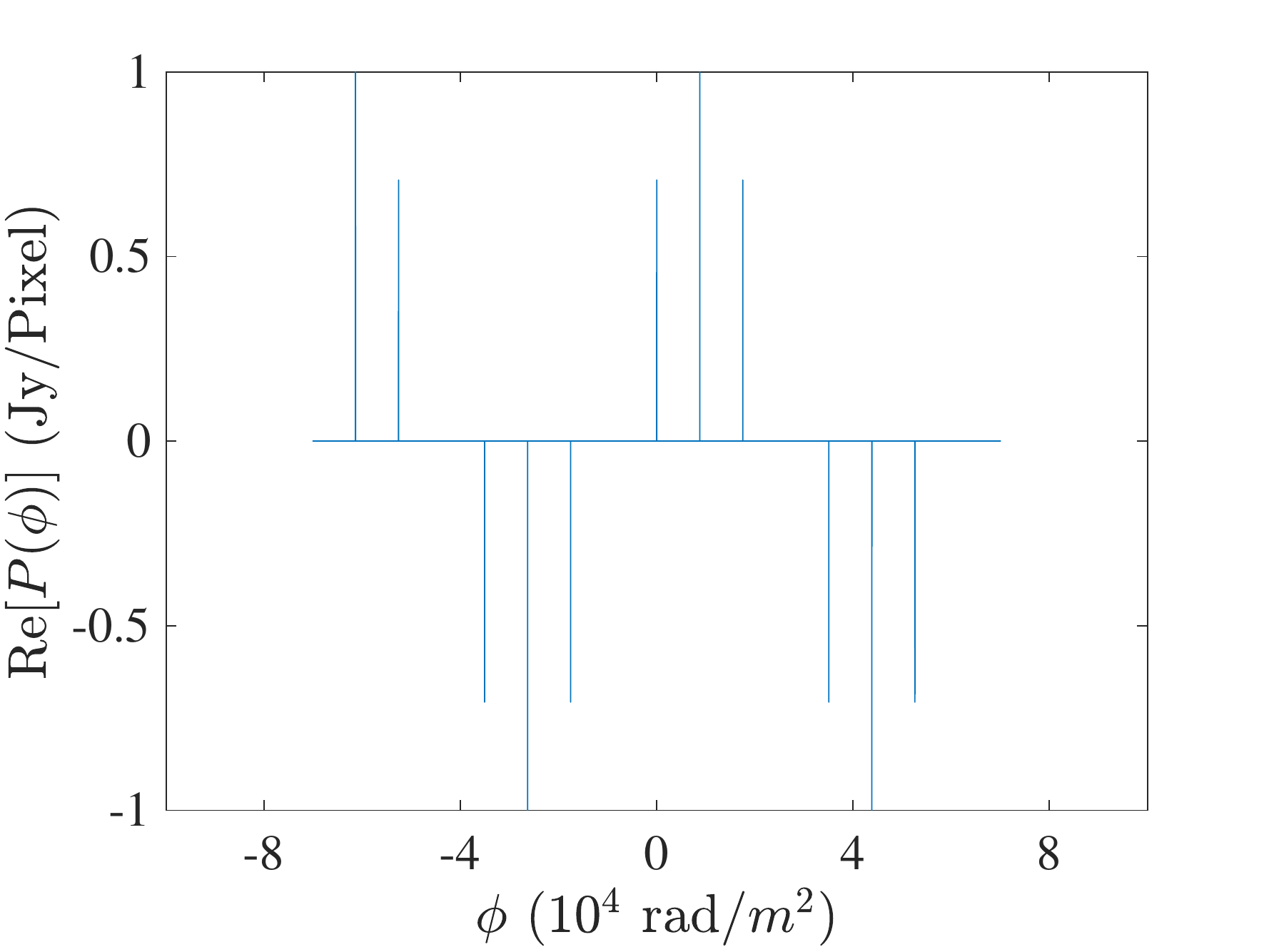}
	\includegraphics[width=5.75cm]{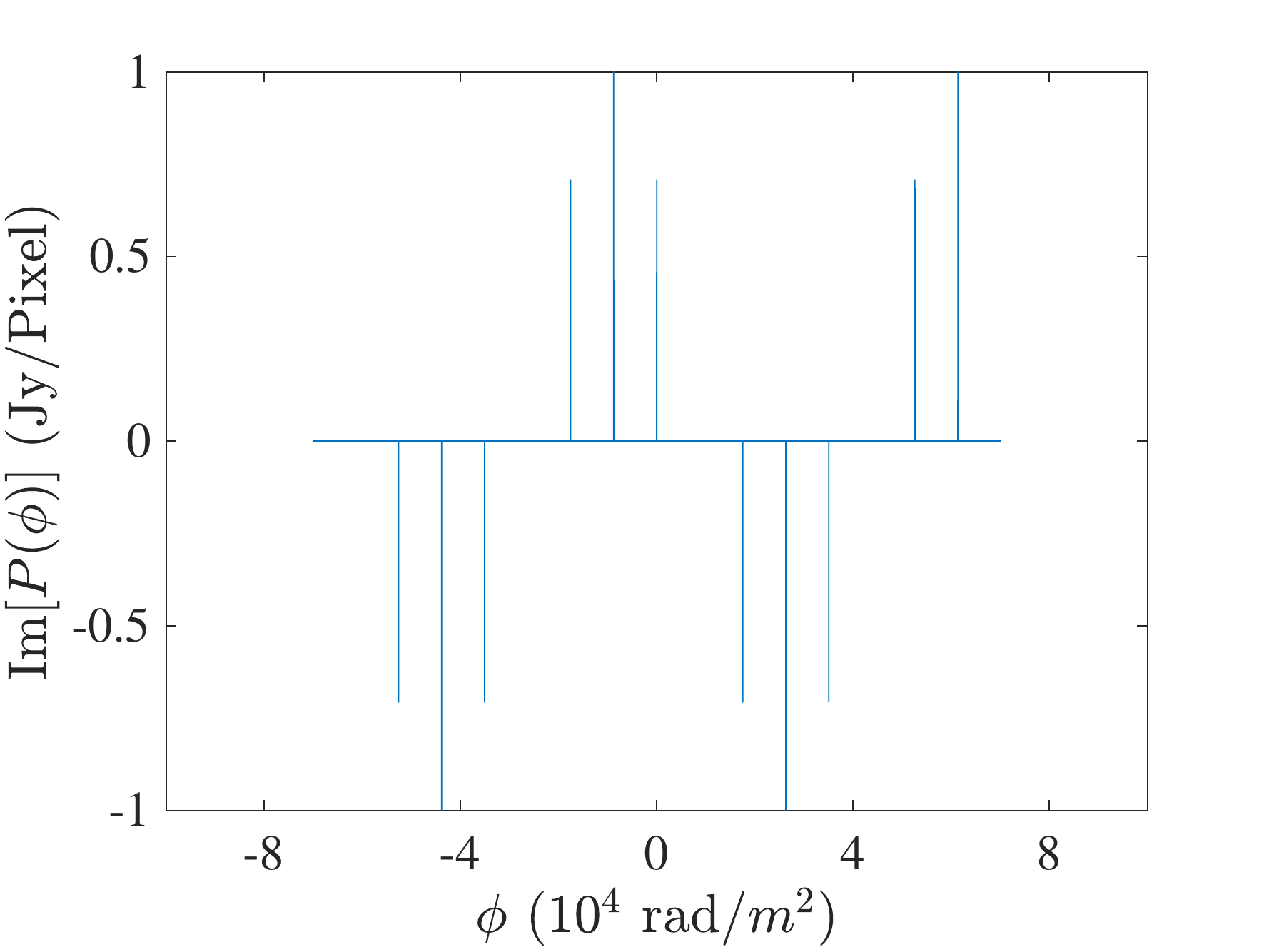}
	\includegraphics[width=5.75cm]{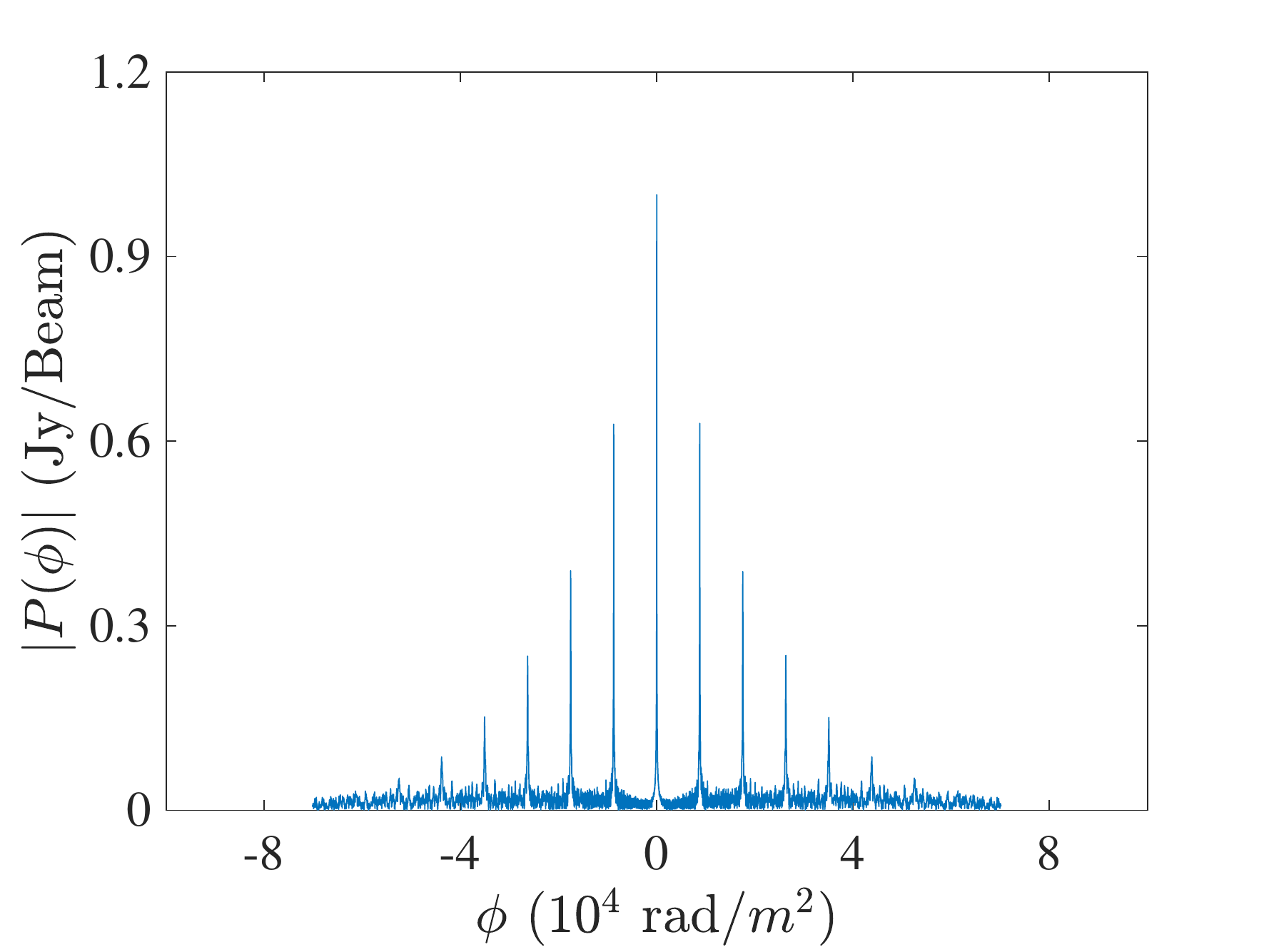}
	\includegraphics[width=5.75cm]{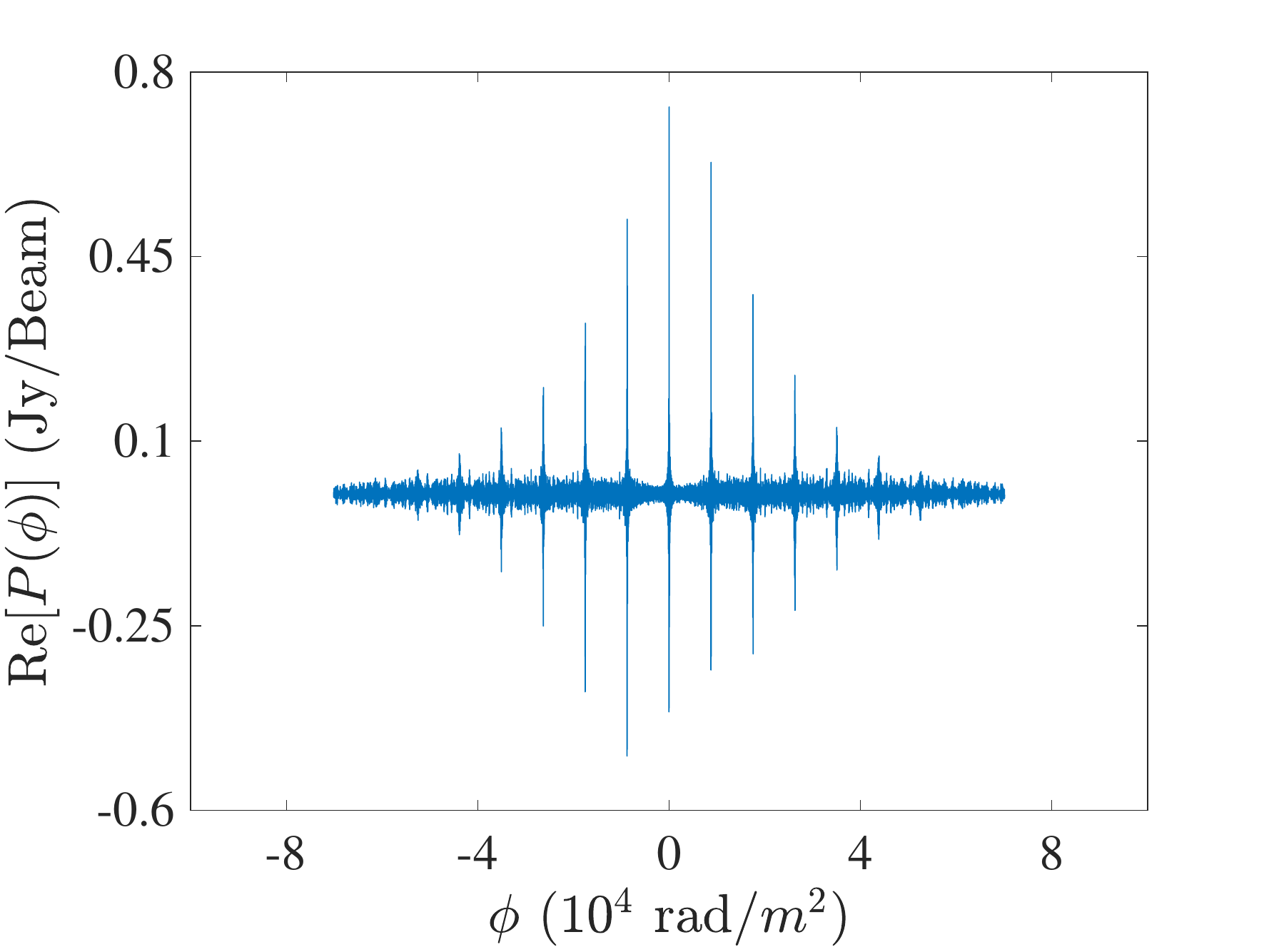}
	\includegraphics[width=5.75cm]{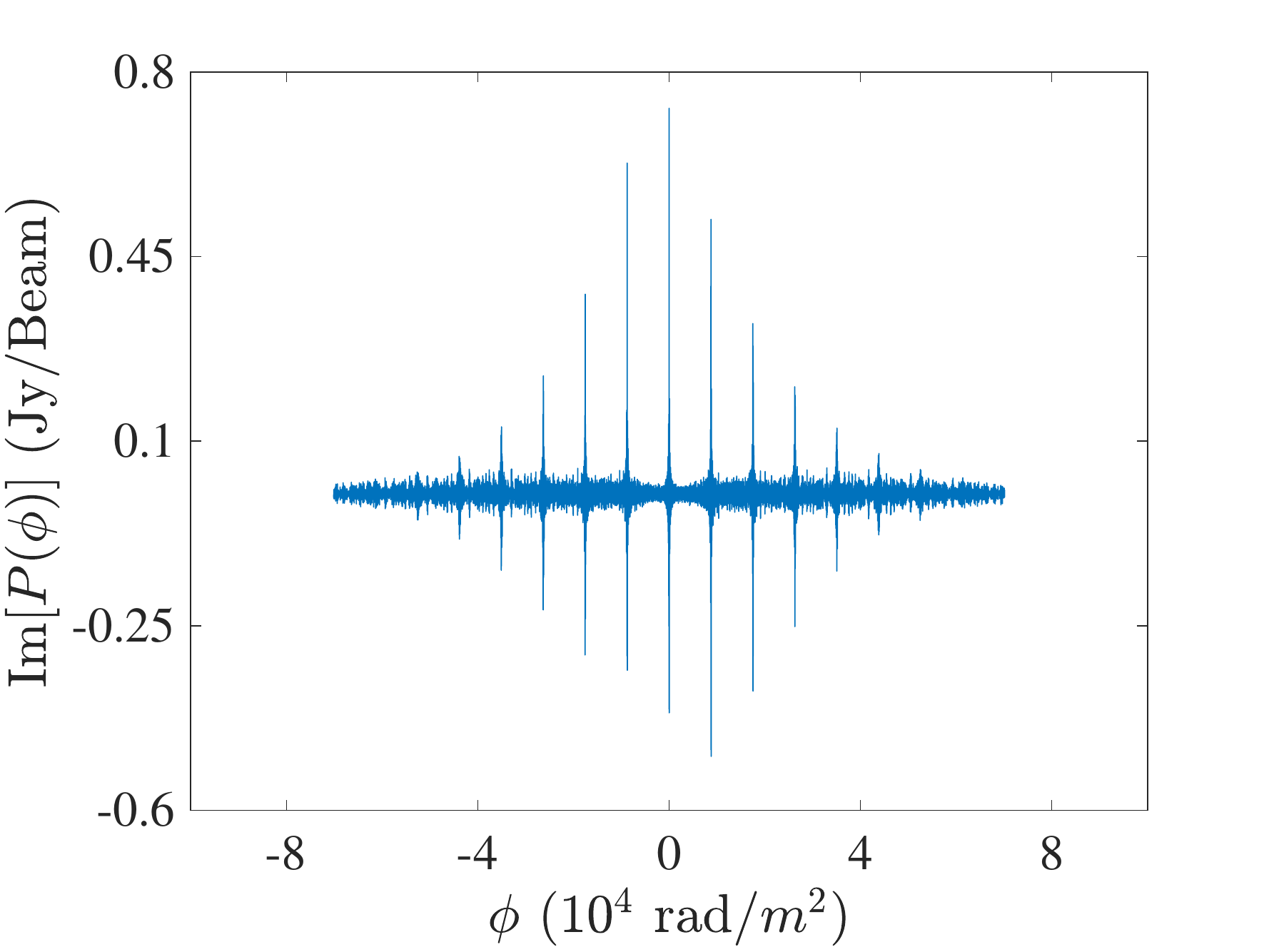}
	\caption{The top row is the ground truth signal and the bottom row is the dirty map signal in Faraday depth for POSSUM. Right to left are the absolute value, real and imaginary values of the signals in Faraday depth. The ground truth signal is a series of 1 Jy point sources. From the dirty map, it is clear to see that the signal at large values of $|\phi|$ becomes attenuated and close to the noise level.}
	\label{fig:input_signal}
\end{figure*}

The ground truth signal and dirty map can be see in Figure \ref{fig:input_signal}.

\subsubsection{Case II: Low Frequency Polarization}

\begin{figure}
\center
	\includegraphics[width=9cm]{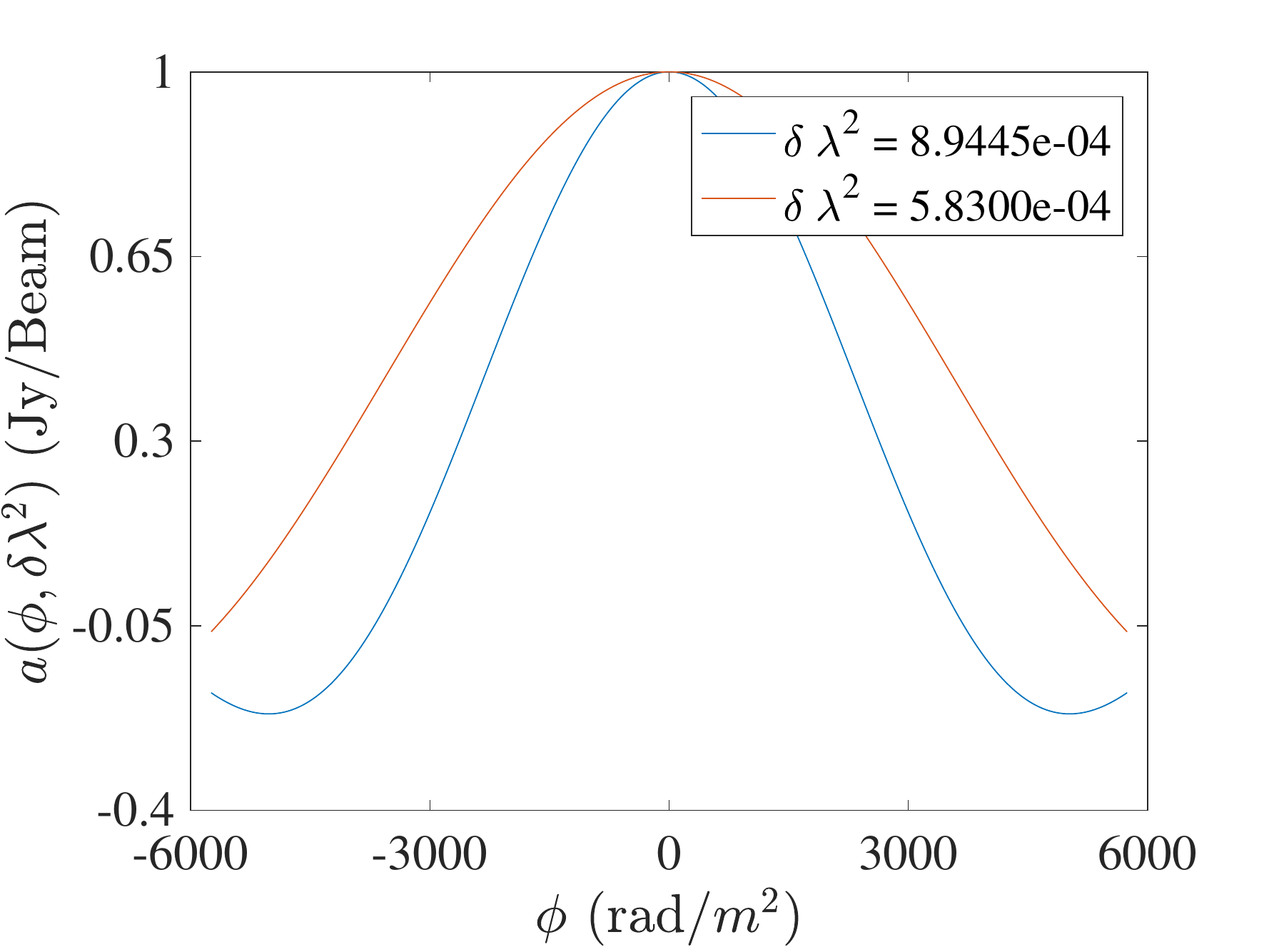}
	\caption{The sensitivity in Faraday depth between two extremes of the POGS band due to channel averaging, at  200.31 MHz (red) and 231.02 MHz (blue) with 40 kHz channel width.}
	\label{fig:low_windows_over_range}
\end{figure}

We repeat the same simulation as above for the frequency range of the POlarisation GLEAM Survey (POGS; \citealp{rise18}) on the MWA. The frequency coverage is 768 evenly separated 40 kHz channels between 200.32 MHz to 231.04 MHz, which is the top end of the GLEAM Survey band \citep{Way15}. Here the FWHM of the sinc functions in Faraday depth for each channel ranges from $\pm 2,118.61$ rad/$m^2$ to $\pm 3,250.42$ rad/$m^2$. This can be seen in Figure \ref{fig:low_windows_over_range}.

\begin{figure*}
	\center
	\includegraphics[width=5.75cm]{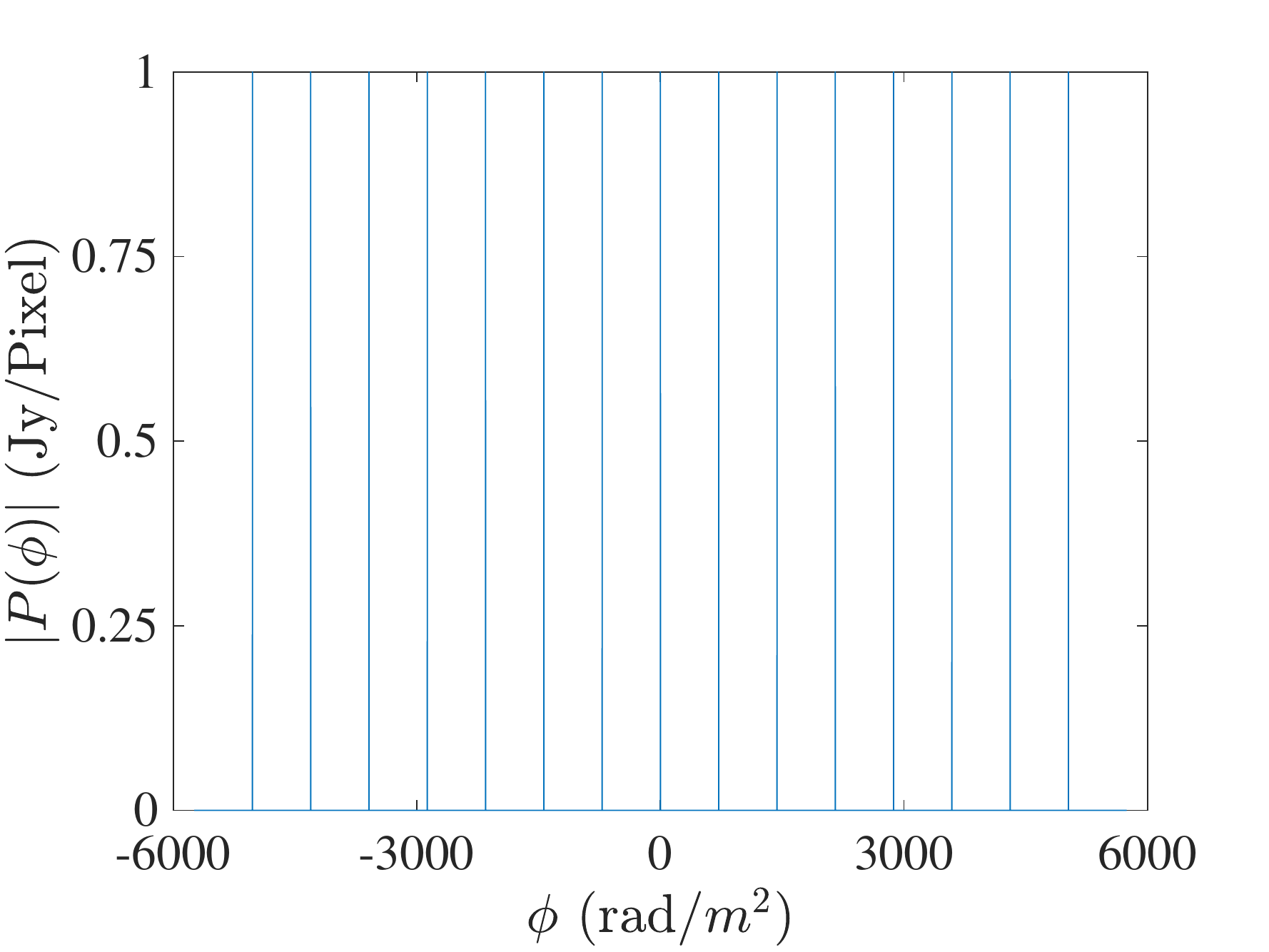}
	\includegraphics[width=5.75cm]{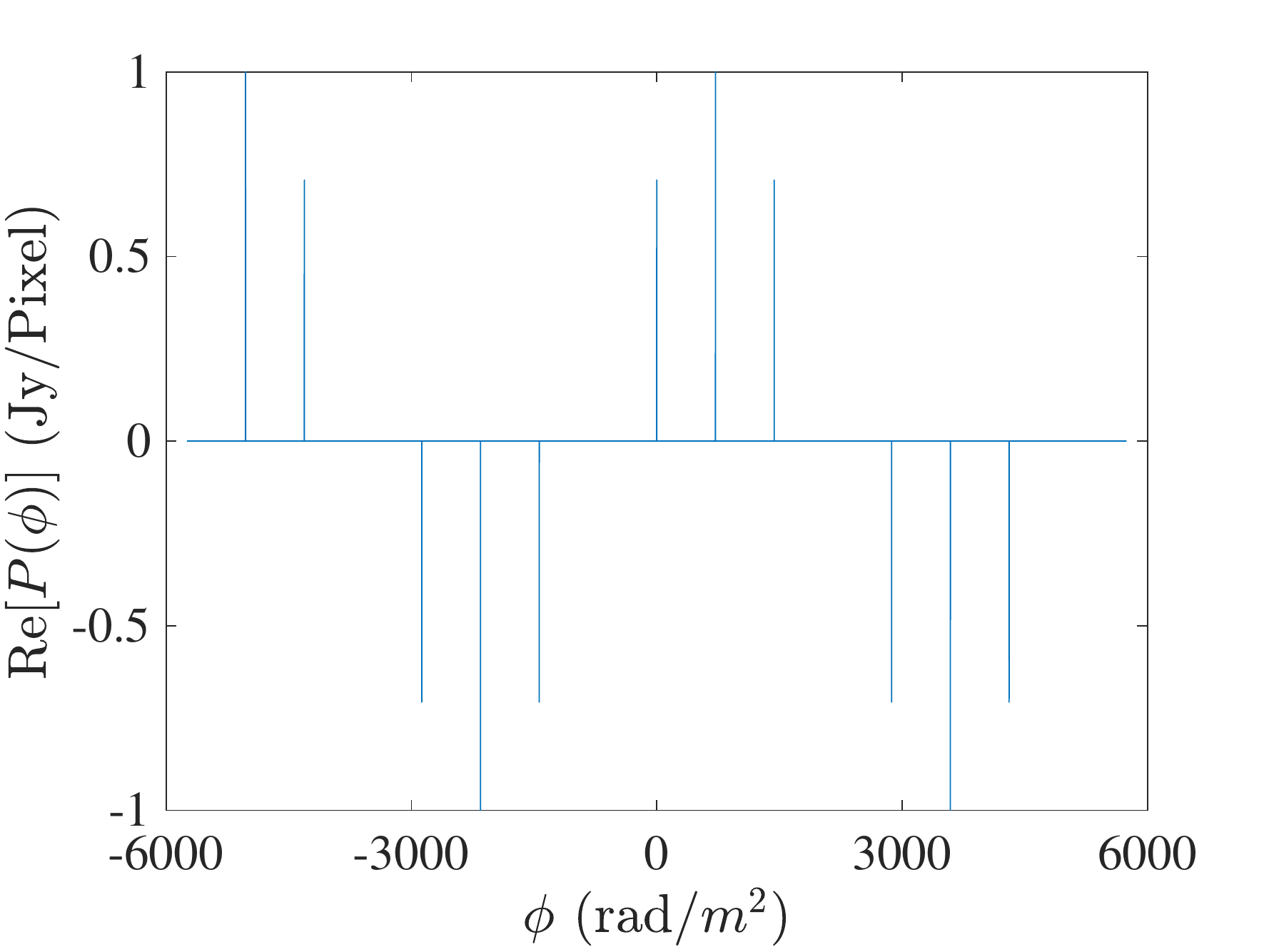}
	\includegraphics[width=5.75cm]{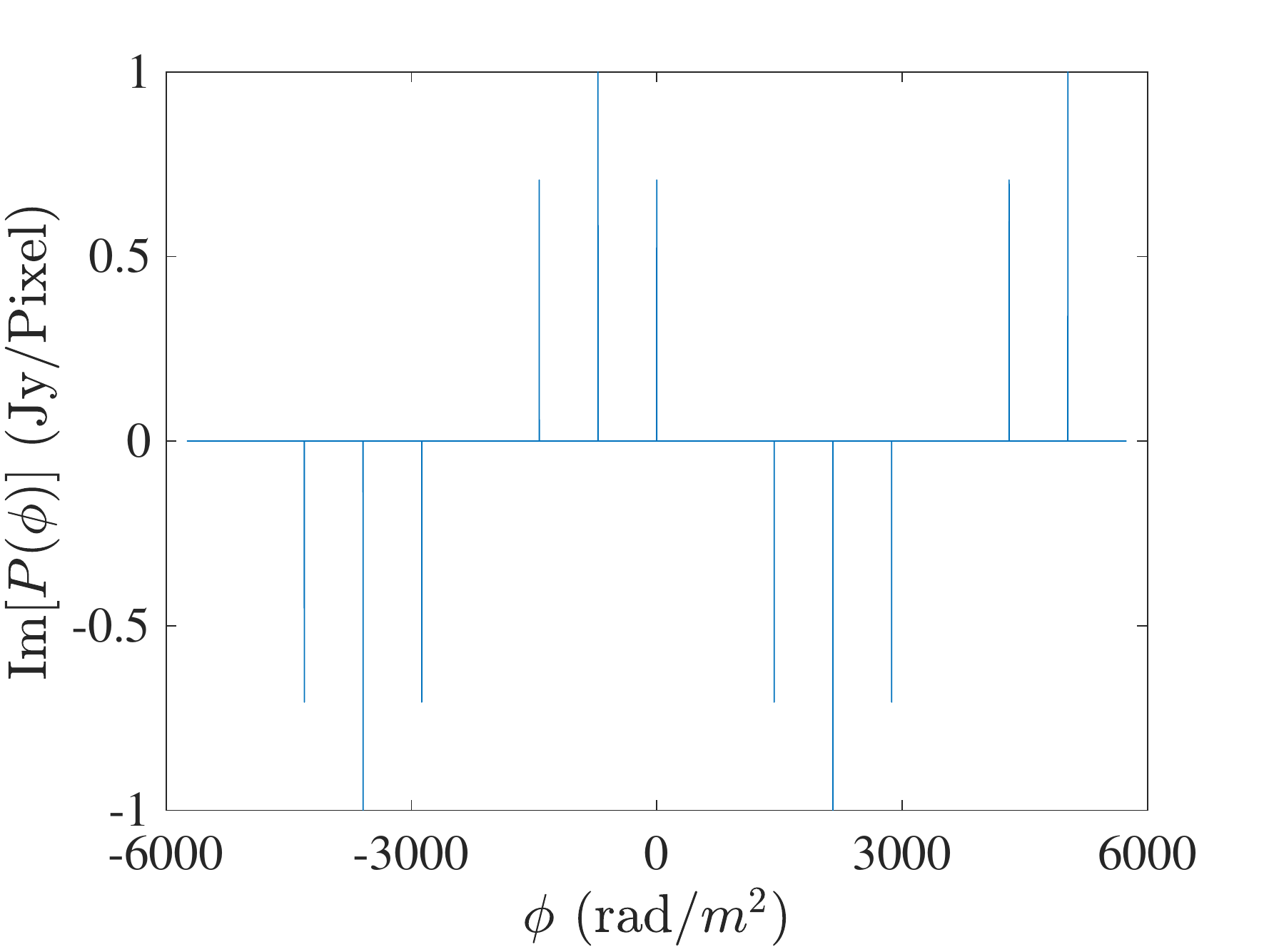}
	\includegraphics[width=5.75cm]{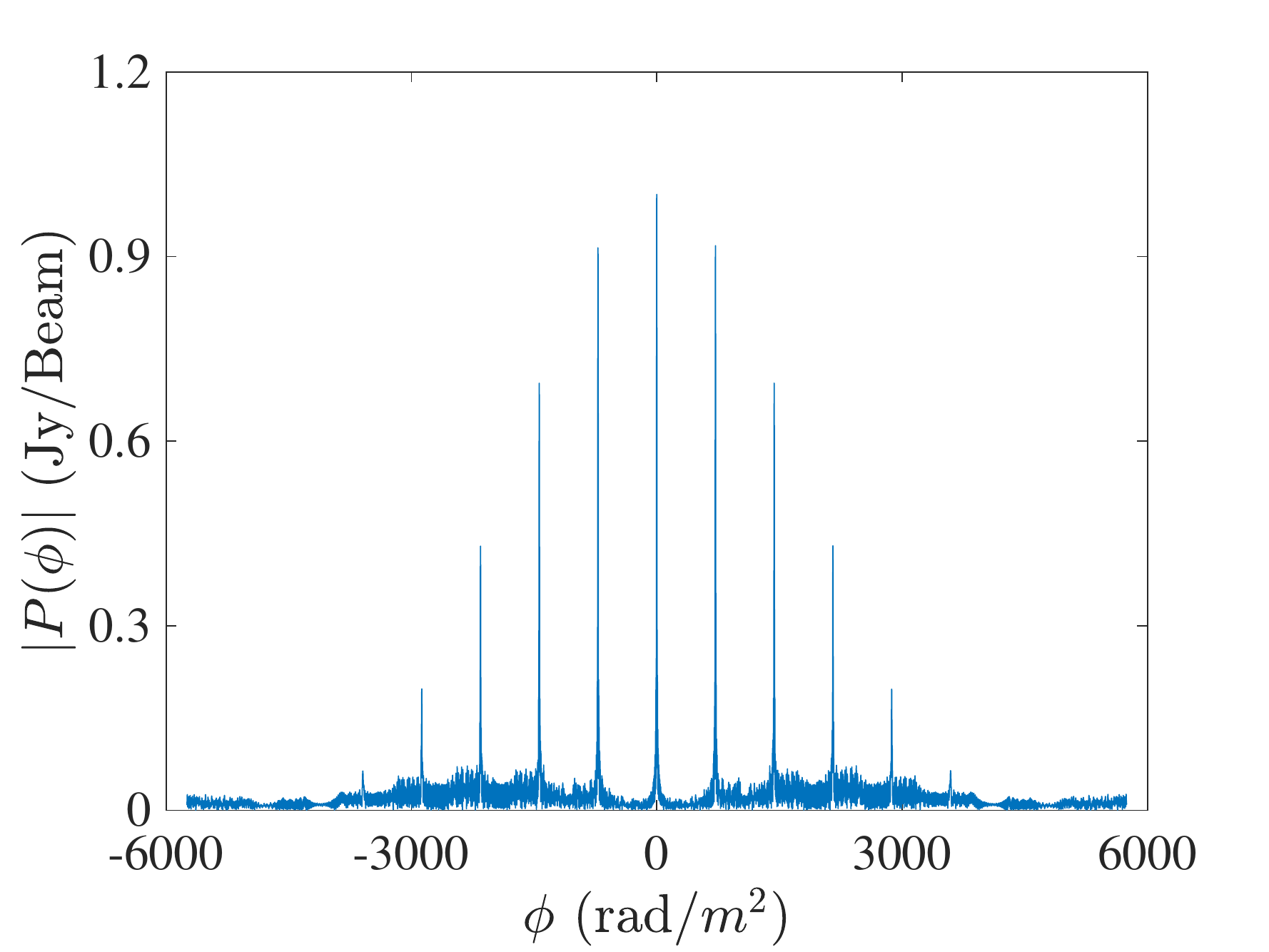}
	\includegraphics[width=5.75cm]{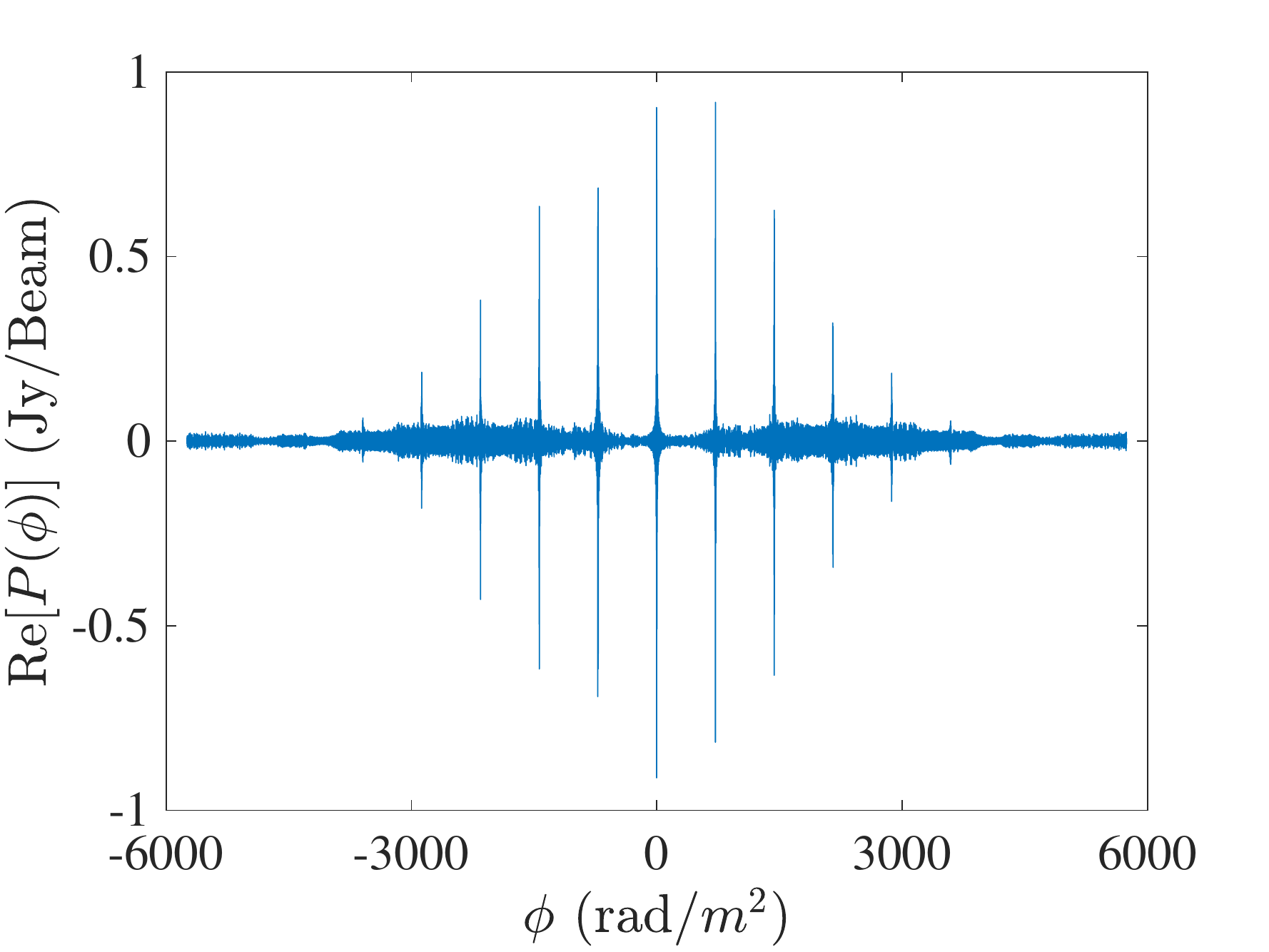}
	\includegraphics[width=5.75cm]{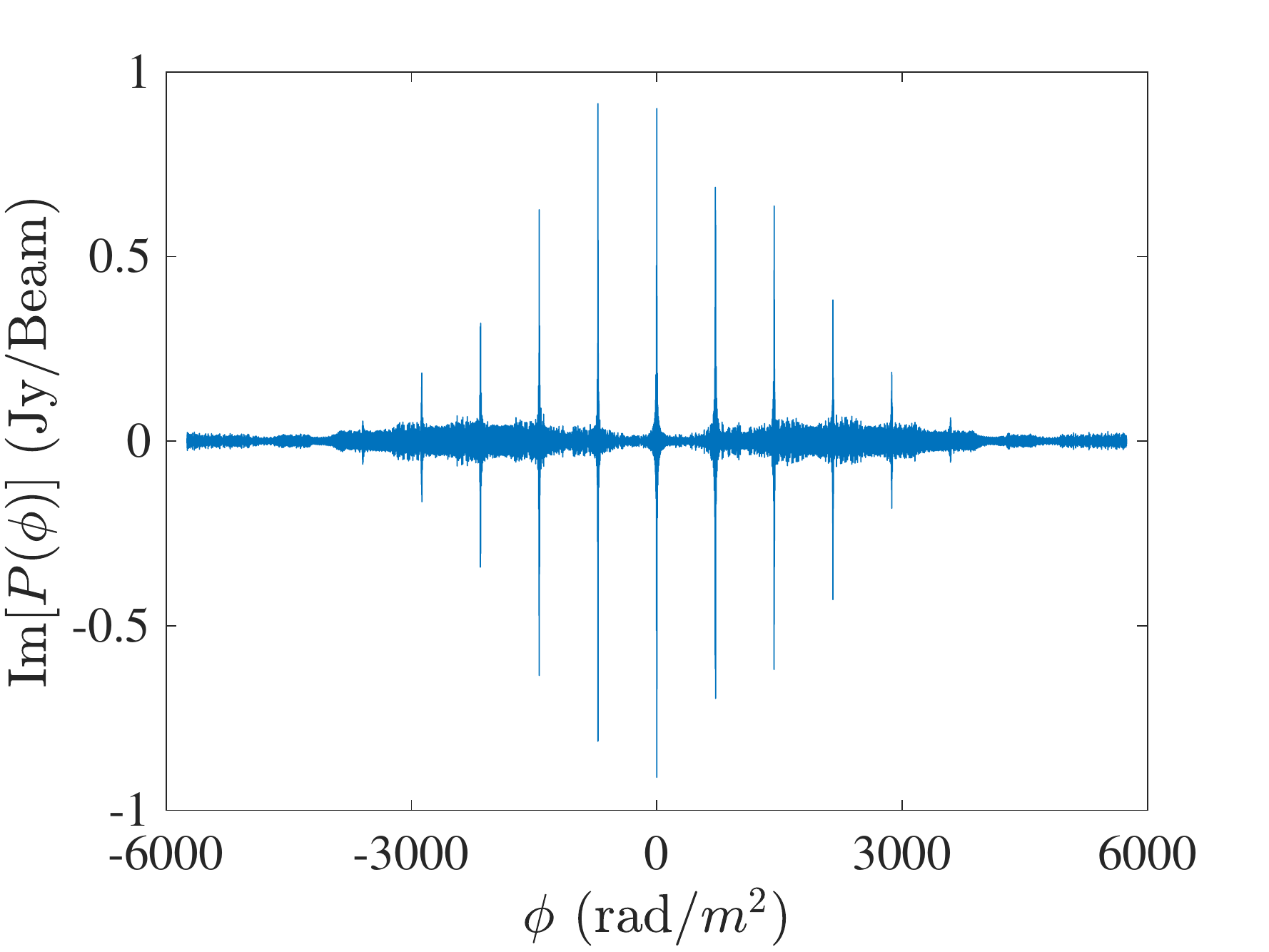}
	\caption{The top row is the ground truth signal and the bottom row is the dirty map signal in Faraday depth for POGS. Right to left are the absolute value, real and imaginary values of the signals in Faraday depth. The ground truth signal is a series of 1 Jy point sources. From the dirty map, it is clear to see that the signal at large values of $|\phi|$ becomes attenuated and close to the noise level.}
	\label{fig:low_input_signal}
\end{figure*}

Again we simulate observation of 15 Faraday point sources evenly spread in Faraday depth, over a range of $\pm 5,745.3$ rad/$m^2$ with an image size of 16,384 pixels and pixel size of 0.70138 rad/$m^2$, which we define as the ground truth $\bm{x}_{\rm Ground Truth}$. Each of the sources has a complex phase added, which determines the intrinsic polarization angle of the source. This can be seen in Figure \ref{fig:low_input_signal}, with the input simulation and dirty maps.

\subsection{Reconstruction}
To perform reconstruction, we use the software package PURIFY \citep{LP18} in conjunction with the Sparse OPTimization (SOPT) \citep{pra19} software package. While reconstruction of signals in Faraday depth is not officially supported in current releases, the similarities of interferometric image reconstruction and Faraday imaging mean that it was possible to build the required reconstruction tools and a working example using these software packages. Because we are reconstructing discrete sources and not extended sources, we use a Dirac basis. However, it is simple to extend this to a dictionary using more than one wavelet basis as done in interferometric imaging, and more accurately reconstruct both discrete and extended Faraday structures. Specifically, we use the alternating direction method of multipliers (ADMM) algorithm to find a solution to the constrained minimization problem
\begin{equation}
    \min_{\bm{x}} \|\bm{x} \|_{\ell_1}\quad {\rm such\, that} \quad \|\bm{y} -\mathsf{\bm{\Phi}}\bm{x}\| \leq \varepsilon\, ,
    \label{eq:analysis}
\end{equation}
where the above imposes the structure that we expect to find discrete point sources while obtaining a model that is within the Gaussian uncertainty. $\varepsilon$ can be set using the value of $\sigma\sqrt{2M + 2\sqrt{4M}}$, which is determined from a $\chi^2$-distribution. The details of the implemented ADMM algorithm can be found in \citet{pra19}. We also note that the use of sparse image reconstruction (c.f. compressed sensing) methods is not new to rotation measure synthesis, similar methods have been applied previously \citep{Li11}. Furthermore, RM CLEAN methods exist that can also be used to restore Faraday structures \citep{hea08,bel12}. For a more general comparison of RM recovery methods, please see \citealp{Sun15}.

\subsubsection{Case I: Wide-band Polarization}
\begin{figure*}
	\center
	\includegraphics[width=5.75cm]{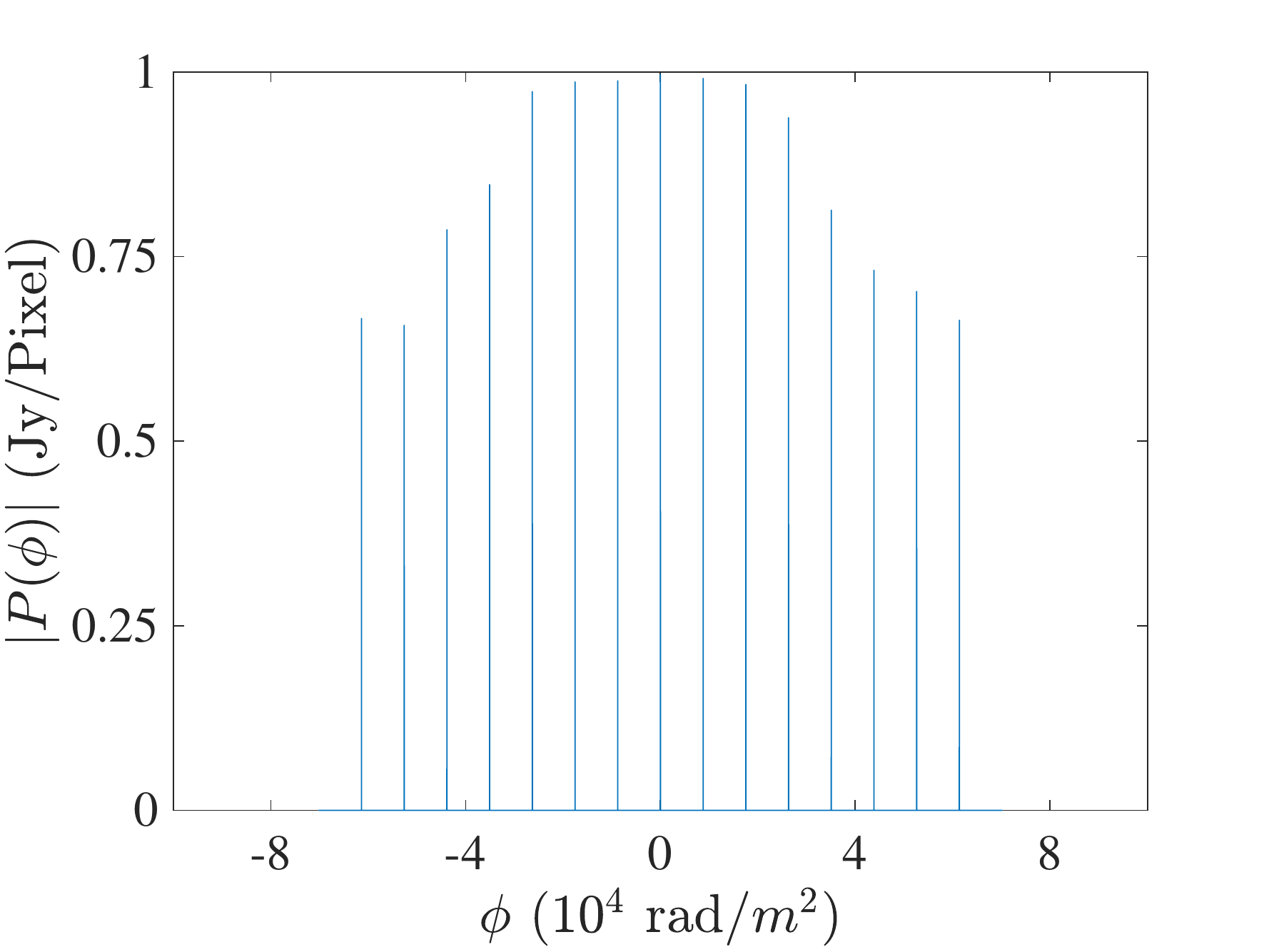}
	\includegraphics[width=5.75cm]{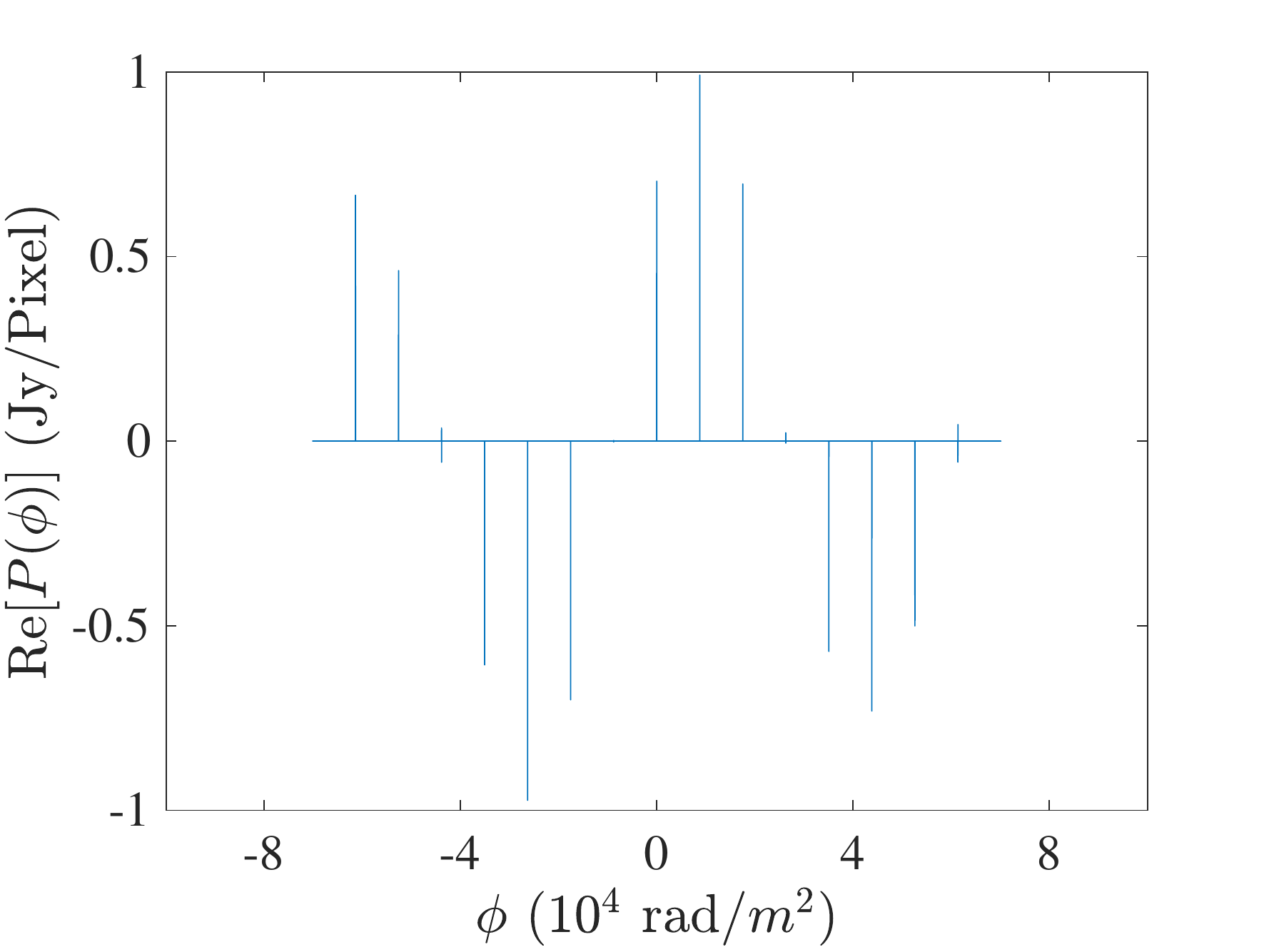}
	\includegraphics[width=5.75cm]{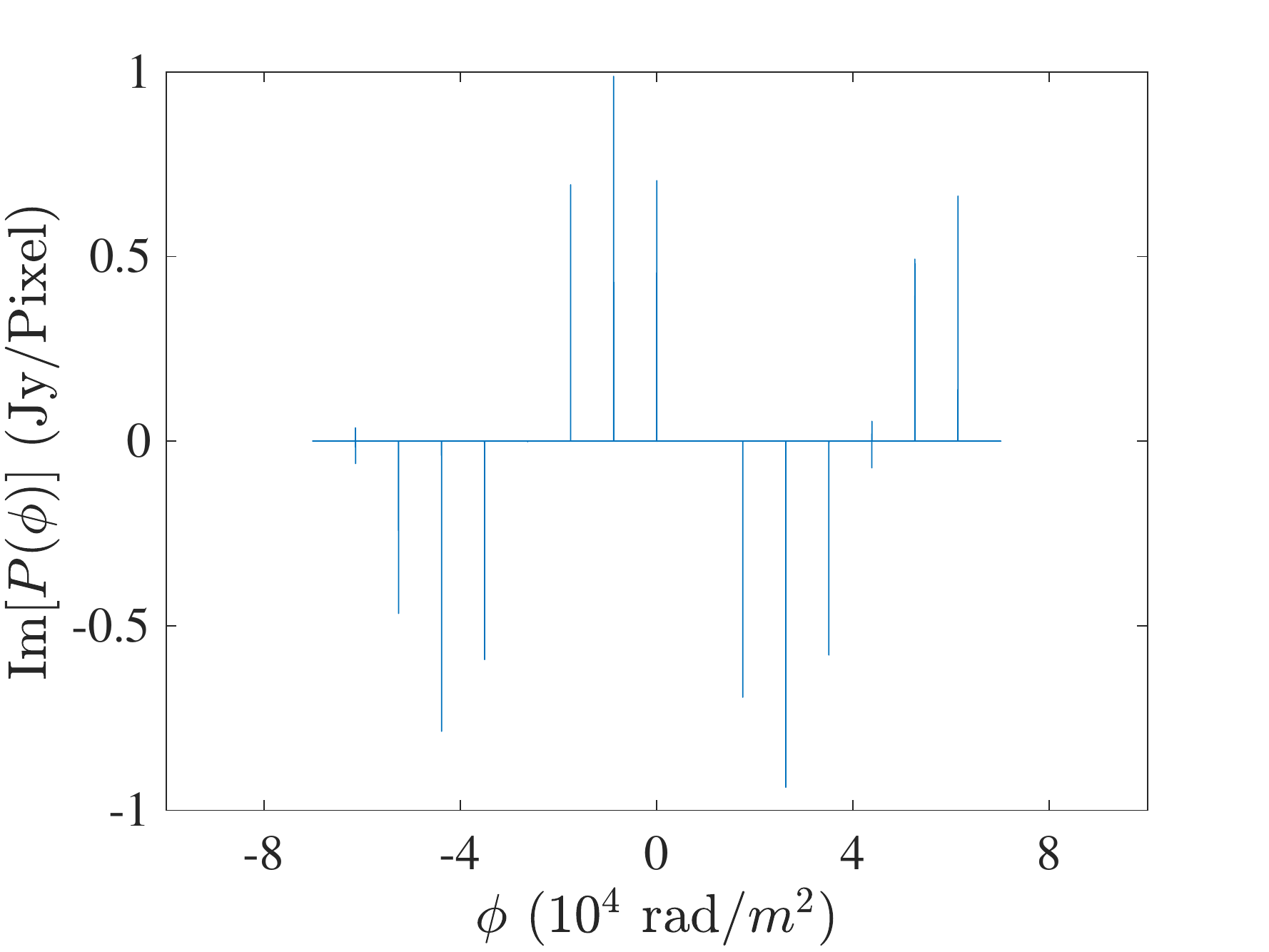}
	\includegraphics[width=5.75cm]{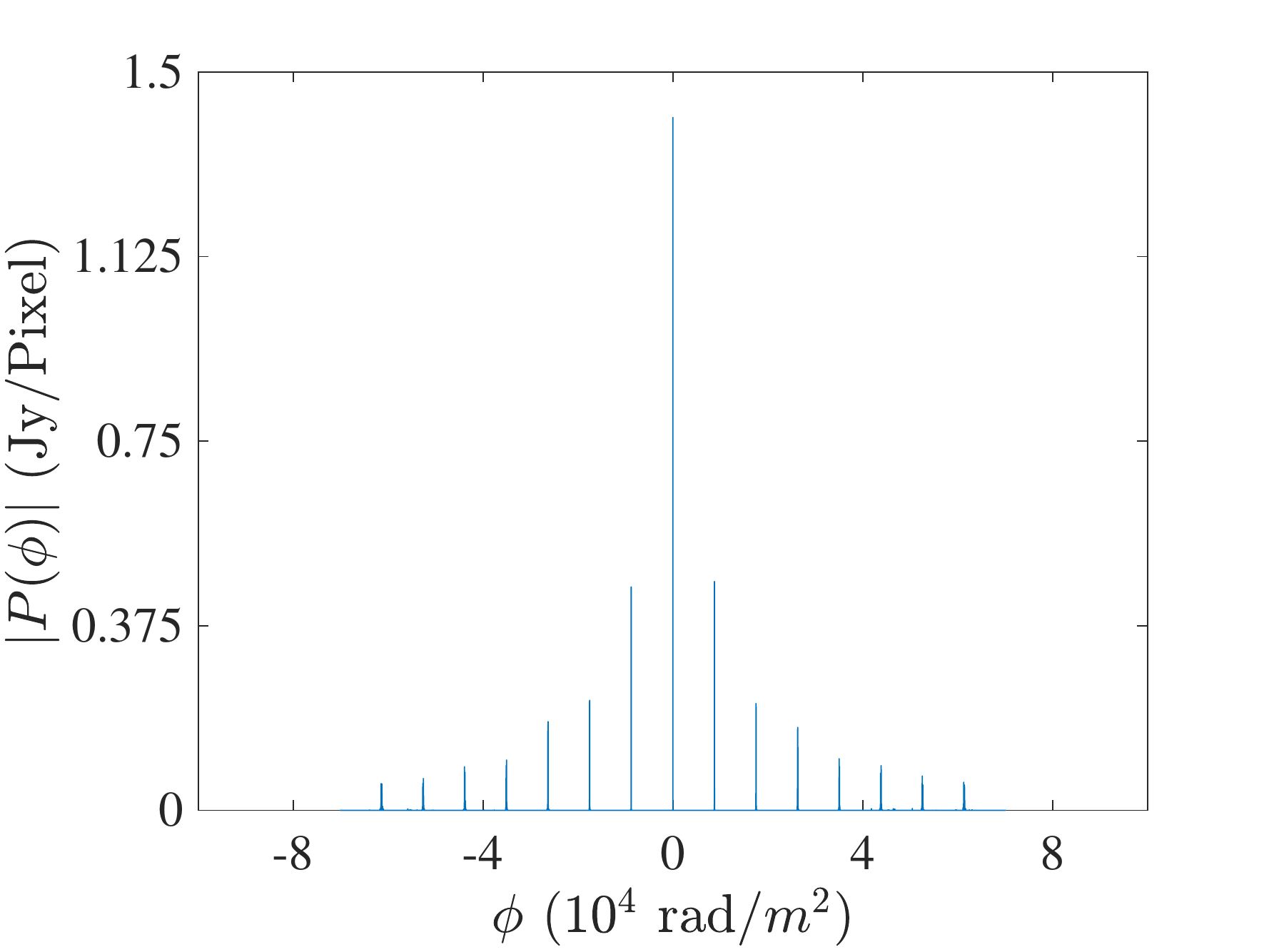}
	\includegraphics[width=5.75cm]{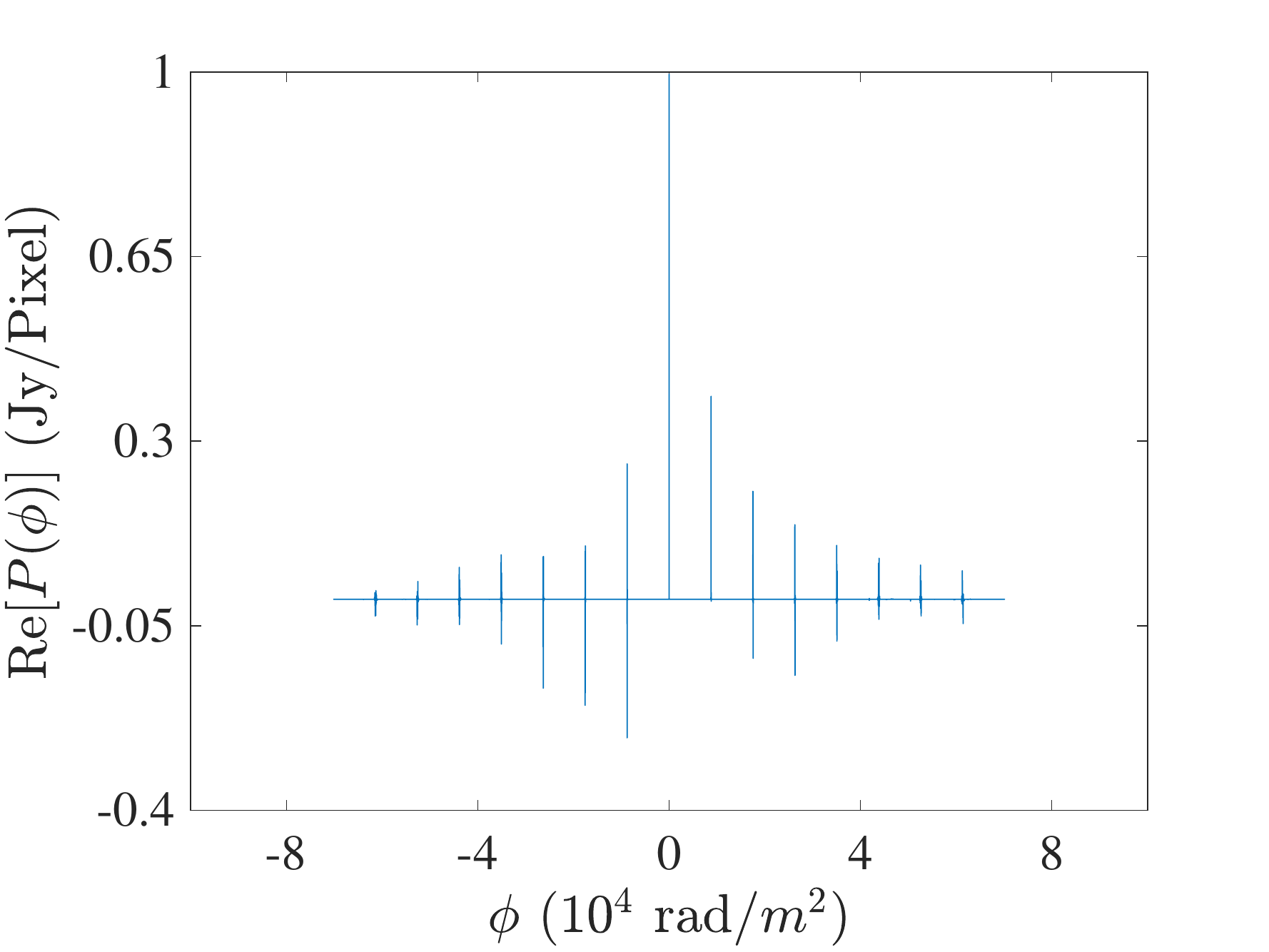}
	\includegraphics[width=5.75cm]{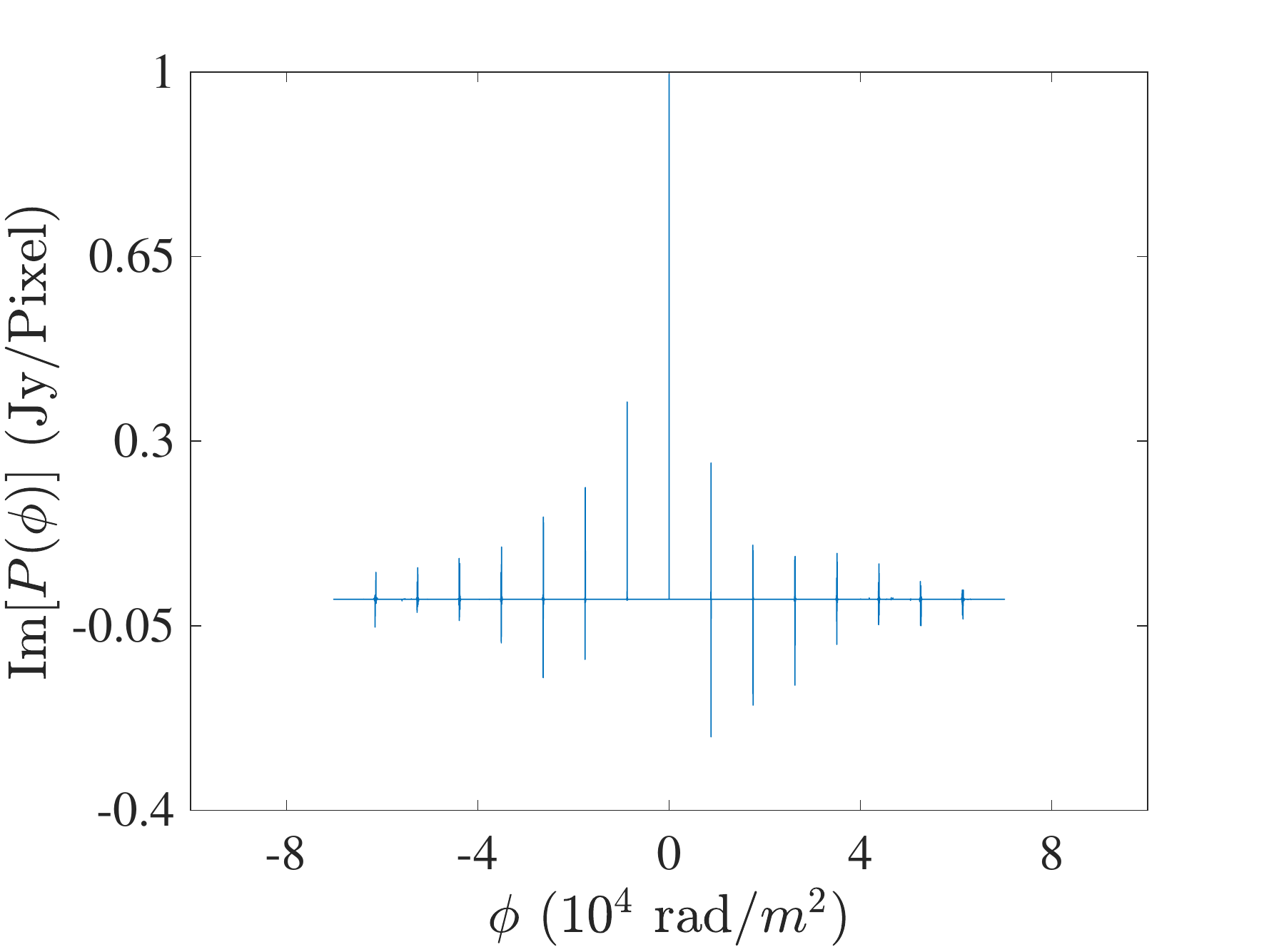}
	\caption{(Top row) shows the reconstruction that corrects for channel averaging, with the absolute, real, and imaginary values from right to left. (Bottom row) is the same as the top row without channel averaging corrected during reconstruction with the POSSUM coverage. It is clear that channel averaging needs to be modelled during reconstruction to obtain more accurate flux values for both real and imaginary components.}
	\label{fig:reconstructions}
\end{figure*}

\begin{figure*}
\center
	\includegraphics[width=5.75cm]{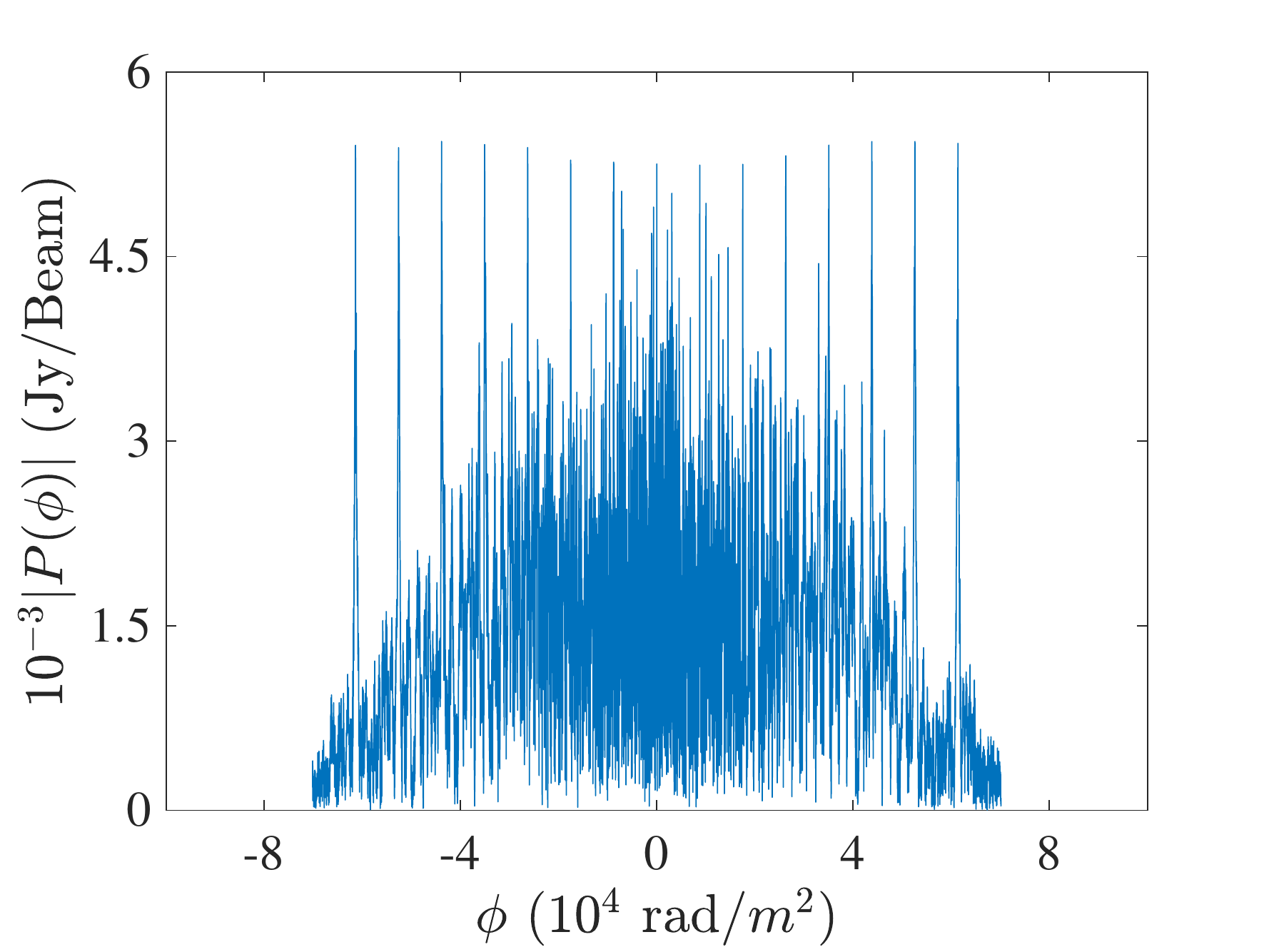}
	\includegraphics[width=5.75cm]{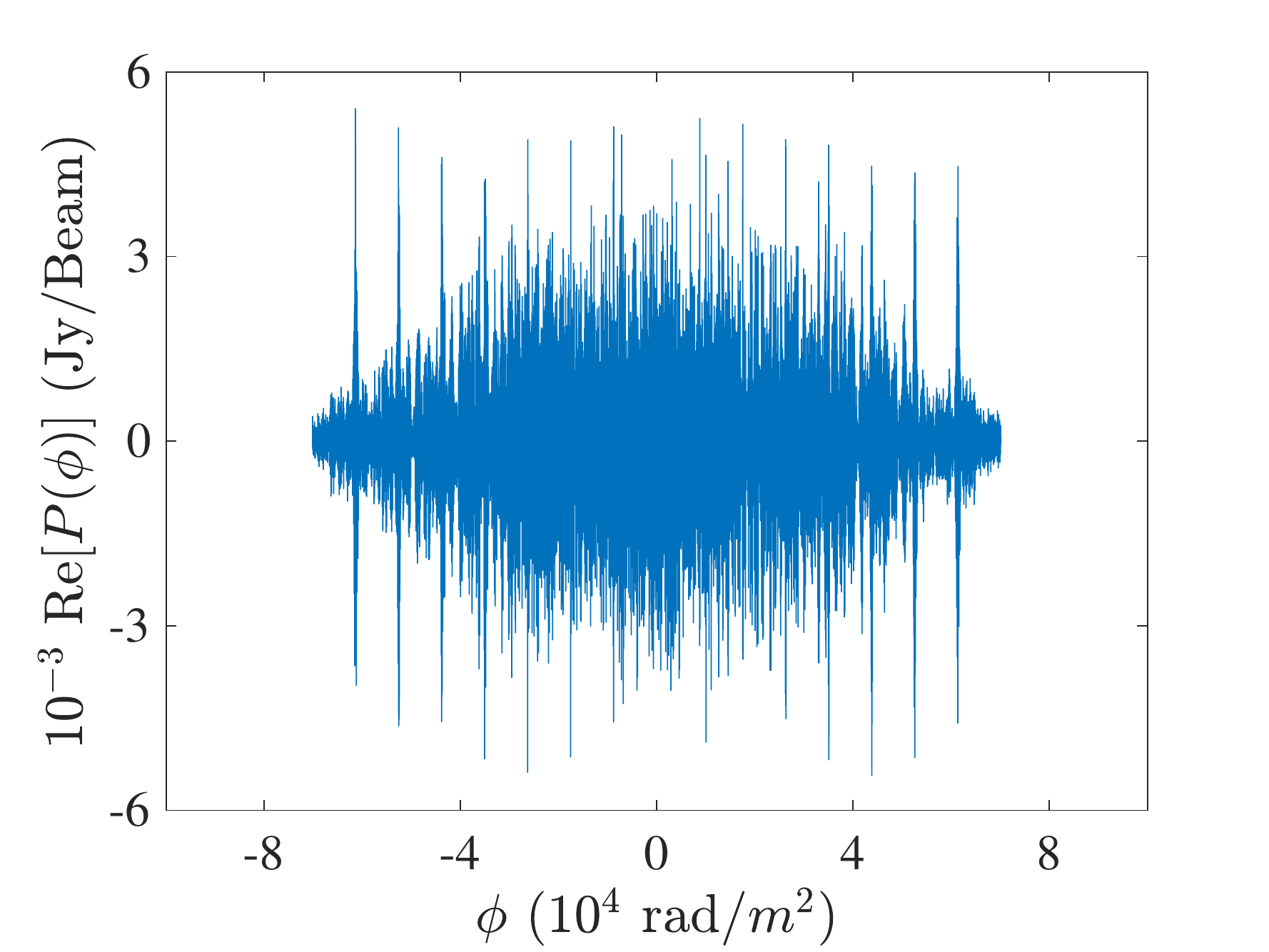}
	\includegraphics[width=5.75cm]{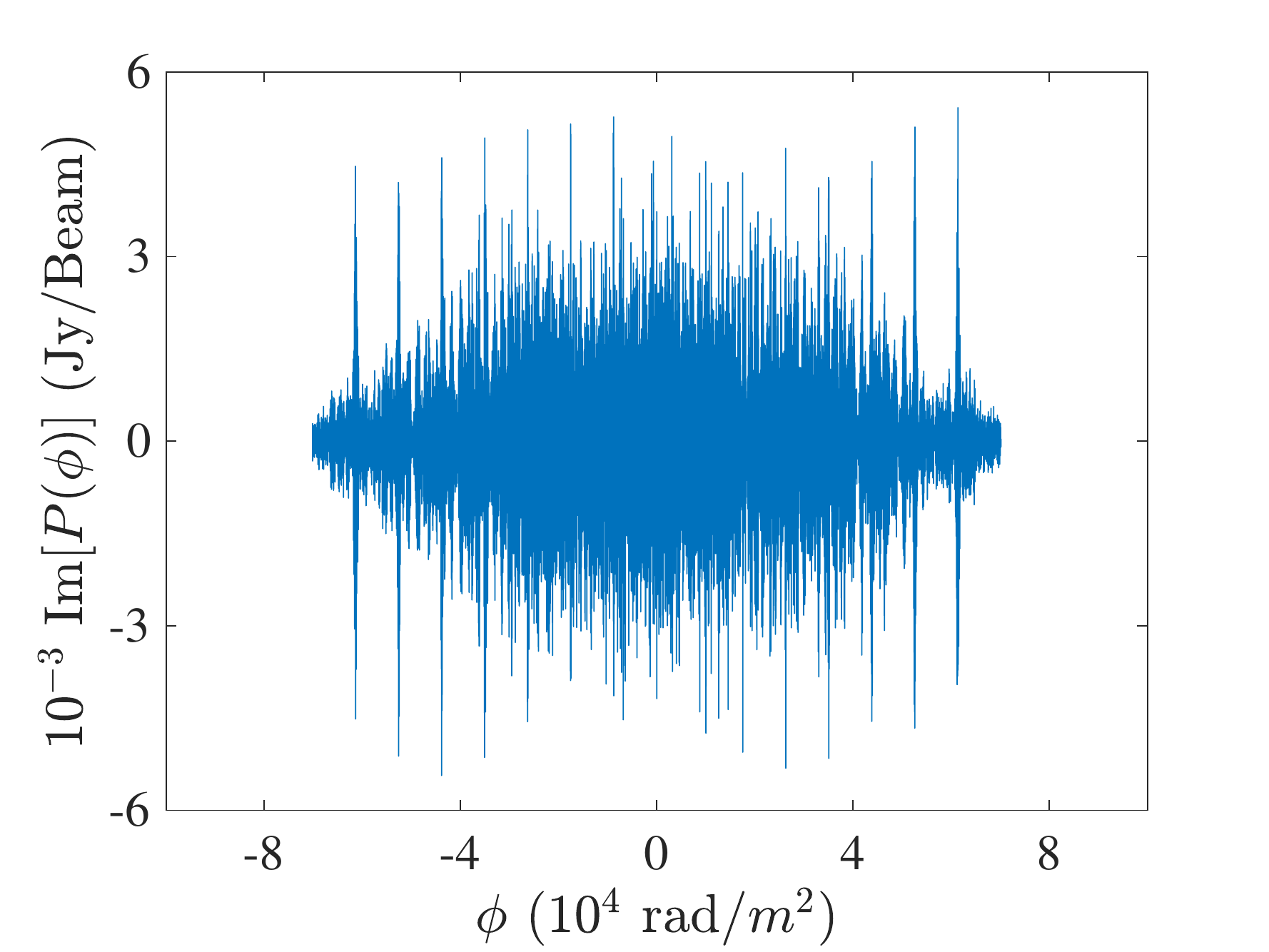}
	\includegraphics[width=5.75cm]{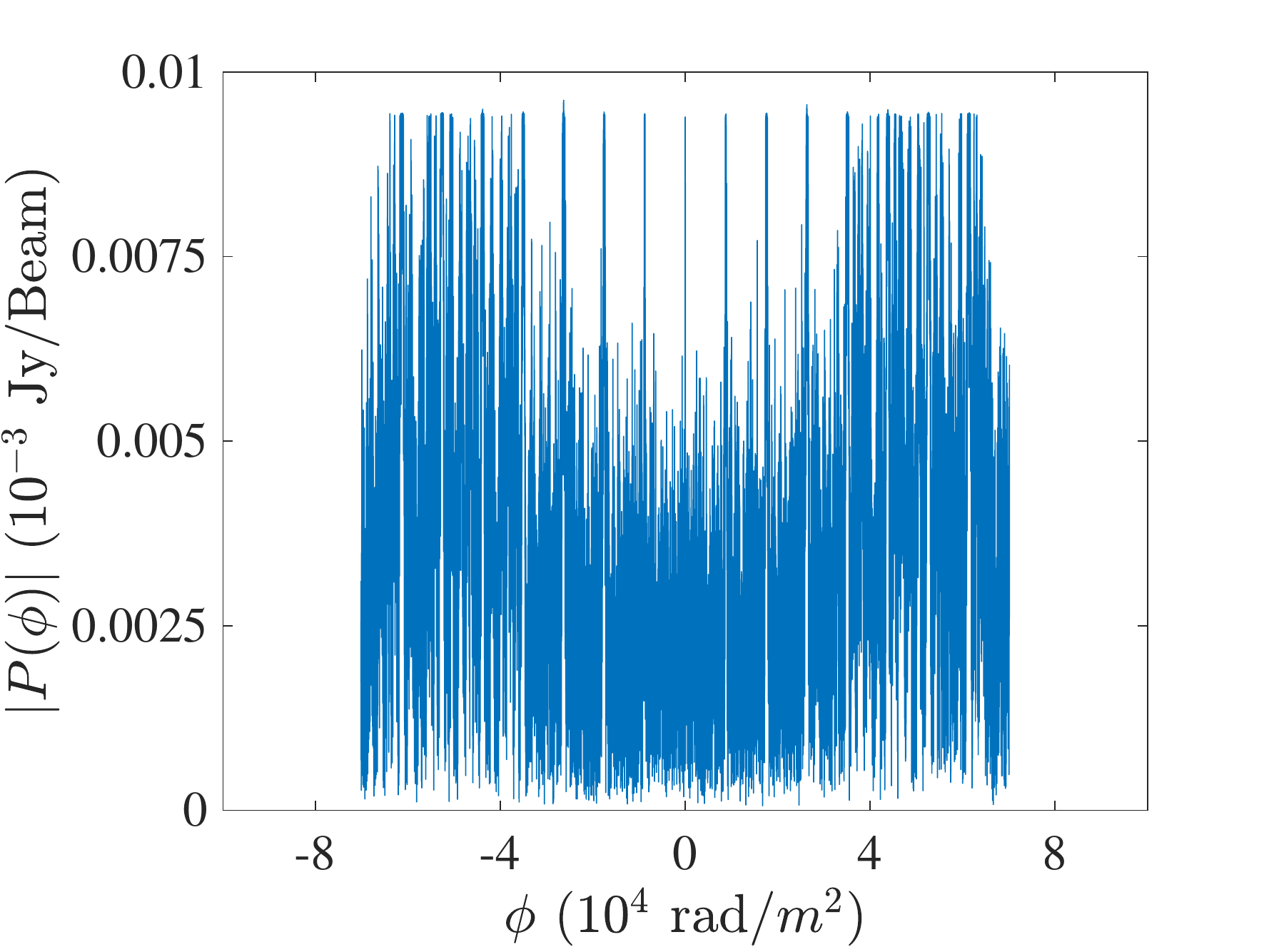}
	\includegraphics[width=5.75cm]{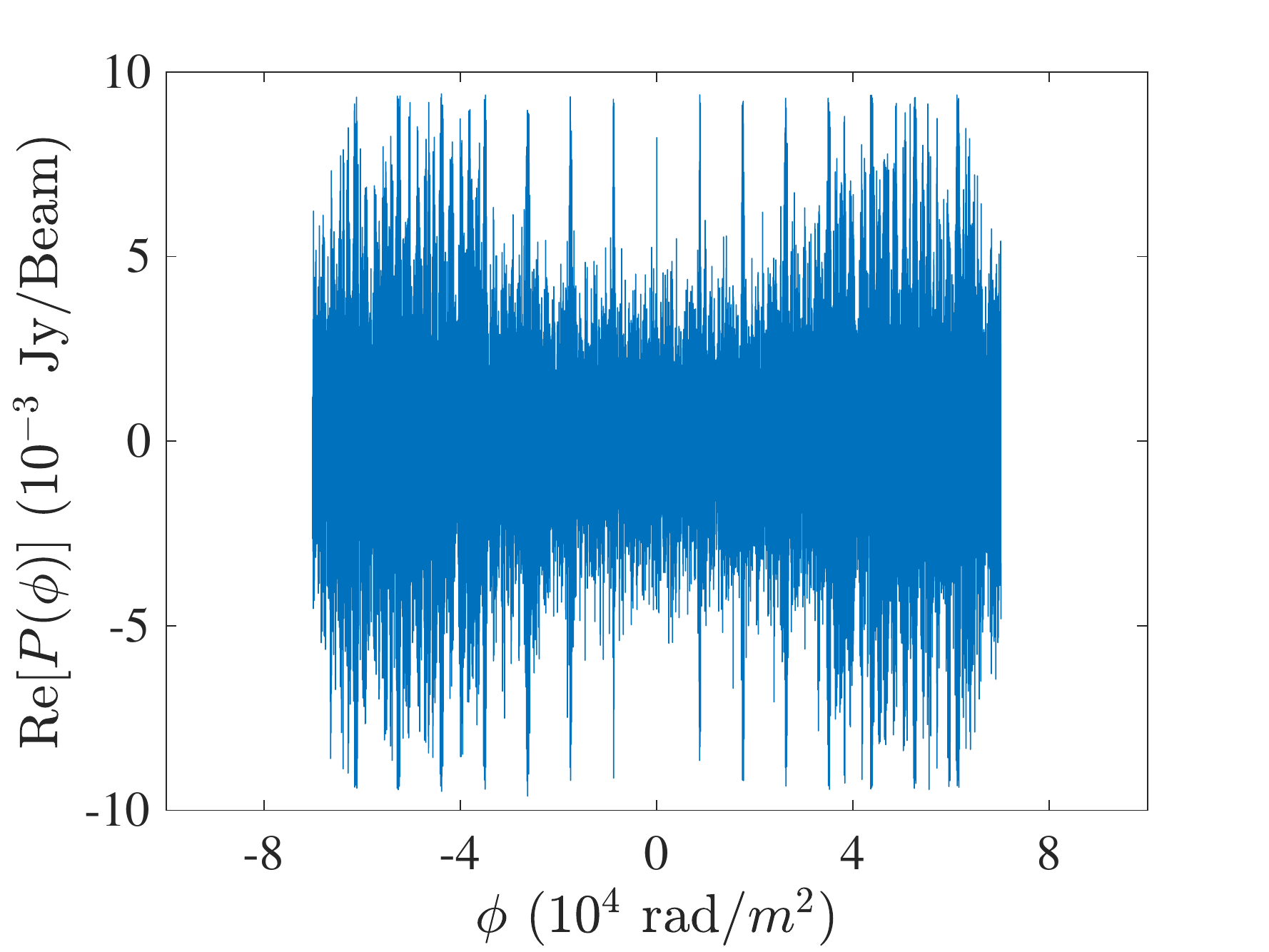}
	\includegraphics[width=5.75cm]{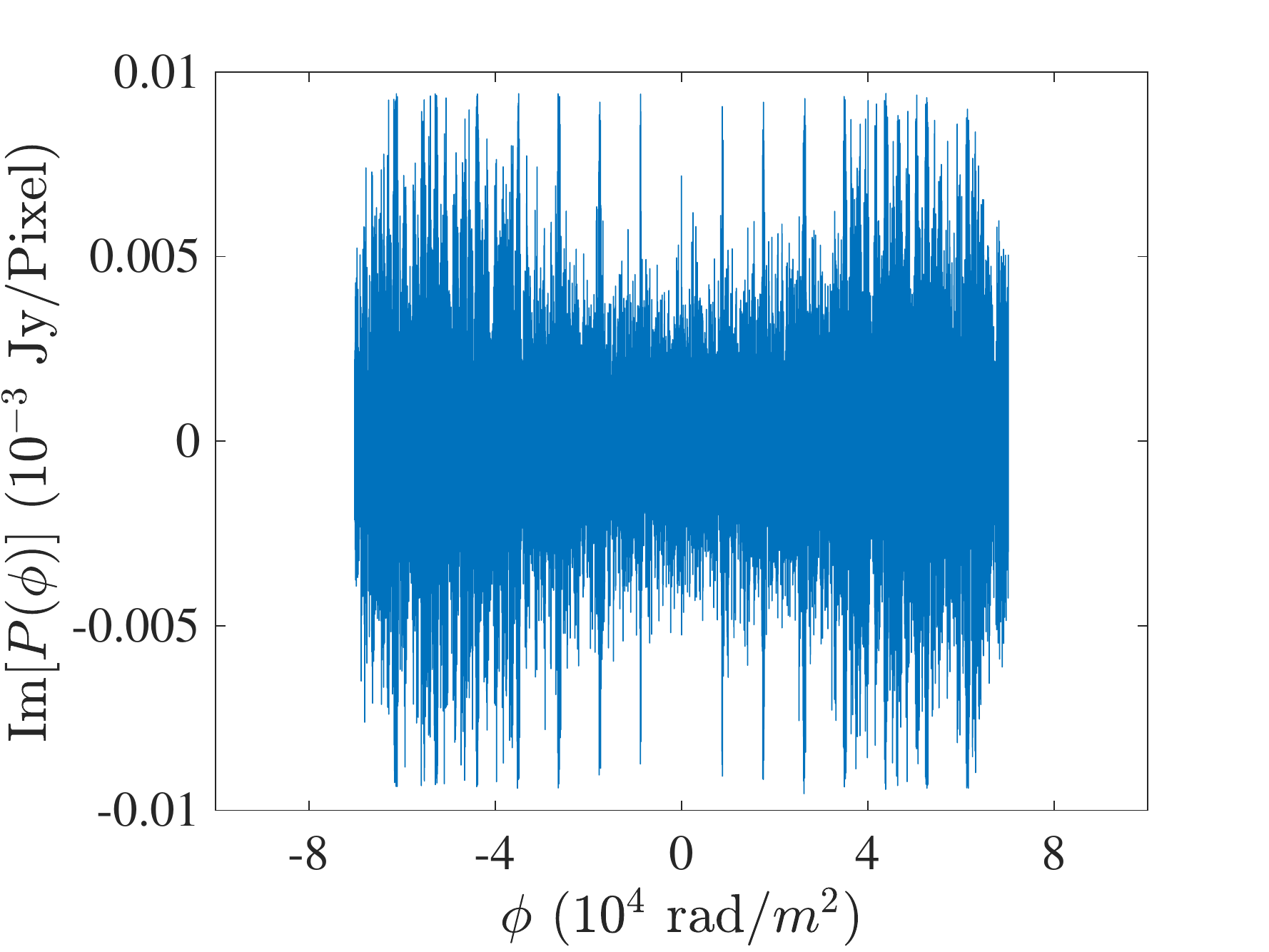}
	\caption{Here we show the residuals corresponding to Figure \ref{fig:reconstructions}. (Top row) shows the residuals where channel averaging was corrected in reconstruction using the POSSUM coverage, with the absolute, real, and imaginary values from right to left. (Bottom row) is the same as the top row without channel averaging corrected during reconstruction. We find that although both residual maps are dominated by noise, most of the signal has been modelled by the fit using both methods. This is to be expected, as it's not only the quality of the modelling fitting that determines the accuracy of the reconstruction. We note that the attenuation is more accurately corrected within the null of the widest sinc function.}
	\label{fig:residuals}
\end{figure*}

In this subsection we show the results of correcting for channel averaging in the Faraday depth reconstruction process for POSSUM, we then compare this to when channel averaging is not corrected, as seen in Figure \ref{fig:reconstructions}. Figure \ref{fig:residuals} then provides the residuals in Faraday depth.

Figure \ref{fig:reconstructions} shows that channel averaging needs to be included during the reconstruction process to correct for the attenuated polarization intensity of sources with large rotation measure values. Whereas Figure \ref{fig:residuals} shows that we have accurately fit the data sufficiently in each case. 

\begin{figure*}
\center
	\includegraphics[width=5.75cm]{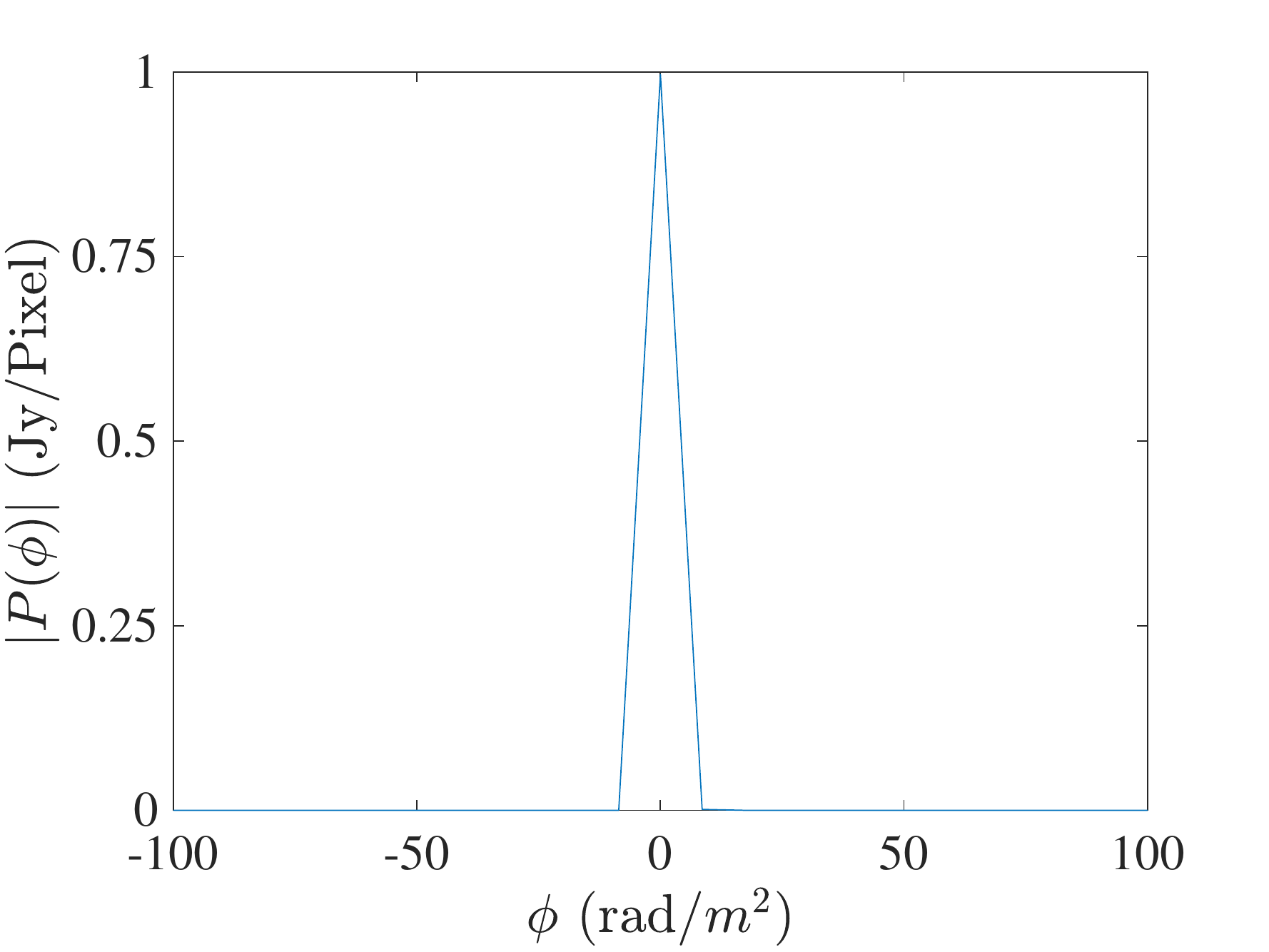}
	\includegraphics[width=5.75cm]{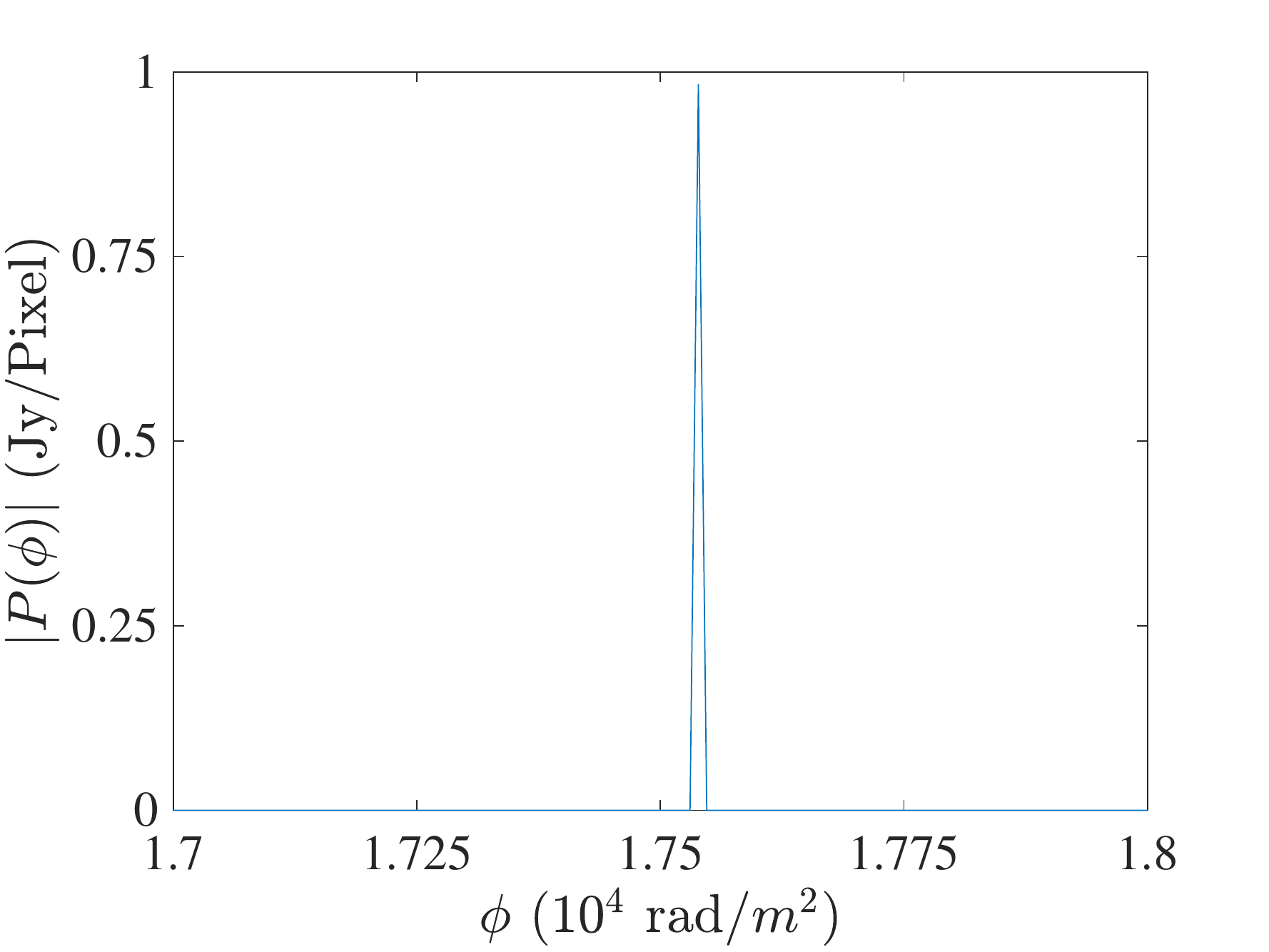}
	\includegraphics[width=5.75cm]{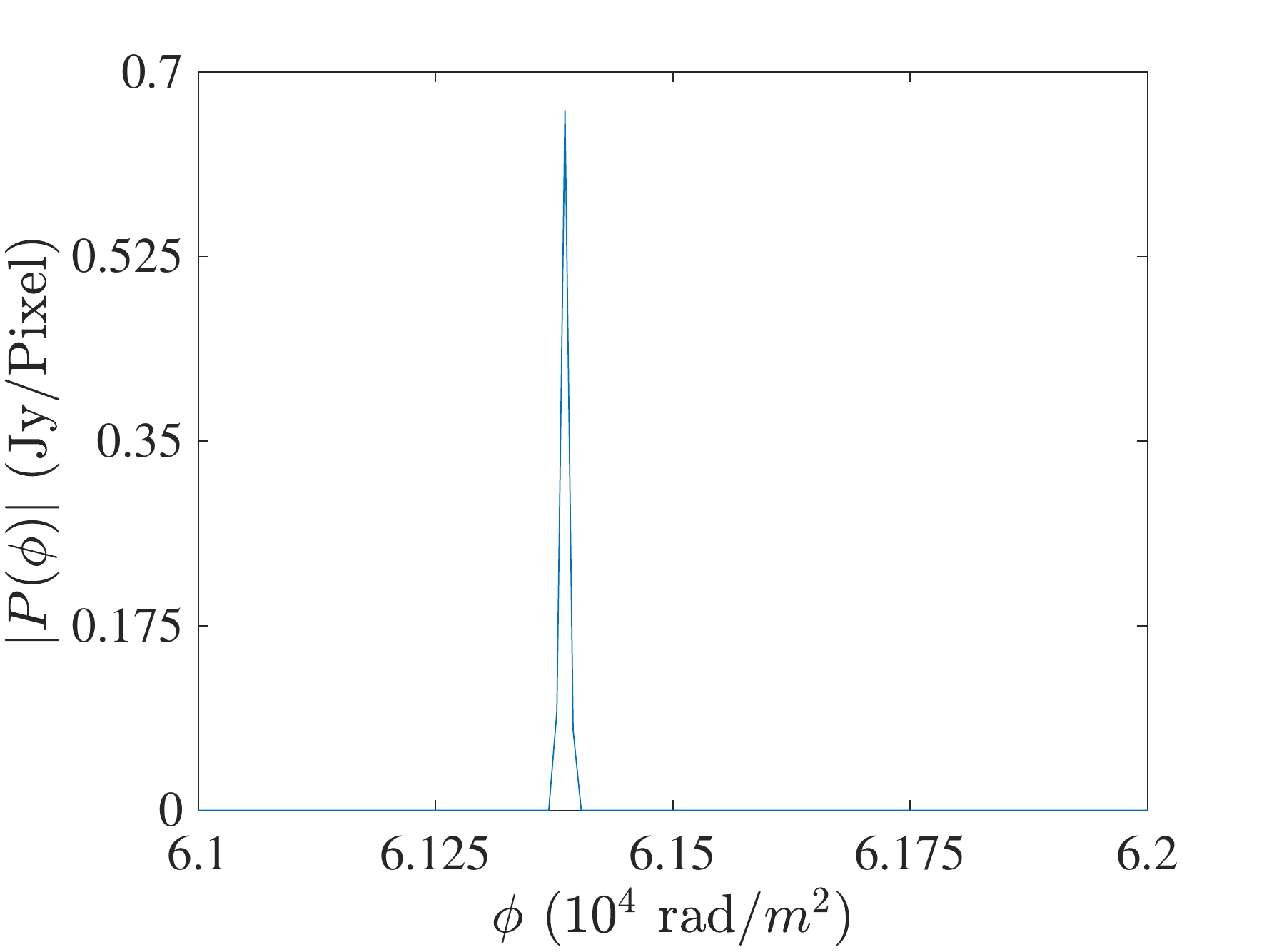}
	\includegraphics[width=5.75cm]{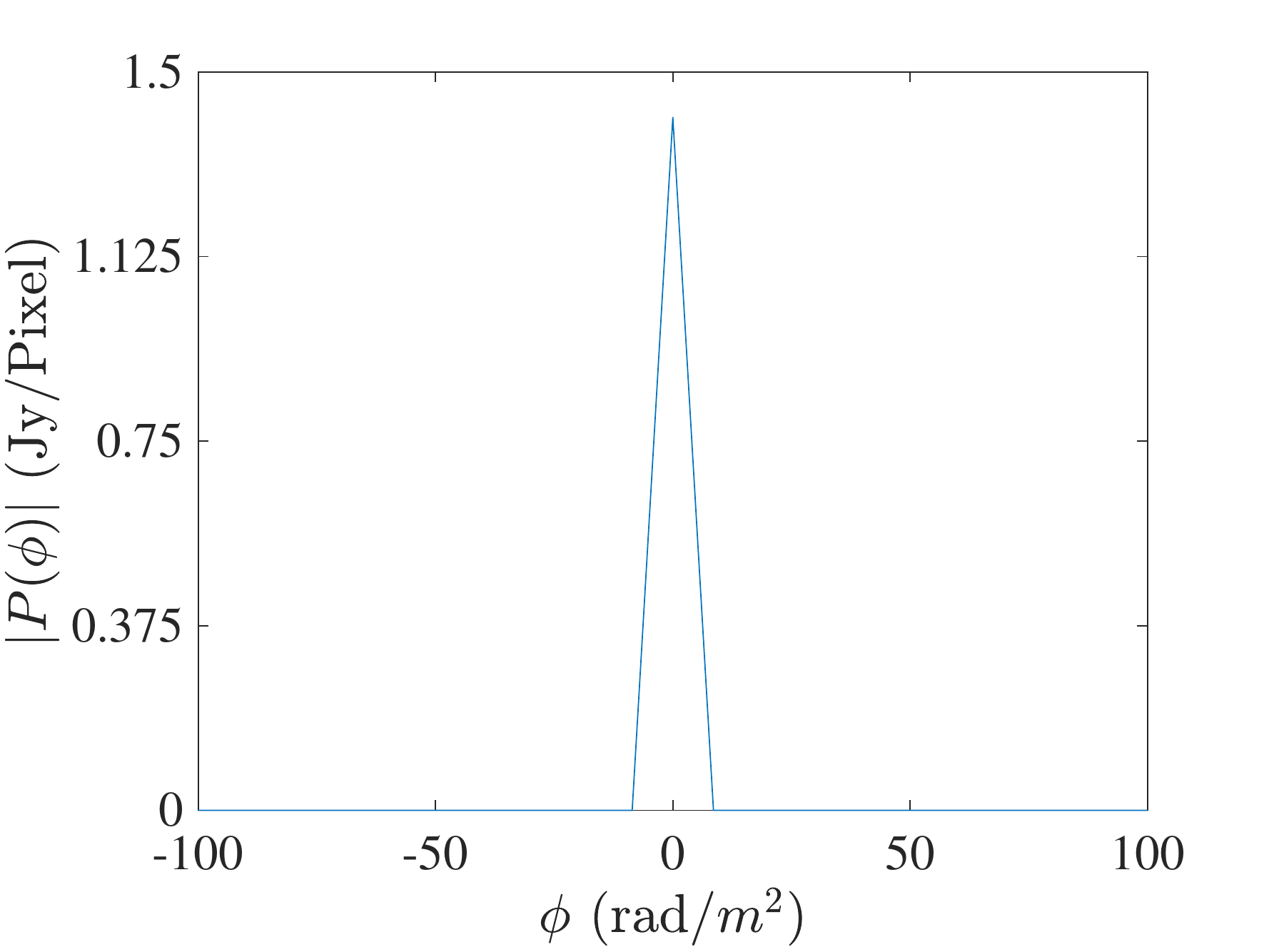}
	\includegraphics[width=5.75cm]{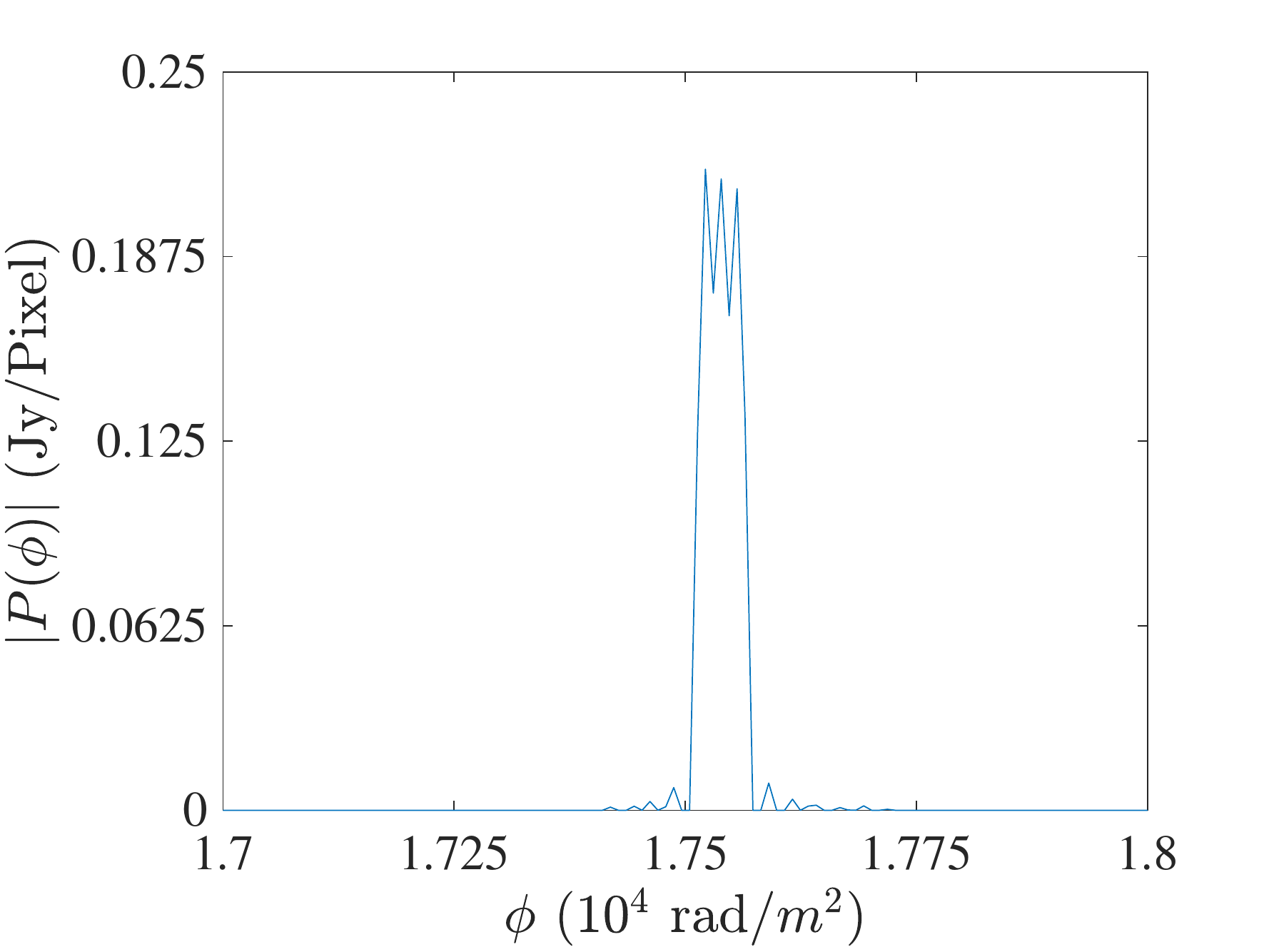}
	\includegraphics[width=5.75cm]{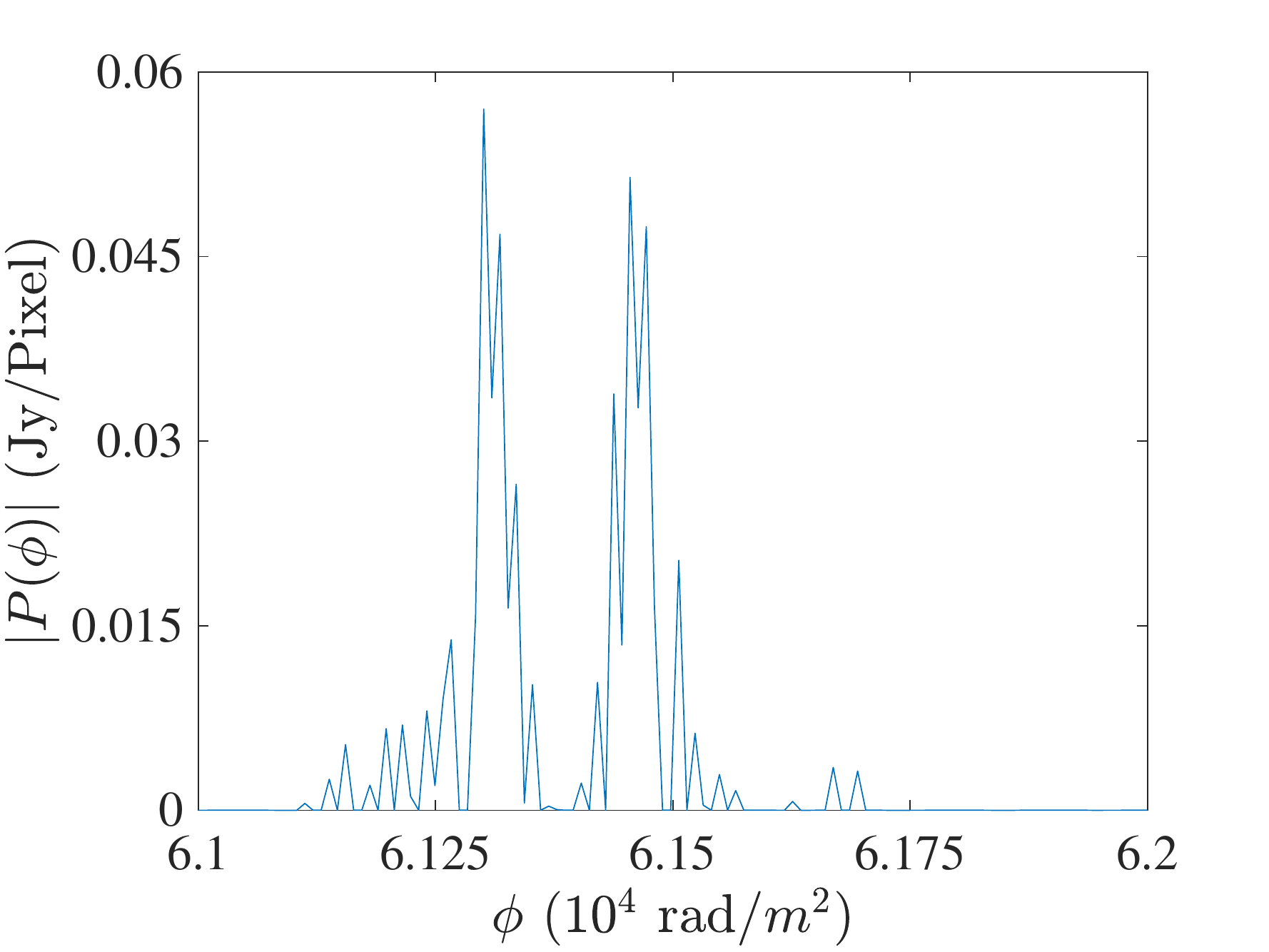}
	\caption{Here we zoom in on selected Faraday sources in absolute polarization intensity, with increasing rotation measure from left to right. The top and bottom rows show these sources reconstructed with and without channel averaging using the POSSUM coverage. At $\phi = 0$ we find that both Faraday sources remain point sources. However, it is clear at much large rotation measure values that not correcting for channel averaging can introduce non-physical structure that is not present in the ground truth or corrected images. In the worst case we find that the non-channel corrected sources can appear as two Faraday sources. This makes it clear that it is not just the attenuation that is corrected by the $\delta \lambda^2$-projection method, it is required for obtaining the correct Faraday depth morphology.}
	\label{fig:comparison}
\end{figure*}

Note that in Figure \ref{fig:comparison} we show a closer look at some of the sources in Faraday depth, where it is clear that sources with large rotation measure have non-physical structure introduced when channel averaging has not been corrected for during reconstruction. This makes it clear that the $\delta \lambda^2$-projection algorithm is not just required for correcting the amplitude, but to ensure the correct characteristics of sources in Faraday depth, and hence the correct rotation measure values. However, the accuracy of the $\delta \lambda^2$-projection algorithm is limited when correcting a Faraday rotation measure signal that is close the noise level. This is the results in greater uncertainty on the corrected flux value. This can be seen where the Faraday rotation is large enough that the channel averaging can depolarize sources to the point of being close to the noise level. 

These results show that sources with large rotation measure values without correction are instrumentally depolarized, causing an under estimate in the linear polarization. Furthermore, such sources lay closer to or below the noise level of the observations, in an analogous way to sources far from the phase center of a primary beam prior to primary beam correction. This makes it clear that instrumental depolarization due to channel averaging of rotation measures needs to be accounted for when predicting and measuring both polarized source counts and flux density values.

Additionally, these results show that unless channel averaging is corrected for during reconstruction, sources with large rotation measure values will appear to have more complex Faraday structure introduced due to the varying channel sensitivity over a wide band. This will artificially inflate the number of complex Faraday sources measured in the current literature.

Moreover, being able to correct for the sensitivity loss due to channel averaging suggests that it is possible to recover larger rotation measure signals when increased channel averaging is unavoidable. This is especially useful for next generation radio interferometric telescopes, where channel averaging is necessary to reduce the amount of data.

Finally, wide-band rotation measure synthesis provides the ability to combine wide-band polarimetric data from different telescopes and observations. The channel averaging response $A$ is tailored for each channel, allowing for arbitrary channel widths $\delta \lambda^2_i$ for each channel to be corrected for in Faraday depth.

\subsubsection{Case II: Low Frequency Polarization}
\begin{figure*}
\center
	\includegraphics[width=5.75cm]{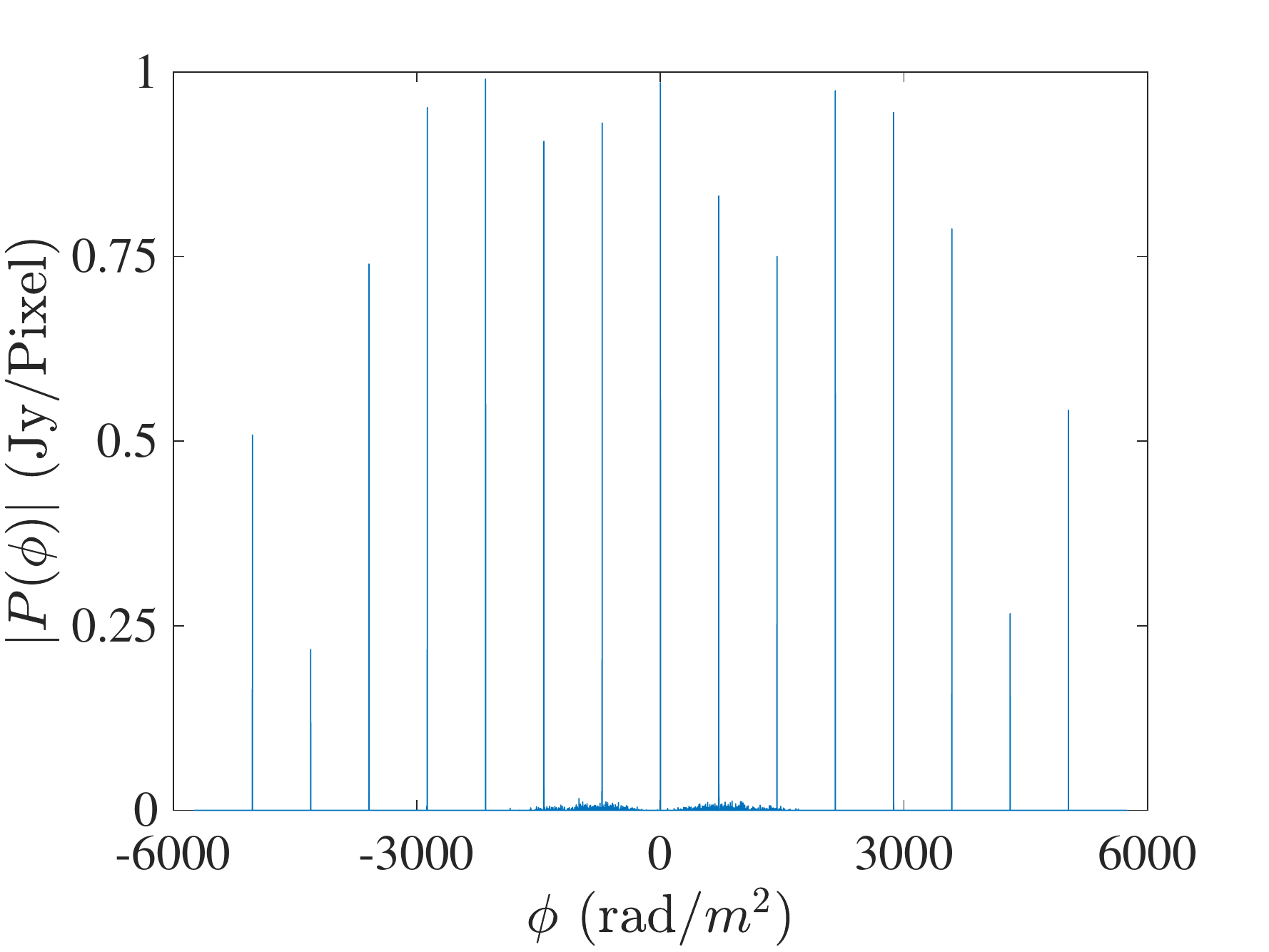}
	\includegraphics[width=5.75cm]{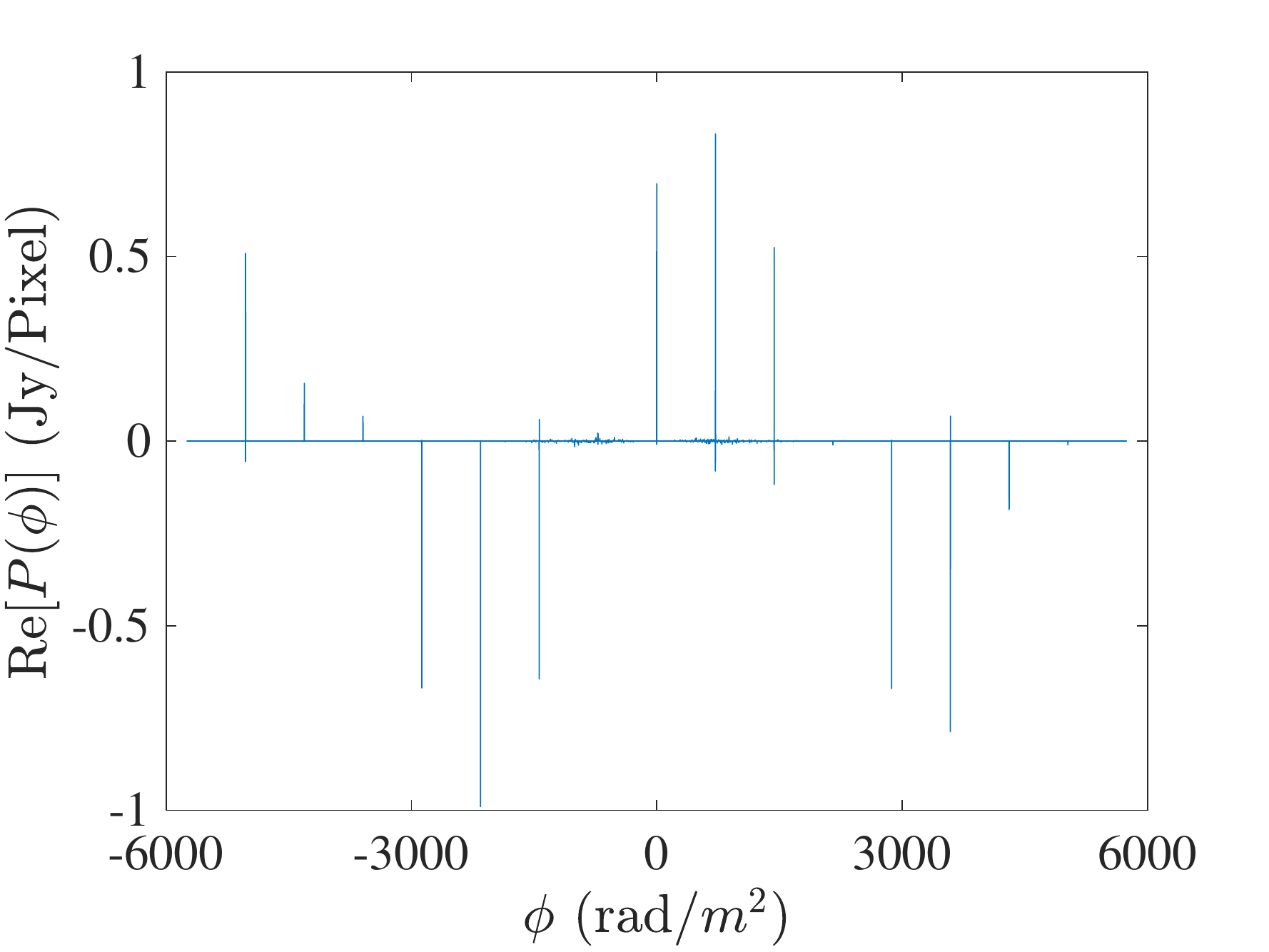}
	\includegraphics[width=5.75cm]{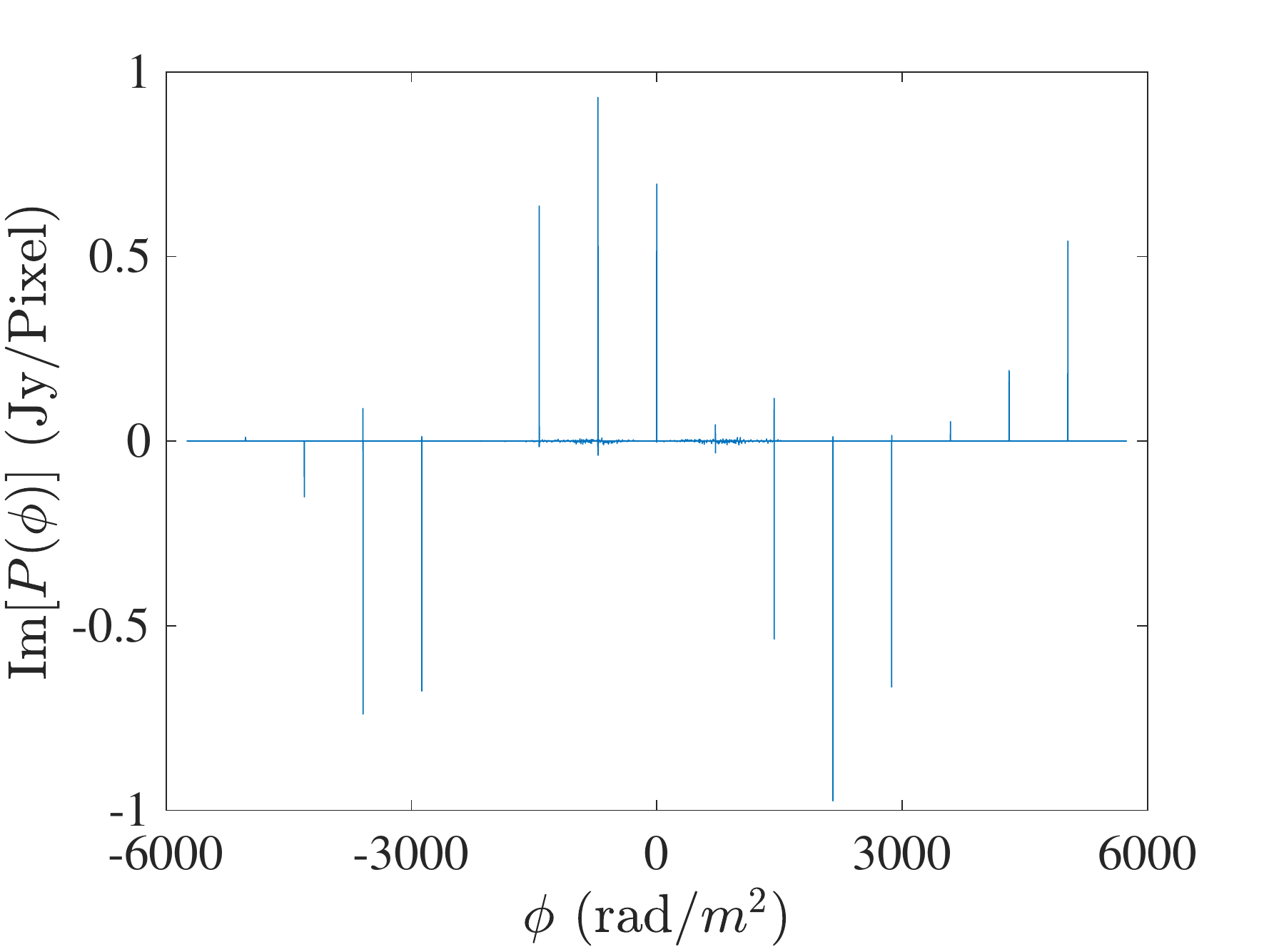}
	\includegraphics[width=5.75cm]{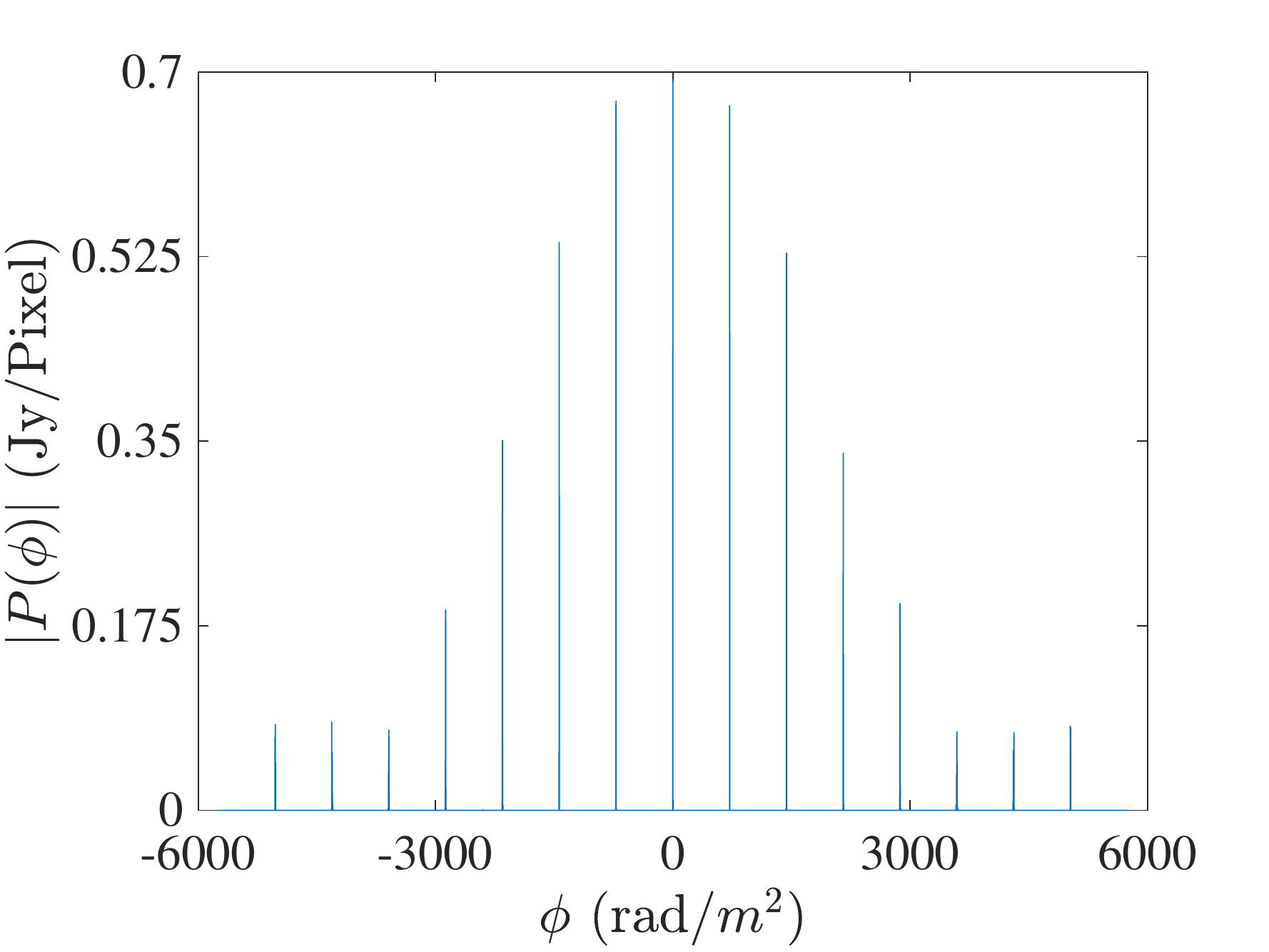}
	\includegraphics[width=5.75cm]{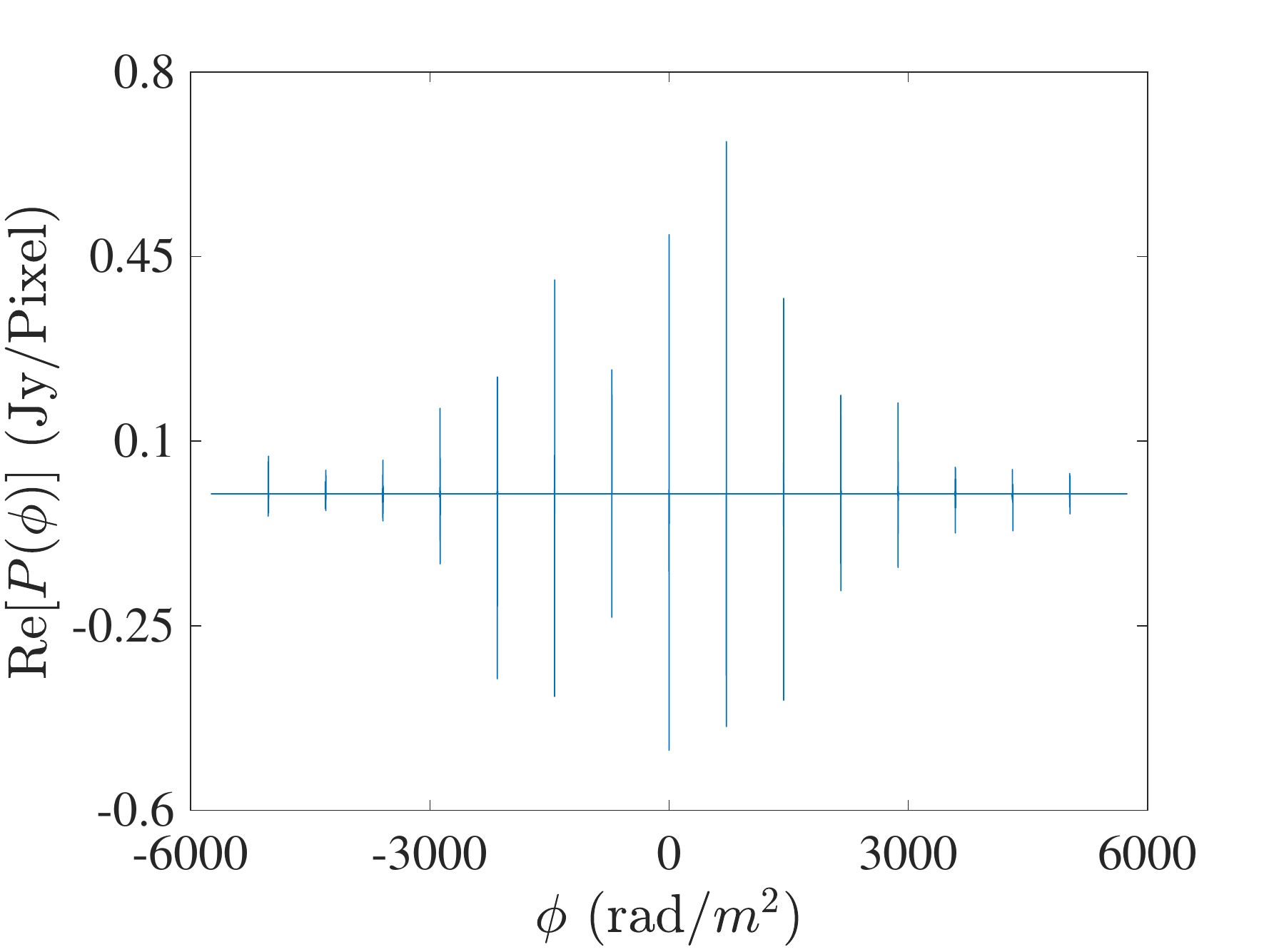}
	\includegraphics[width=5.75cm]{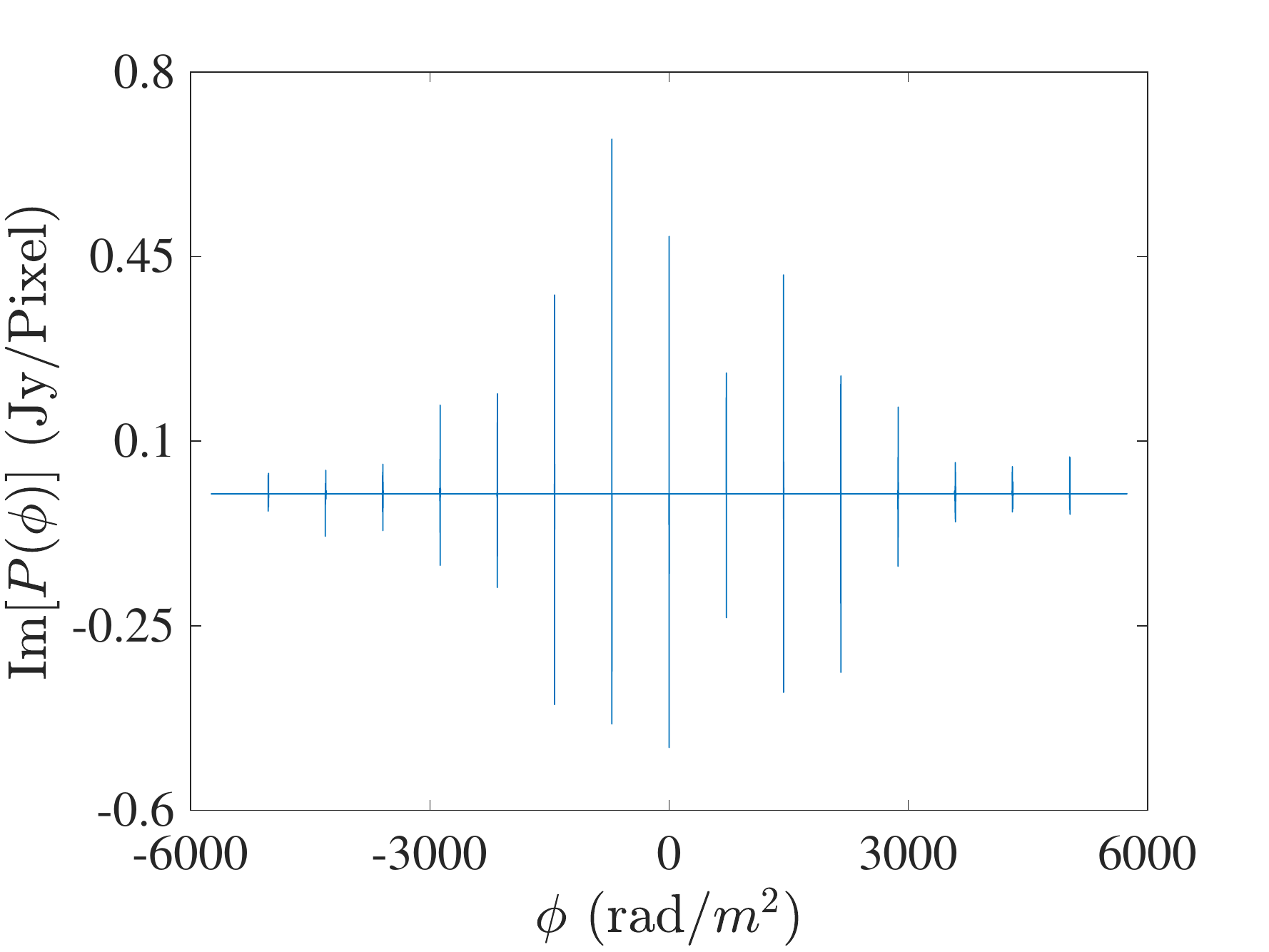}
	\caption{(Top row) shows the reconstruction that corrects for channel averaging in POGS, with the absolute, real, and imaginary values from right to left. (Bottom row) is the same as the top row without channel averaging corrected during reconstruction. It is clear that channel averaging needs to be modelled during reconstruction to obtain more accurate flux values for both real and imaginary components.}
	\label{fig:low_reconstructions}
\end{figure*}

\begin{figure*}
\center
	\includegraphics[width=5.75cm]{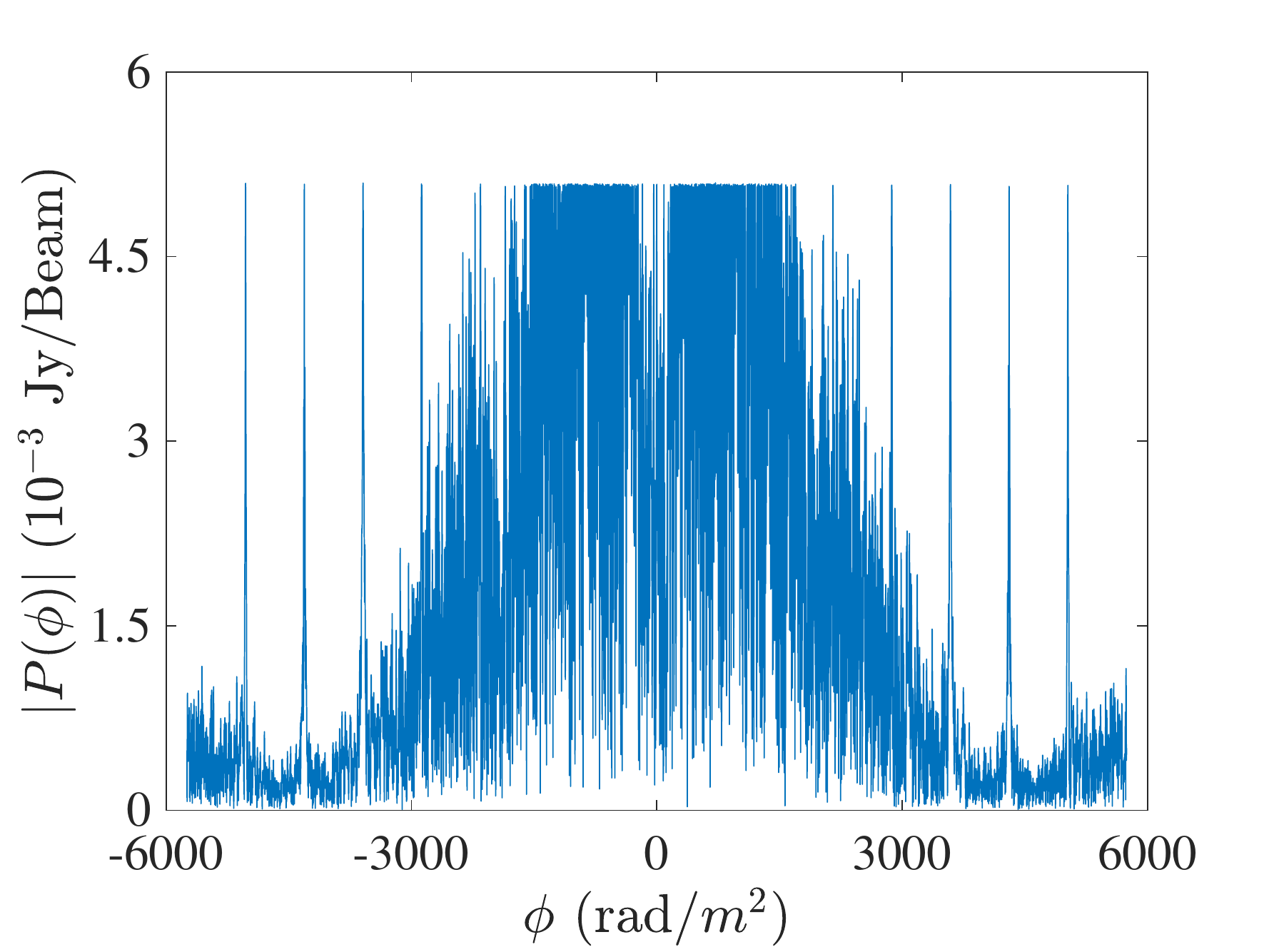}
	\includegraphics[width=5.75cm]{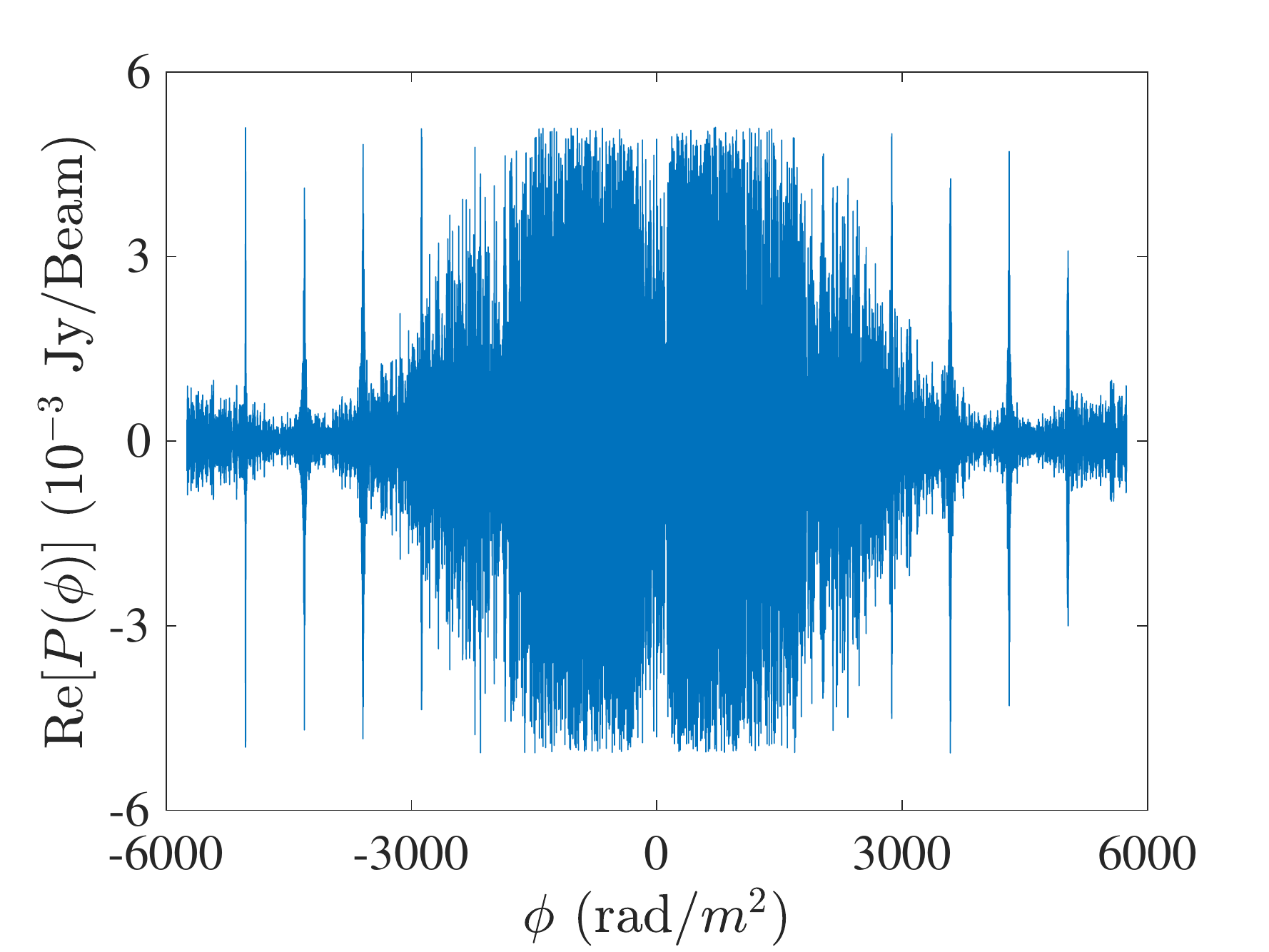}
	\includegraphics[width=5.75cm]{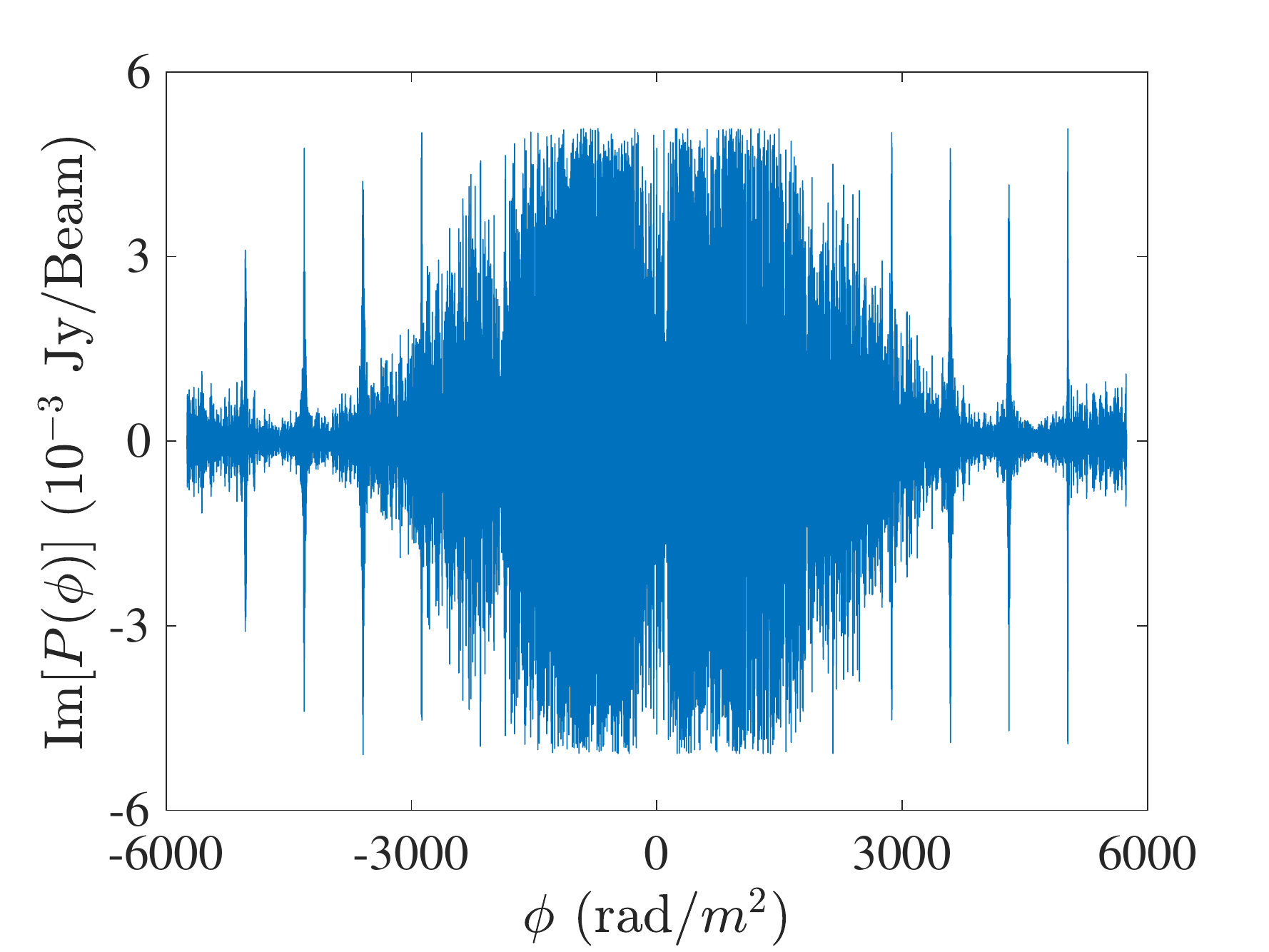}
	\includegraphics[width=5.75cm]{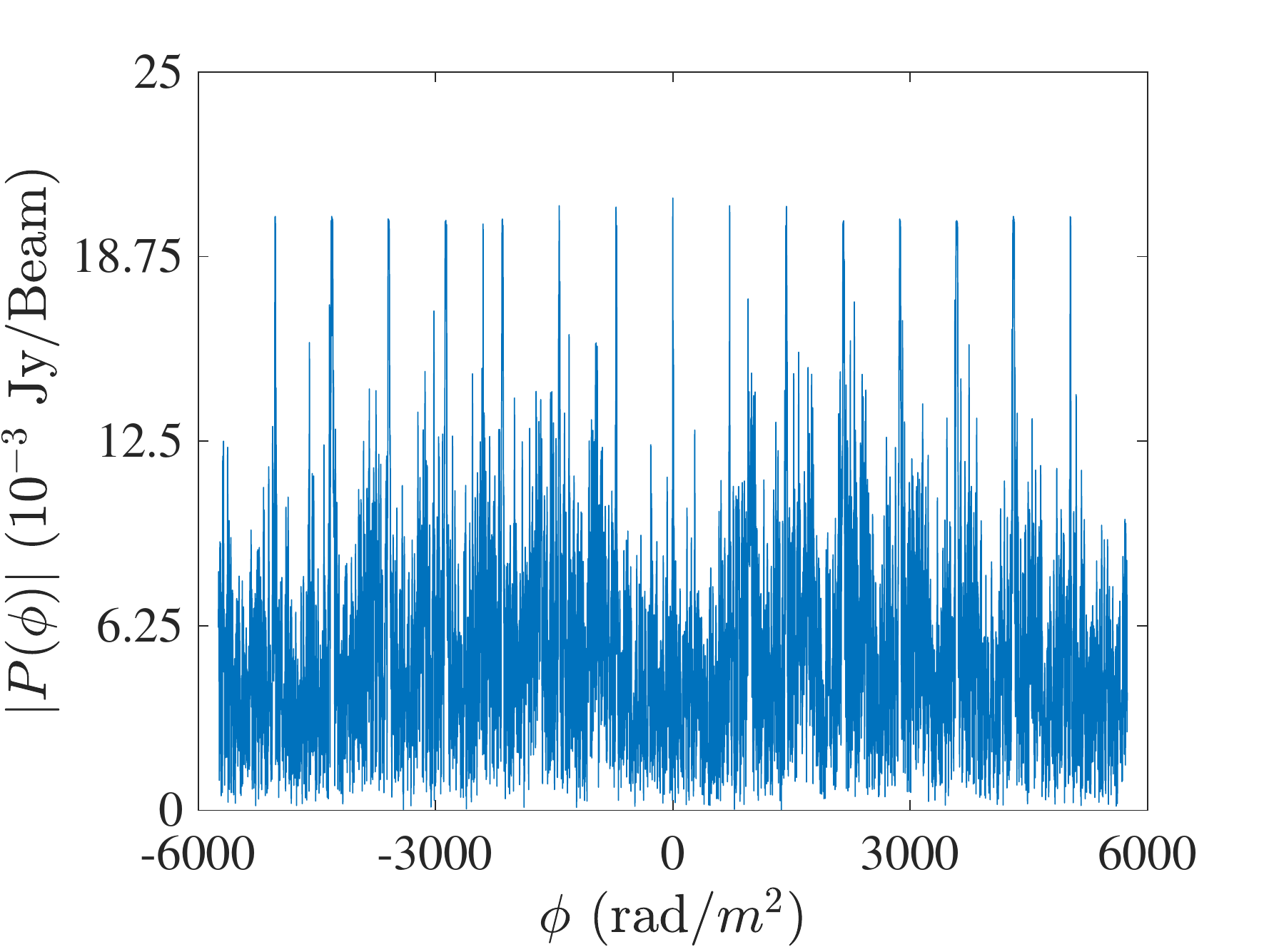}
	\includegraphics[width=5.75cm]{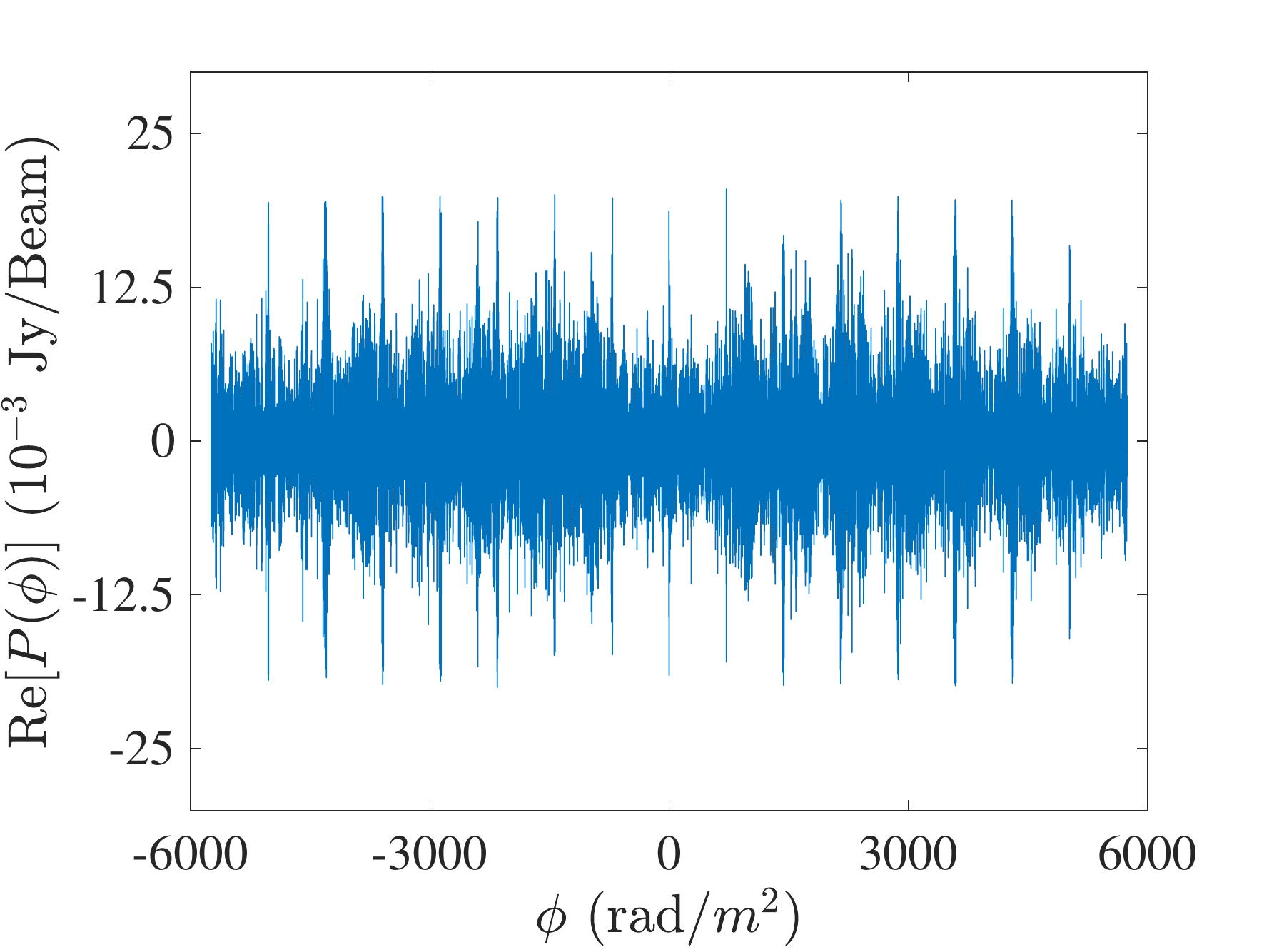}
	\includegraphics[width=5.75cm]{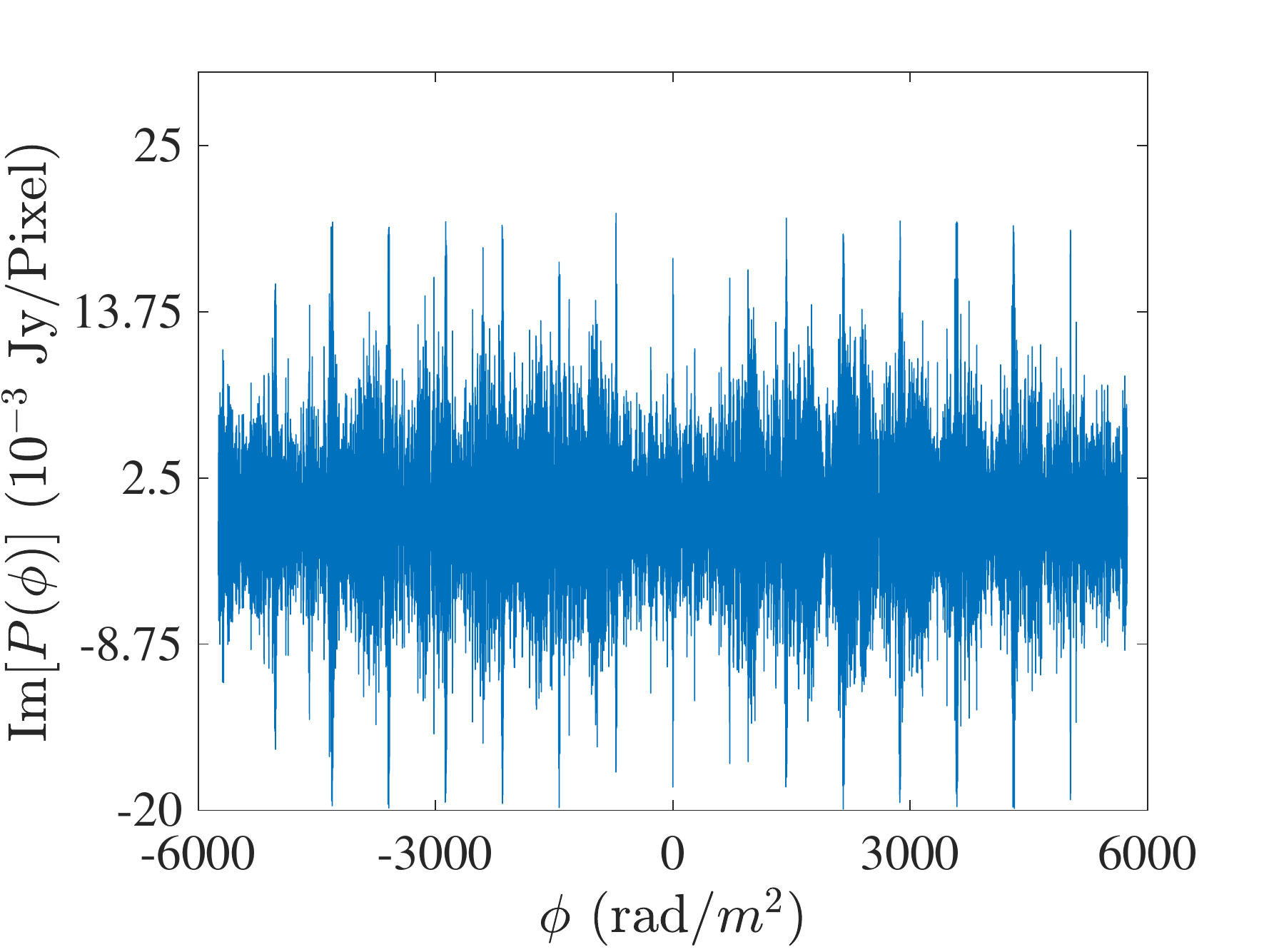}
	\caption{Here we show the residuals corresponding to Figure \ref{fig:low_reconstructions}. (Top row) shows the residuals where channel averaging was corrected in reconstruction, with the absolute, real, and imaginary values from right to left for POGS. (Bottom row) is the same as the top row without channel averaging corrected during reconstruction. We find that although both residual maps are dominated by noise, most of the signal has been modelled by the fit using both methods. This is to be expected, as it's not only the quality of the modelling fitting that determines the accuracy of the reconstruction.}
	\label{fig:low_residuals}
\end{figure*}

The reconstruction with and without channel averaging for the POGS simulation can be seen in Figure \ref{fig:low_reconstructions} and the residuals can be seen in Figure \ref{fig:low_residuals}. The results again show that without correction there is: i) an underestimate of the total linear polarized intensity for large rotation measure values, meaning fewer sources will be detected, and ii) sources with large RMs will have artificially complicated RM structure. We find that the large RM sources become artificially broaden at larger Faraday depth.

Additionally, if we compare the accessible RM range for the non-corrected and corrected observations we see that corrections allows a considerably larger RM regime to be probed. For example, in the traditional method \citet{rise18} reports POGS is sensitive to $|\phi_{\rm max}|$ = 1,937 rad/m$^{2}$ whereas here we can probe RMs out to the first null described in Equation \ref{eq:chan_sensivity}, which is 5,389 rad/m$^{2}$ for the top end of the band. 

We note that the units of flux density in the reconstructions are in Jy/Pixel, meaning that if a Faraday thin source is not localized due to lack of sampling in $\lambda^2$, lack of signal, or lack of reconstruction algorithm iterations, the peak will appear lower than the ground truth because the flux is spread across multiple pixels is Faraday depth. Reconstruction algorithms have the ability to determine localization better than the dirty map through deconvolution, but this limited by the signal-to-noise ratio of the source. For bright signals the localization can be limited to a single pixel, but for faint sources the total flux is spread across many pixels. However, the total flux integrated across the source should remain the same within the noise. We find that the total flux of the reconstructed map is consistent with the ground truth.

\section{Shifting Faraday Phase Centre and Faraday Mosaicking}
\label{sec:mosaic}
In this section we discuss how performing a phase shift before channel averaging can be used to shift the sensitivity window in Faraday depth of each channel and how this could lead to mosaicking in Faraday depth.

Before averaging (down-sampling) of frequency channels is performed with the use of an averaging (anti-aliasing) kernel $A$, a phase shift can be applied to the kernel to shift the sensitivity window in Faraday depth. This can be done be understood through the use of the convolution theorem and the translation (operator) relation
\begin{equation}
    y(u) \star \left [A(u, \delta u) {\rm e}^{-2\pi i u \phi_0}\right] \Leftrightarrow x(\phi) a(\phi - \phi_0, \delta u) \, .
\end{equation}

\begin{figure}
\center
	\includegraphics[trim=30 0 0 0,clip,width=9.5cm]{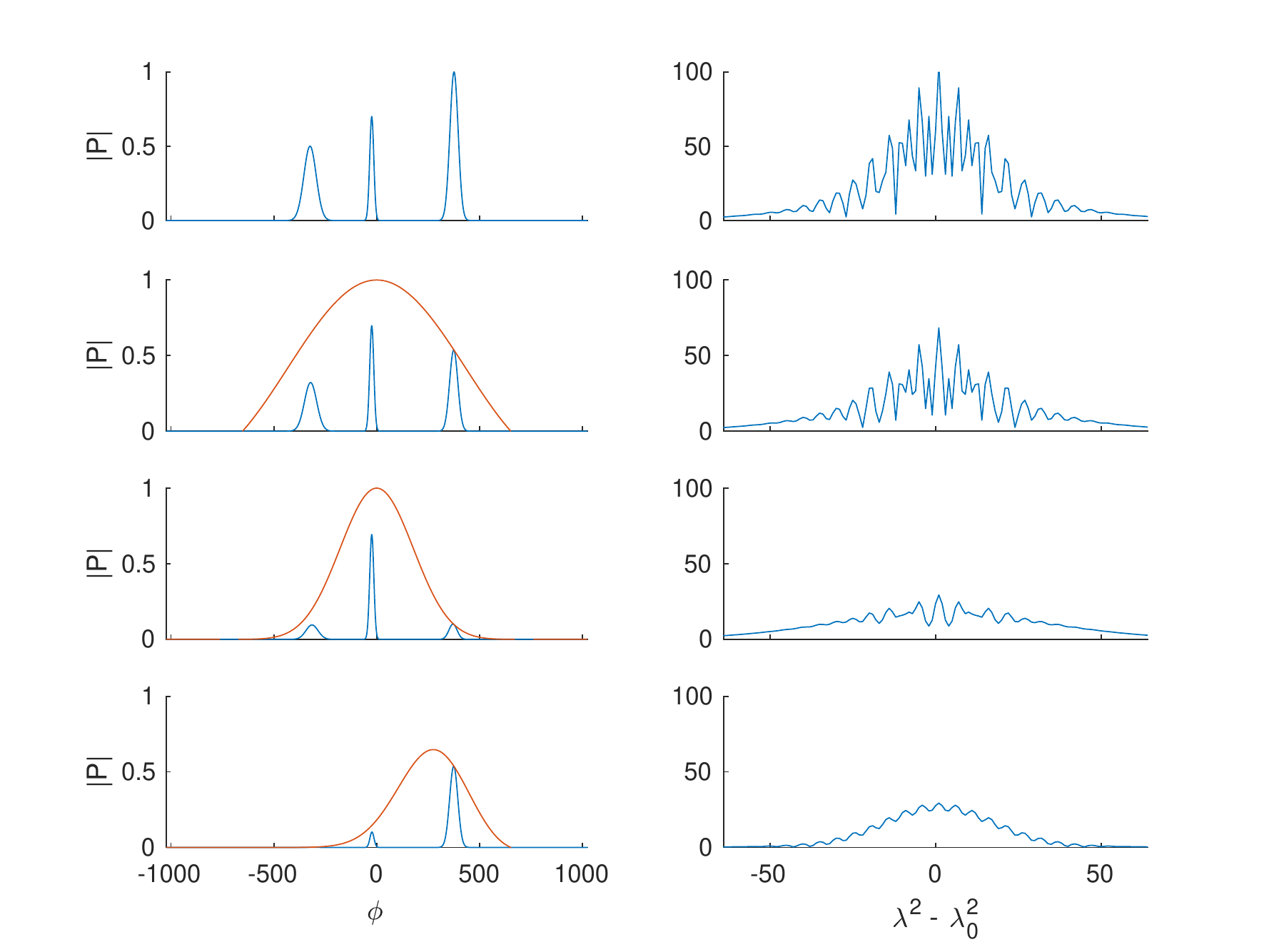}
	\caption{(Left column) Faraday spectrum with different window functions. (Right column) Corresponding spectrum in $\lambda^2$. (Top) shows the effect of no instrumental window, providing high resolution in $\lambda^2$ and no attenuation at high $\phi$. (Top Middle) A sinc window is applied due to instrumental channel averaging, which is equivalent to convolving the spectrum with a box function in $\lambda^2$-space. (Bottom Middle) An anti-aliasing kernel is convolved in $\lambda^2$, simulating extra channel averaging. The channels are averaged by a factor of 24, this additional channel averaging greatly reduces sensitivity to high values of $|\phi|$. (Bottom) A phase shift can be applied to the convolution function used for averaging the channels in $\lambda^2$, translating the sensitivity to the scientific region of interest at high $|\phi|$ values. However, instrumental channel averaging still limits sensitivity at large $|\phi|$. This is because the sensitivity is determined by both the instrumental averaging and additional channel averaging. In this figure, channel width is constant as a function of $\lambda^2$ for demonstration purposes.}
	\label{fig:rm_mosaicking}
\end{figure}

If this is done at the correlator stage of averaging it is possible to shift the kernel's window in Faraday depth before down-sampling. The window $a$ is expected to be enveloped by a wider window that is due to the raw sampling channel width before down-sampling, this larger window sets the limit on how far we can shift $\phi_0$ before all signal is lost. However, this allows one to maximize sensitivity for a given region of Faraday depth while channel averaging, as shown in Figure \ref{fig:rm_mosaicking}. This could be extremely useful when data needs to be averaged or reduced from observations of next generation surveys, allowing one to increase the amount of channels averaged to reduce data but still maintain sensitivity to large rotation measure values. This averaging can then be corrected during reconstruction using the $\delta \lambda^2$-projection algorithm.

This naturally leads to Faraday mosaicking of sensitivity windows. This idea is again analogous to one of the standard techniques in interferometry used overcome the limits of an antenna's field of view, where one can combine images from different regions of the sky (known as pointings) via mosaicking \citep{Ekers79}. In the case of interferometric imaging, by arranging the points of a mosiac such that they perform a Nyquist sampling of the sky \citep{Sault96}, mosaicing provides the capacity to recover spatial information which would otherwise be lost via the deconvolution of temporally and spatially sensitive pointings. 

In the RM context, it is possible to combine different observations with different spectral channels and also to phase shift to different regions of Faraday depth. In the same way as one would select pointing centres in an image mosaic to Nyquist sample the sky, if the instrument is configured to have channelisation which correctly Nyquist samples in Faraday depth, one can then recover information about the Faraday depth spectrum at higher frequencies than would otherwise be possible. This could be critical for dealing with big data challenges in next generation polarimetric surveys, where channel averaging could be unavoidable.

\section{Conclusion}
\label{sec:conclusion}
A number of wide-area radio polarimetric surveys are currently planned for the next generation of radio telescopes including POSSUM on ASKAP \citep{gae10}, VLASS on the JVLA \citep{Mao14}, the Global Magneto-Ionic Medium Survey (GMIMS; \citealp{wol19}) on Parkes, and the RM Grid experiment for the SKA \citep{mjh15}. In addition, polarization products are being derived from current intensity surveys such as POGS from GLEAM \citep{rise18} or the point source polarization catalogues from the LOFAR Two-metre Sky Survey \citep{Osul18,vanE19}. All of these efforts correspond to either wide bands or low frequencies or both. 

Here we have demonstrated that without channel-averaging corrections the RMs derived from these surveys will fail to reach the full potential of the data by: i) detecting fewer high RMs sources due to a reduction in polarized intensity, ii) not probing the full range of high RMs accessible to the data, and iii) not correctly recovering the correct RMs and generating false complexity in Faraday depth space. This has already likely resulted in an underestimate of both polarized source counts and the RM distribution in historical polarization work, particularly at low frequencies \citep{Lenc17,Osul18,vanE19}, whilst simultaneously likely overestimating the number of Faraday complex sources with high RMs \citep{Pas18}.

Moreover, it is not just the science of polarimetric source counts and Faraday complexity that will be enhanced by the methods presented here. Such techniques are vital both to the study of extreme environments with high magnetic fields, such as those seen in some pulsars and Fast Radio Bursts (FRBs) \citep{2018Natur}, and the use of these objects as environmental probes. FRBs, in particular, provide one of the best methods to constrain magnetic field properties in the cosmic web and intra-cluster medium \citep{aka16,2016Sci,vaz18}, and without accurately accounting for channel-averaging, we diminish our opportunity to use such sources to investigate cosmological magnetic fields. 

This work thus establishes the first full framework to perform rotation measure synthesis of truly wide-band signals via the introduction of the $\delta \lambda^2$-projection algorithm which allows efficient modelling of channel dependent sensitivity in Faraday depth. We further demonstrate how polarimetric data from different telescopes can be co-added to increase sensitivity and coverage to RMs.

As we enter the big data era of radio astronomy, it is vital to understand the effects of channel averaging and instrumental depolarization if we are to realize the science goals of ambitious new instruments like the SKA. With both new wide-band and low frequency polarimetric surveys coming online, the methodology presented here going to be critical for the analysis of cosmic magnetic fields of all scales.

% unnumbered section
\section*{Acknowledgements}
LP and MJ-H thank the anonymous referee for their careful reading of the manuscript and comments which improved the clarity of Section 5, in particular.
MJ-H thanks Dave Gahan, Martin Gore, and Andy Fletcher for providing the Faith \& Devotion vital for the completion of this manuscript. The Dunlap Institute is funded through an endowment established by the David Dunlap family and the University of Toronto. 
\bibliographystyle{aasjournal}
\bibliography{refs}

\end{document}